\documentclass[fleqn,usenatbib]{mnras}

\usepackage{newtxtext,newtxmath}
\usepackage{tablefootnote}

\usepackage[T1]{fontenc}

\DeclareRobustCommand{\VAN}[3]{#2}
\let\VANthebibliography\thebibliography
\def\thebibliography{\DeclareRobustCommand{\VAN}[3]{##3}\VANthebibliography}

\usepackage{graphicx}	
\usepackage{amsmath}	
\usepackage[flushleft]{threeparttable}

\title[Bar Properties as a Function of Wavelength]{Bar Properties as a Function of Wavelength: A Local Baseline with S$^4$G for High-Redshift Studies}

\author[Men\'{e}ndez-Delmestre et al.]{Kar\'{i}n Men\'{e}ndez-Delmestre,$^{1}$\thanks{kmd@astro.ufrj.br}
Thiago S. Gon\c{c}alves\,$^{1}$
Kartik Sheth,$^{2}$
Tom\'{a}s D{\"u}ringer Jacques de Lima,$^{1,3}$
\newauthor Taehyun Kim,$^{4}$
Dimitri A. Gadotti,$^{5}$
Eva Schinnerer,$^{6}$
E. Athanassoula,$^{7}$,
Albert Bosma,$^{7}$
\newauthor Debra Meloy Elmegreen,$^{8}$
Johan H. Knapen,$^{9,10}$
Rubens E.~G.~Machado$^{11}$
and Heikki Salo$^{12}$.
\\
$^{1}$Universidade Federal do Rio de Janeiro, Observatorio do Valongo, Ladeira Pedro Ant\^{o}nio, 43, Sa\'{u}de CEP 20080-090 Rio de Janeiro, RJ, Brazil\\
$^{2}$NASA Headquarters, 300 E. Street SW, Washington DC 20546, USA\\
$^{3}$WeWork c/o Michael Wagstaff (Spectora), 1550 Wewatta St, Suite 4007, Denver CO 80202, USA\\
$^{4}$Department of Astronomy and Atmospheric Sciences, Kyungpook National University, Daegu, 41566, Republic of Korea\\
$^{5}$Centre for Extragalactic Astronomy, Department of Physics, Durham University, South Road, Durham DH1 3LE, UK\\
$^{6}$Max-Planck-Institute for Astronomy, Königstuhl 17, 69117, Heidelberg, Germany\\
$^{7}$Aix Marseille Univ, CNRS, CNES, LAM, Marseille, France\\
$^{8}$Dept. of Physics \& Astronomy, Vassar College, Poughkeepsie, NY 12604, USA\\
$^{9}$Instituto de Astrofísica de Canarias, La Laguna, Spain\\
$^{10}$Departamento de Astrofísica, Universidad de La Laguna, La Laguna, Spain\\
$^{11}$Departamento Acad\^emico de F\'isica, Universidade Tecnol\'ogica Federal do Paran\'a, Rua Sete de Setembro 3165, Curitiba, Brazil\\
$^{12}$University of Oulu, Astronomy Research Unit, Oulu, Finland\\
}

\date{Accepted XXX. Received YYY; in original form ZZZ}

\pubyear{2023}

\begin{document}
\label{firstpage}
\pagerange{\pageref{firstpage}--\pageref{lastpage}}
\maketitle

\begin{abstract}
The redshift evolution of bars is an important signpost of the dynamic maturity of disk galaxies. To characterize the intrinsic evolution safe from band-shifting effects, it is necessary to gauge how bar properties vary locally as a function of wavelength. We investigate bar properties in 16 nearby galaxies from the Spitzer Survey of Stellar Structure in Galaxies (S$^4$G) at ultraviolet, optical and mid-infrared wavebands. Based on the ellipticity and position angle profiles from fitting elliptical isophotes to the two-dimensional light distribution, we find that both bar length and ellipticity -- the latter often used as a proxy for bar strength -- increase at bluer wavebands. Bars are 9\% longer in the B-band than at 3.6 $\mu$m. Their ellipticity increases typically by 8\% in the B-band, with a significant fraction ($>$40\%) displaying an increase up to 35\%. We attribute the increase in bar length to the presence of star forming knots at the end of bars: these regions are brighter in bluer bands, stretching the bar signature further out. The increase in bar ellipticity could be driven by the apparent bulge size: the bulge is less prominent at bluer bands, allowing for thinner ellipses within the bar region. Alternatively, it could be due to younger stellar populations associated to the bar. The resulting effect is that bars appear longer and thinner at bluer wavebands. This indicates that band-shifting effects are significant and need to be corrected for high-redshift studies to reliably gauge any intrinsic evolution of the bar properties with redshift.
\end{abstract}


%
%

\section{INTRODUCTION}

Stellar bars are present in roughly 2/3 of all nearby galaxies (\citealt{devaucouleurs63, eskridge00, knapen00, whyte02, md07, marinova07, sheth08, erwin18}). Their formation, evolution and potential destruction (the latter still under debate; e.g., \citealt{bournaud02, bournaud05, athanassoula13}) can dramatically affect their host galaxies. This is due to the fact that a bar introduces a non-axisymmetric component to the gravitational potential of the host galaxy disk, inducing large-scale streaming motions in the gas that may have a major impact in the chemical, dynamic and structural evolution of the host galaxy itself \citep{athanassoula92b, martinroy94, zaritsky94, sheth03, sheth08, grand15}. Simulations have shown that a galaxy disk will succumb to the bar instability (i.e., form a bar) within a Hubble time unless it is dynamically hot or its potential is dominated by dark matter \citep{athanassoula03}. Work by \citet{sheth12} provided strong observational evidence of this by showing that bars are preferentially found in massive, rotation-dominated galaxies, indicating that mass and dynamic coldness of a disk are necessary -- albeit not sufficient -- conditions for bar formation. For cosmological studies, the redshift evolution of bars is thus an important signpost on the growth and dynamic maturity of disk galaxies \citep{sheth08, kim+21}. 

Early claims of a sudden reduction of the bar fraction beyond z$\sim0.8$ \citep{abraham99} suscitated an immediate concern for the impact that band-shifting effects -- the progressive shift of the photometric band to bluer rest-frame wavebands --  could have on the reported bar fraction \citep{sheth03}. Could we be losing track of existing bars simply by observing them at bluer bands? The measure of the local bar fraction had been extended to longer wavelengths with the advent of large near-infrared (near-IR) surveys, including the Ohio State University Bright Spiral Galaxy Survey (OSUBSGS; \citealt{eskridge02}) and the Two Micron All Sky Survey \citep{skrutskie06} Large Galaxy Atlas \citep{jarrett03}. Although bars are structures primarily comprising the older stellar population (\citealt{gadotti+06, sanchez-blazquez+11, deSaFreitas+22}), better traced at longer wavelengths, it has been well established that -- although individual weak bars in the optical become more conspicuous in the near-IR -- the overall bar fraction remains roughly the same from the B-band to the near-IR \citep{eskridge02, md07, barway+11, buta+15}. Hence, when observations are limited to the optical-through-IR wavebands, band-shifting effects do not have a significant impact on the measured bar fraction.

%
%

\begin{figure*}
\centering
\includegraphics[trim=0mm 0mm 0mm 0mm, clip, scale=0.7, angle=90]{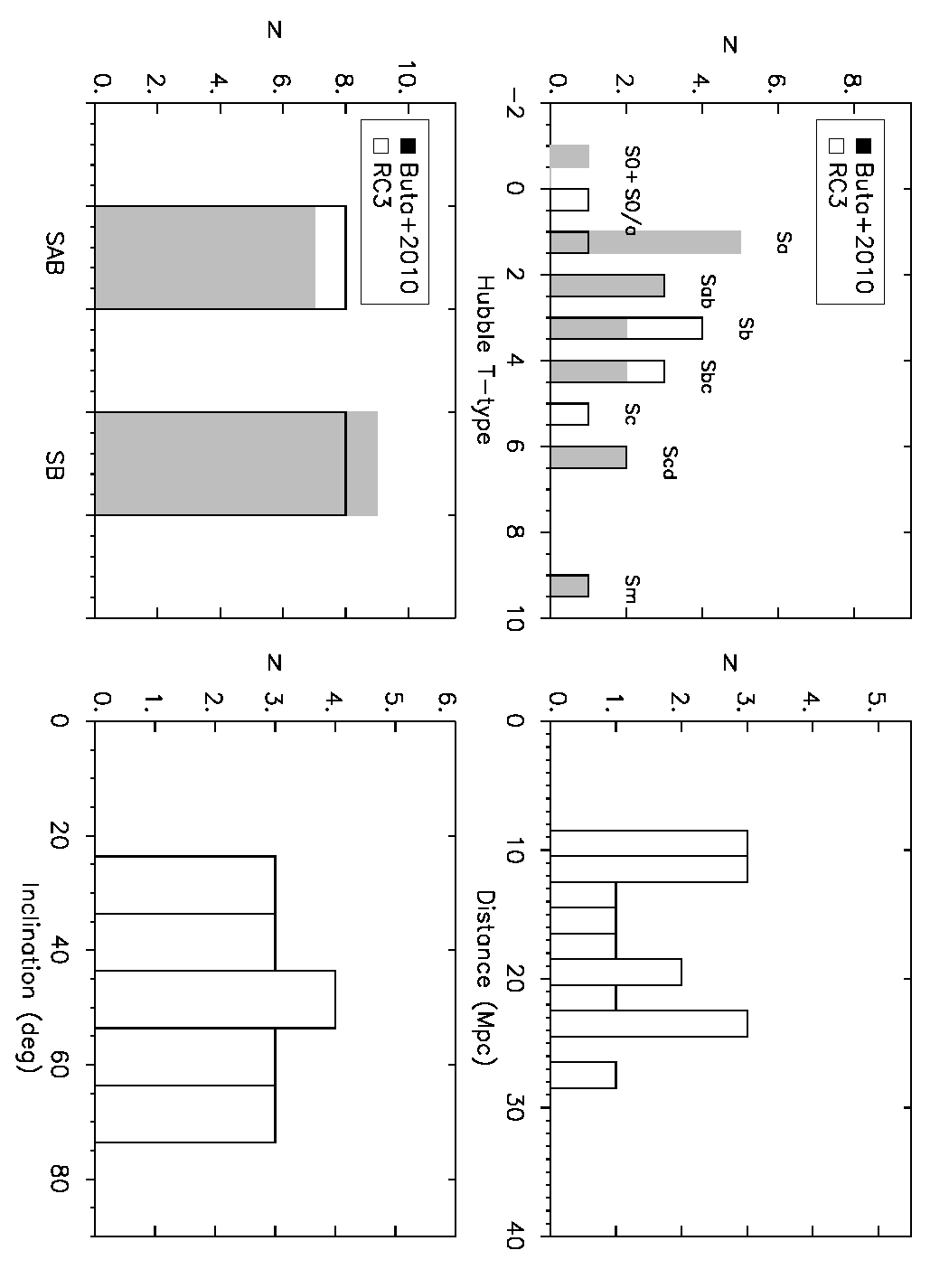}
\caption{Distribution of properties in our galaxy sample, including: ({\it top left panel}) Hubble T-type and ({\it bottom left panel}) bar type (SAB/SB) as given by RC3 and \citet{buta+10}; and ({\it top right panel}) distance and ({\it bottom right panel}) inclination. These figures show that our sample spans a wide range in galaxy properties.
\label{sample}}
\end{figure*}

 With the advent of the Advanced Camera for Surveys (ACS) on the Hubble Space Telescope in 2002, the availability of high-resolution optical imaging motivated the detailed exploration of stellar structure of galaxies at high redshift, including a characterization of the bar fraction out to $z\sim0.8-1$ (e.g., \citealt{abraham99, sheth03, elmegreen04, jogee04, sheth08, melvin14}).  Even with  detailed correction for magnitude survey limits, stellar mass, inclination and galaxy type selection, \citet{sheth08} showed that reports on the bar fraction evolution beyond $z\sim0.85$ based on optical data (e.g., \citealt{jogee04, elmegreen04}) fall prey to band-shifting effects; beyond $z\sim0.85$, even the red ACS I-band traces ultraviolet (UV) light, at which point bars are difficult to recognize \citep{sheth03}. More recent studies exploiting Wide-Field Camera 3 (WFC3) near-IR data pushed the redshift limit of the bar fraction characterization beyond $z\sim1$ based on rest-frame optical images \citep{simmons14}. With the advent of the JWST, a handful of barred galaxies have been detected beyond $z=2$ \citep{guo+23, costantin+23}. Furthermore, \citet{LeConte+23} find that the fraction of barred galaxies beyond z$\sim$1 as measured with JWST at rest-frame near-IR is $\sim$3-4 times that found earlier in studies based on HST WFC3 data.  These results suggest the importance of bar studies at rest frame near- and mid-IR bands (see also \citealt{mendez-abreu+23} and \citealt{Bland-Hawthorn+23}).

Considering local bar properties is of critical importance when assessing the completeness of high-redshift bar fraction studies, as weak bars are more difficult to pick out \citep{sheth03} and small bars become undetectable as a consequence of resolution limits (see Figure 6 in \citealt{md07}). Beyond an interest in characterizing the completeness of high-redshift studies, studying the distribution of  bar properties allows us to test the accuracy of bar formation models \citep{athanassoula13}. Theoretical models predict that when the bar evolves, it becomes longer and stronger (e.g., \citealt{athanassoula03, villa-vargas+10, athanassoula13}). Hydrodynamical simulations have started to look at the redshift evolution of bar properties showing this very behavior and -- more importantly -- have started to establish predictions of how bar properties change over cosmological time (e.g.,  \citealt{okamoto+15}). The evolution of bar fraction and other bar properties have been studied using recent cosmological hydrodynamical simulations of galaxy formation such as: EAGLE \citep{algorry+17}, Illustris TNG100 \citep{rosas-guevara+20, zhao+20, zhou+20}, Illustris TNG50 \citep{rosas-guevara+22, zana+22, izquierdo-villalba+22}, NewHorizon \citep{reddish+21} and Auriga \citep{fragkoudi+21}. In such simulations, bars typically start developing after redshift z$\sim$1 although massive disks are already present earlier than that. This redshift of formation depends on mass, size and other selection criteria. Theoretical predictions can thus be made for the evolution of bar properties, but simulation studies generally rely on stellar mass rather than light to measure structural properties of the bar.

A number of optical studies have looked at the distribution of bar properties in optically-selected local bars \citep{kormendy79, elmegreen85, martin95, erwin05b, aguerri09, barazza09, marinova09, gadotti2011, zhaoyuli11, hoyle11}, while others have extended the analysis to the near- and mid-IR based on statistically-large samples (e.g., \citealt{md07,erwin18, erwin19}). A few studies have started to venture out to intermediate redshifts to characterize bar properties based on optical data \citep{abraham99,barazza09, kim+21}, but the role of band-shifting effects has not yet been well established in the interpretation of their results. With ample optical through near-IR space-based data with sufficient spatial resolution now available (e.g., ACS/WFPC2/WFC3) and within the context of new high-resolution mid-IR imaging  (e.g., JWST), we are at a stage that to reliably characterize the redshift evolution of bar properties, a local baseline is necessary to establish how bar properties vary with wavelength. Such a baseline is crucial for comparison, as it allows high-redshift studies to determine if observed changes in bar properties with redshift are simply band-shifting effects or genuine signs of secular evolution. Accurately understanding how bars evolve with redshift will provide valuable information about the astrophysical processes that lead to bar formation and an observational confirmation of what simulations point as necessary conditions for a galaxy to succumb to the bar instability (e.g., \citealt{athanassoula13}). The interactions between bars and galactic disks will, in turn, give us a better understanding of how galaxies evolve over cosmological timescales.  

In this paper we present a detailed study of bar properties  -- bar {\it length} and bar {\it strength}, as given by the bar's ellipticity -- as a function of wavelength for a sample of 16 large nearby barred galaxies, spanning the wavelength range from the UV to the mid-IR (MIR). The MIR waveband mitigates dust-obscuration effects and provides the optimal window to  probe the details of stellar structure in galaxies (see Fig. 23 in \citealt{driver+16}). By extending our investigation to the UV light we project the applicability of our results to ongoing and future studies of bar properties beyond $z\sim0.8$ with the ample high-resolution optical data provided by current space telescopes. With this work we further envision preparing for the MIR inspection of bar properties at sub-kpc scales out to cosmic noon, in preparation for new and future space-based observing facilities (e.g., JWST, Euclid). 

We describe our sample selection and observations in Section \ref{obs}, summarize our analysis in Section \ref{processing} and state our results in Section \ref{results}.  In Section \ref{discussion} we discuss our results and give our main conclusions in Section \ref{conclusions}. We assume a $\Lambda$CDM cosmology, with $H_0 = 71$\,km\,s$^{-1}$\,Mpc$^{-1}$, $\Omega_M = 0.27$ and $\Omega_\Lambda = 0.73$.

%
%

\section{Sample Selection and Observations}\label{obs}

The rest frame MIR light in normal galaxies is dominated by the old stellar population; by mitigating dust-obscuration effects, it provides the optimal window to  probe the details of stellar structure in galaxies, in particular bars and their properties. We thus base our sample selection on the Spitzer Survey of Stellar Structure in Galaxies (S$^4$G; Sheth et al. 2010), a volume- (D$<40$~Mpc), magnitude- ($m_b<15.5$), and size-limited ($D_{25}>1$\arcmin) survey of 2,331 galaxies in the nearby Universe with the 3.6 $\mu$m and 4.5 $\mu$m bands of the Spitzer Space Telescope's Infrared Array Camera (IRAC; \citealt{fazio04}). Our sample comprises spiral galaxies with clear bar signatures from the optical through the MIR: classified as either barred (SB) or intermediately barred (SAB) in the optical by \citet{devaucouleurs63} and confirmed as bars in the near-IR \citep{md07} and the MIR \citep{buta+10}. We required the availability of high-quality galaxy images in the UV, optical and MIR to ensure adequate comparison. Taking into consideration that the galaxy database from the Spitzer Infrared Galaxy Survey (SINGS; \citealt{kennicutt03}) includes excellent optical coverage in the B- and R-bands, we selected S$^4$G galaxies that were also part of the SINGS sample. We exclude barred spirals with high disk inclinations, i$>70^{\circ}$. These selection criteria resulted in a sample of 16 nearby barred spirals, listed in Table \ref{resultstab}. We show the distribution in RC3 bar type (SAB/SB), Hubble T-type, inclination and distance in Fig.\,\ref{sample}, taken from the Hyperleda Extragalactic database \citep{makarov+14}\footnote{http://leda.univ-lyon1.fr/}. 

For our MIR analysis, we use {\it Spitzer} IRAC 3.6~$\mu$m images. These were first reduced by the {\it Spitzer} Science Center S18.5 pipeline and further processed by the S$^4$G pipeline (see \citealt{munozmateos+15} for details). The resulting images, with a pixel scale of $0.75\arcsec$ and spatial resolution of $\sim1.5\arcsec$, reach a $1\sigma$ brightness limit of $27$ mag/arcsec$^{2}$ and are part of the S$^4$G data products publicly available at http://irsa.ipac.caltech.edu/data/SPITZER/S4G.

We used B- and R-band images from the SINGS fifth data release\footnote{https://irsa.ipac.caltech.edu/data/SPITZER/SINGS/}. Originating from observations made with the 2.1m telescope at Kitt Peak National Observatory (KPNO) and the 1.5m telescope at Cerro Tololo Inter-American Observatory (CTIO) down to a uniform depth of $\sim25$ mag/arcsec$^{2}$, these images present pixel scales of $0.434\arcsec$ and $0.304\arcsec$, respectively, with spatial resolutions dictated by the seeing at the time of observations, in the range of $\sim1-2\arcsec$.

In order to cover the UV emission for the galaxies in our sample, we rely on GALEX images in the far- and near-UV bands ($\lambda_{\rm{FUV, NUV}}=1516$\AA, 2267\AA, respectively) from the Nearby Galaxies Survey (NGS; \citealt{gildepaz+04, bianchi+03a, bianchi+03b}) down to a surface brightness of  $\mu_{\rm{FUV, NUV}}=27.25$mag/arcsec$^{2}$, 27.35mag/arcsec$^{2}$, respectively \citep{thilker+07}. With a spatial resolution of $4.0-4.5\arcsec$ (FUV) and $5.0-5.5\arcsec$ (NUV), these images establish our coarsest spatial resolutions, at a pixel scale of $1.5\arcsec$ \citep{morrissey+07}.

%
%

\begin{figure*}
\centering
\includegraphics[scale=0.2]{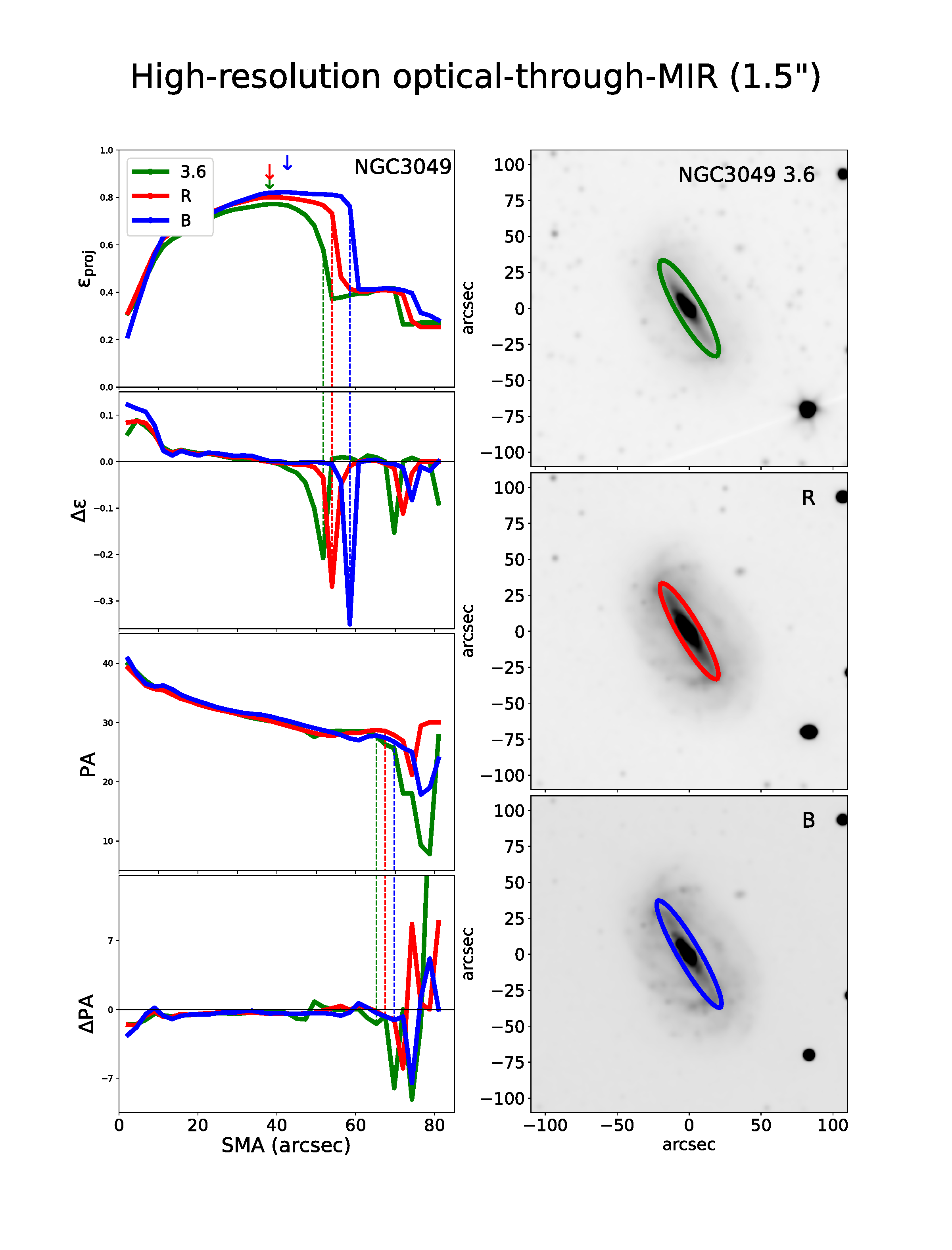}
\includegraphics[scale=0.2]{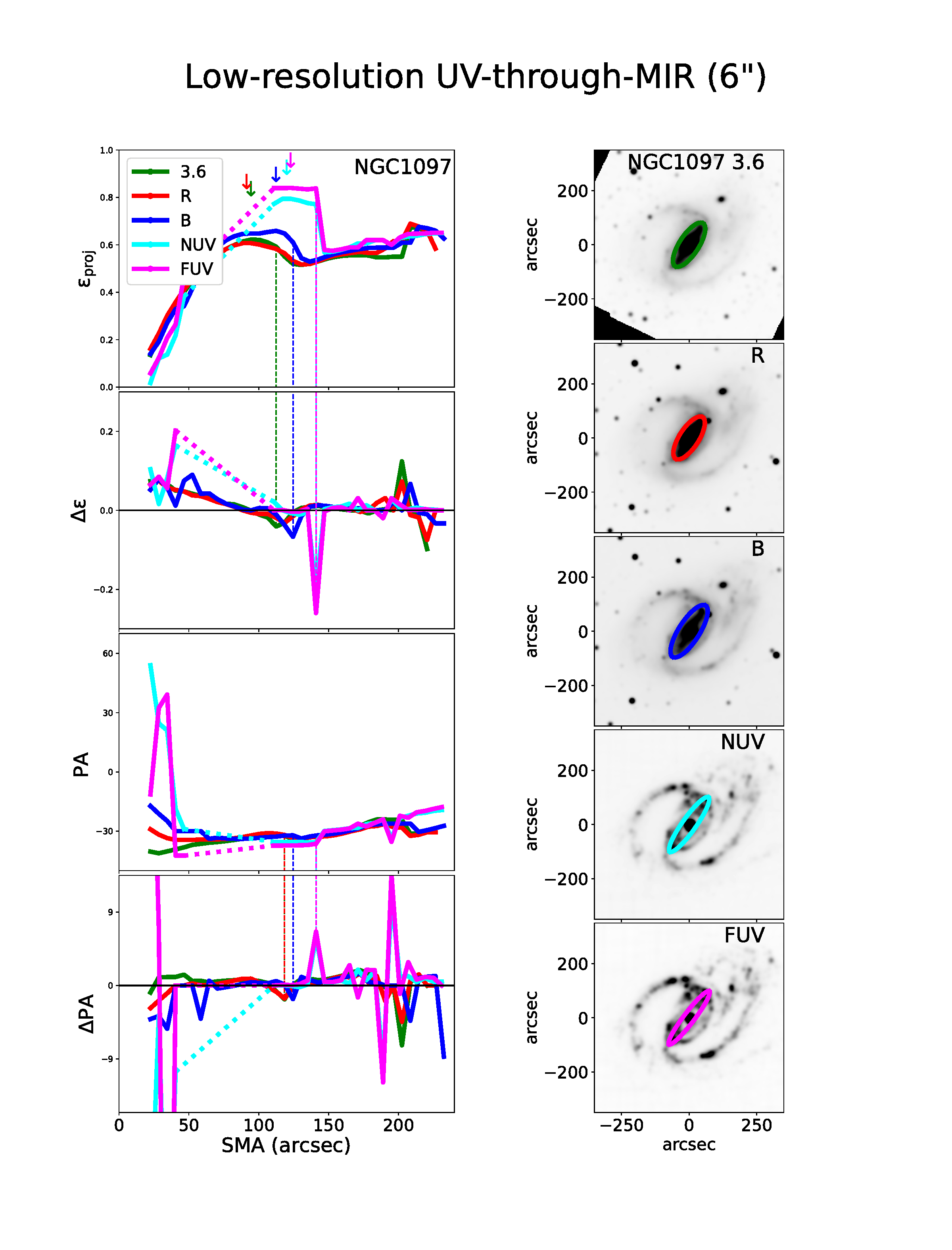}

\caption{Ellipticity and position angle radial profiles for NGC~3049 (left) and NGC~1097 (right), the former exemplifying our methodology for the {\it higher-resolution optical-through-MIR} study (including B-, R-, and 3.6$\mu$m bands) and the latter for the {\it low-resolution UV-through-MIR} study (including FUV-, NUV-, B-, R-, and 3.6$\mu$m bands). In both cases we show radial profiles for ellipticity and position angle, as well as their respective variations at each isophotal fit. The vertical down-pointing arrows indicate the semi-major axis (SMA) of maximum ellipticity, i.e., a$_{\epsilon max}$ for each band.  The vertical dashed lines indicate the radial location of the maximum variation in ellipticity (top two panels) and PA (bottom two panels). The dashed portions of the radial profiles indicate regions where the FUV/NUV emission deviates significantly from a smooth light distribution, consequently impacting our isophotal ellipse-fitting approach (see Section \ref{processing}). From top to bottom, postage stamp images in the 3.6$\mu$m, R- and B-bands for NGC~3049 and in the 3.6$\mu$m, R-, B-, NUV- and FUV-bands for NGC~1097 are shown with an overlaid ellipse showing the isophote of maximum ellipticity. Similar figures for all of our sample are shown as part of Appendix \ref{appendixA_profiles_hires} for the {\it higher-resolution optical-through-MIR} study and Appendix \ref{appendixB_profiles_lores} for the {\it low-resolution UV-through-MIR} study.
\label{profiles}}
\end{figure*}

To accommodate the large range in angular resolutions presented by this combined dataset, we divide our analysis into two parts: a {\it higher resolution optical-through-MIR study}, where we compare B- and R-band optical images with the mid-IR S$^4$G-processed Spitzer/IRAC images; and a {\it low-resolution UV- through-MIR study}, where we compare the mid-IR images with the coarser resolution GALEX images. To ensure an adequate comparison of bar parameters across different bands, we resampled and smoothed all images onto a common pixel scale and resolution within each of our two studies. To set a common pixel scale we used the {\sc IDL} routine {\it congrid} adapted for python\footnote{{\sc congrid} is a publicly available routine distributed as part of the open-source software database at http://wiki.scipy.org/Cookbook/Rebinning}. For the {\it higher resolution optical-through-MIR} study we resampled the B- and R-band optical images onto a pixel grid of $0.75\arcsec$/pix set by the S$^4$G-processed IRAC observations. For the {\it low-resolution UV-through-MIR} study we adopted the coarser $1.5\arcsec$ pixel size characteristic of the GALEX images. We smoothed higher-resolution images down to the limit set by the lowest-resolution images within each study. For this we relied on the {\sc iraf} \footnote{IRAF is distributed by the National Optical Astronomy Observatory, which is operated by the Association of Universities for Research in Astronomy (AURA) under a cooperative agreement with the National Science Foundation.} {\sc gauss} task, which convolves images with a gaussian function parametrized by a user-fed {\it sigma} value in pixel units. In this manner we set a common resolution of $6\arcsec$ for the images within the {\it low-resolution UV-through-MIR} study  and a resolution of  $1.5\arcsec$ for the images within the {\it higher resolution optical-through-MIR} study.

%
%

\begin{figure*}
\centering
\includegraphics[scale=0.5]{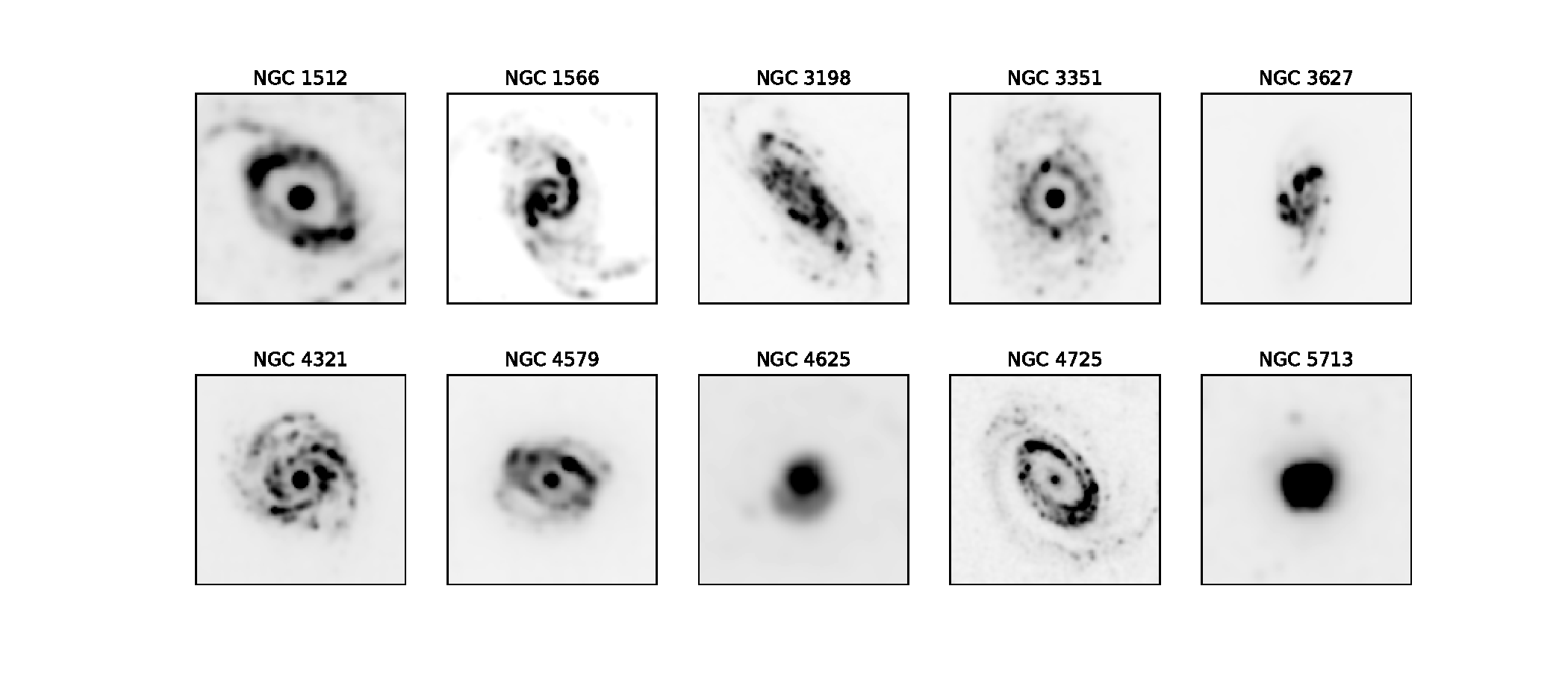}
\caption{GALEX NUV images of the ten galaxies in our sample whose bars become untraceable in the ultraviolet bands.}
\label{LostBarsInUV}
\end{figure*}

%
%

\section{Data Analysis}\label{processing}

To trace the spatial extent of the bar and the degree of the influence that it has on the dynamics of the host galaxy, two bar properties are of fundamental importance: the bar length and the bar strength. Many approaches to measuring the bar length have been proposed in the past (e.g., \citealt{athanassoula&misiriotis02}). The {\it strength} of the non-axisymmetric influence of the bar's gravity on the otherwise axisymmetric gravitational potential of the galactic disk can be quantified in a variety of ways, including the bar-interbar contrast (e.g., \citealt{elmegreen85}, \citealt{regan97}), the Q$_{\rm{b}}$ parameter that measures the maximum gravitational bar torque relative to the galactic disk (e.g., \citealt{combes+81, buta01, laurikainen02,speltincx08}) and a variety of approaches that use a two-dimensional fast Fourier transform method on galaxy images (see \citealt{garcia-gomez+17} and references therein). An alternative,  simple method is based on measuring the maximum ellipticity of the bar isophotes (e.g., \citealt{martin95, aguerri99, shlosman+00, knapen00, laine02, erwin04, md07, marinova07, gadotti2011}).  High ellipticity indicates a {\it thin}, well-defined bar in the equatorial plane of the galaxy, imposing sharp non-axisymmetry, while a low ellipticity is associated with a {\it broad} bar and weaker non-axisymmetrical deviation of the disk potential.  Considering that bar ellipticity has been shown to correlate well with Q$_{\rm{b}}$ measures (e.g., \citealt{laurikainen02}), bar ellipticity represents the simplest, yet accurate approach to characterize bar strength. Its simplicity is crucial for application to large galaxy samples, in particular when dealing with distant barred galaxies, where surface-brightness dimming prevent us from implementing other more elaborate methods (e.g., \citealt{sheth08}). In this work we rely on the bar's ellipticity as a measure of bar strength.

We parametrize the shape of the bar in each galaxy of our sample by applying the widely-used {\it ellipse}-fitting technique \citep{jedrzejewski87}, which consists of tracing the projected light distribution with a series of elliptical isophotes across the full 2D galaxy image. For this we use the {\sc iraf} task {\sc ellipse} to fit elliptical isophotes of increasing semi-major axis (SMA), closely following the implementation presented by \citet{md07}. We used a constant SMA step equal to the resolution of the images ($1.5\arcsec$ and $6\arcsec$ for the high- and low-resolution study, respectively), and allowed both the isophotes' ellipticity ($\epsilon$) and position angle (PA) to vary, while holding the center of fitted ellipses fixed to the center of the galaxy. The great advantage of this method is that while it is very efficient at picking out elliptical deviations in a galaxy's photometry, it is fairly simple to apply to large samples and out to higher redshifts (e.g., \citealt{sheth08}).

The ellipse-fitting algorithm has its limitations: in the presence of bright star-forming knots, it may fail to converge onto a solution. This becomes a great concern at the bluest bands considered in our study, where compact star-forming regions may come to dominate the light distribution within the disk region. To circumvent this issue, we created custom-made masks for the images where the {\it ellipse}-technique would otherwise fail \citep{athanassoula+90}. These masks were produced by hand, where the pixels occupied by either a foreground star or a compact star-forming region were replaced by the median value of the background adjacent to the masked region. In this manner we were able to successfully apply the {\it ellipse}-technique to all the images in our study.

For each object the ensemble of isophote ellipticities ($\epsilon$) and PA provides us with galaxy-wide profiles that allow us to recognize the signature of a bar (see \citealt{md07} for details): a monotonic increase in isophote ellipticity until the end of the bar is reached, at which point ellipticity drops sharply ($\Delta\epsilon\gtrsim0.1$) to settle onto the ellipticity of the disk due to its inclination \citep{regan97}.  Additionally, an abrupt variation in the PA profile ($\Delta$PA$\gtrsim10$ deg) may accompany the signature of the end of a bar, depending on the global geometry of the system; if the disk PA is equal to that of the bar, no such PA change is observed. Example ellipticity and PA profiles for galaxies in our sample are shown in Fig.\,\ref{profiles} for the {\it higher resolution optical-through-MIR} and {\it low-resolution UV-through-MIR} studies; a complete set of figures for our entire sample is shown in Appendix \ref{appendixA_profiles_hires} and Appendix \ref{appendixB_profiles_lores}.

We note that the ellipticity and PA profiles of certain galaxies display a deviation from the ideal bar signature just described. These deviations may be linked to: presence of dust lanes, leading to deviations from a monotonic increase in ellipticity within the bar region and particularly evident in the optical bands (e.g., NGC 1097, NGC 3351, NGC 3627, NGC 4321, NGC 4579, NGC 4725, NGC 7552); presence of nuclear bars, leading to a small increase in ellipticity followed by a drop close to the central regions of the galaxy  (e.g., NGC 1097, NGC 1291, NGC 3198 with SMA$\sim15, 20, 10\arcsec$, respectively); and the presence of prominent spiral arms, delaying the ellipticity drop at the end of the bar and resulting in a bar signature with a broad, flat ellipticity peak (e.g., NGC 4559). 

The maximum ellipticity in the bar signature corresponds to the {\it bar ellipticity} ($\epsilon_{\rm{bar}}$). Although the SMA at which the maximum bar ellipticity is reached (hereafter, a$_{\epsilon \rm{max}}$) is commonly adopted as a measure the bar length \citep{wozniak91, jungwiert97, laine02, sheth03, md07}, a number of alternative definitions of bar length -- also based on the ellipticity and PA profiles of a galaxy -- are found in the literature (e.g., \citealt{athanassoula&misiriotis02,erwin05b, gadotti+07}).

%
%

\begin{figure}
\centering
\includegraphics[scale=0.3]{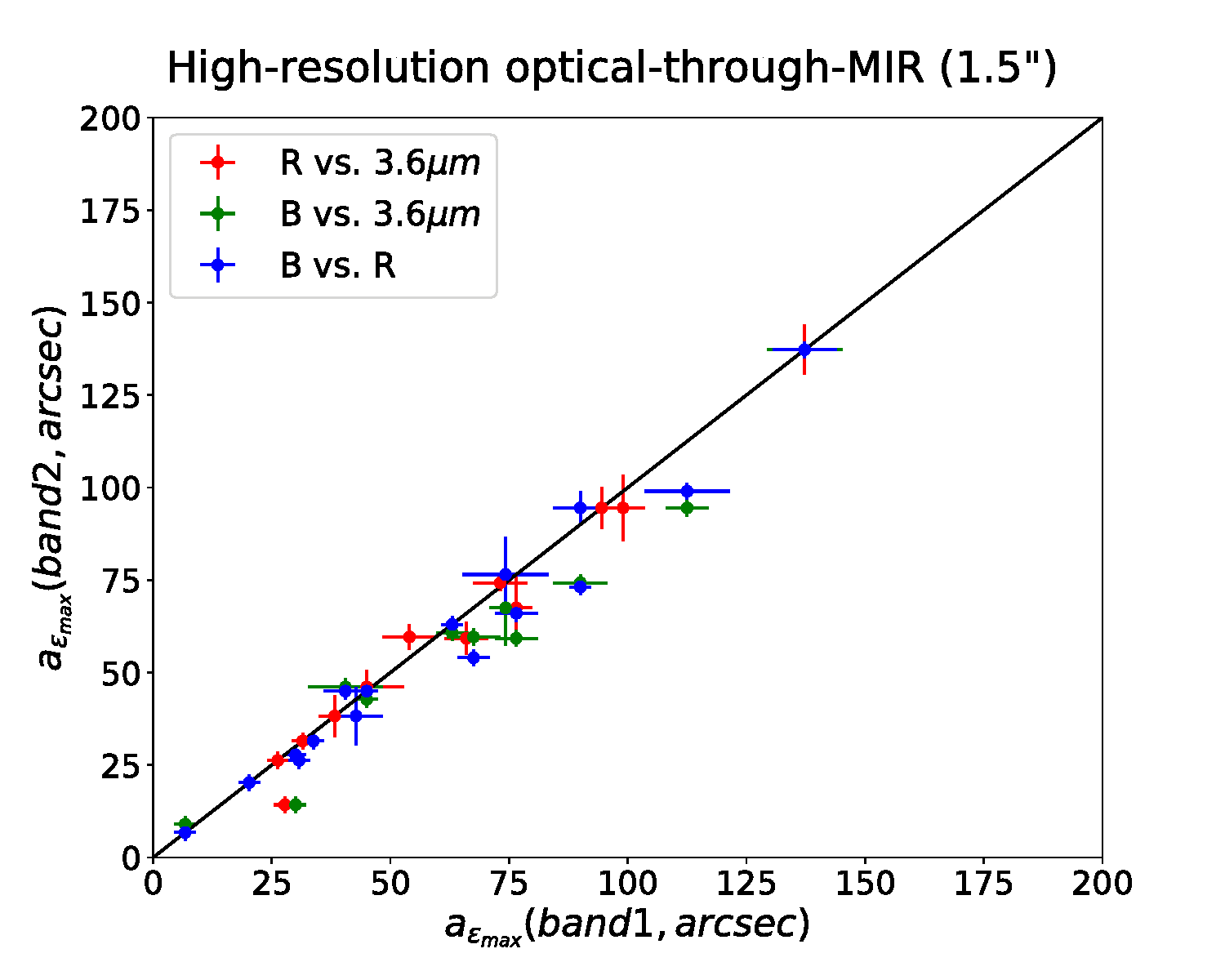}
\includegraphics[scale=0.3]{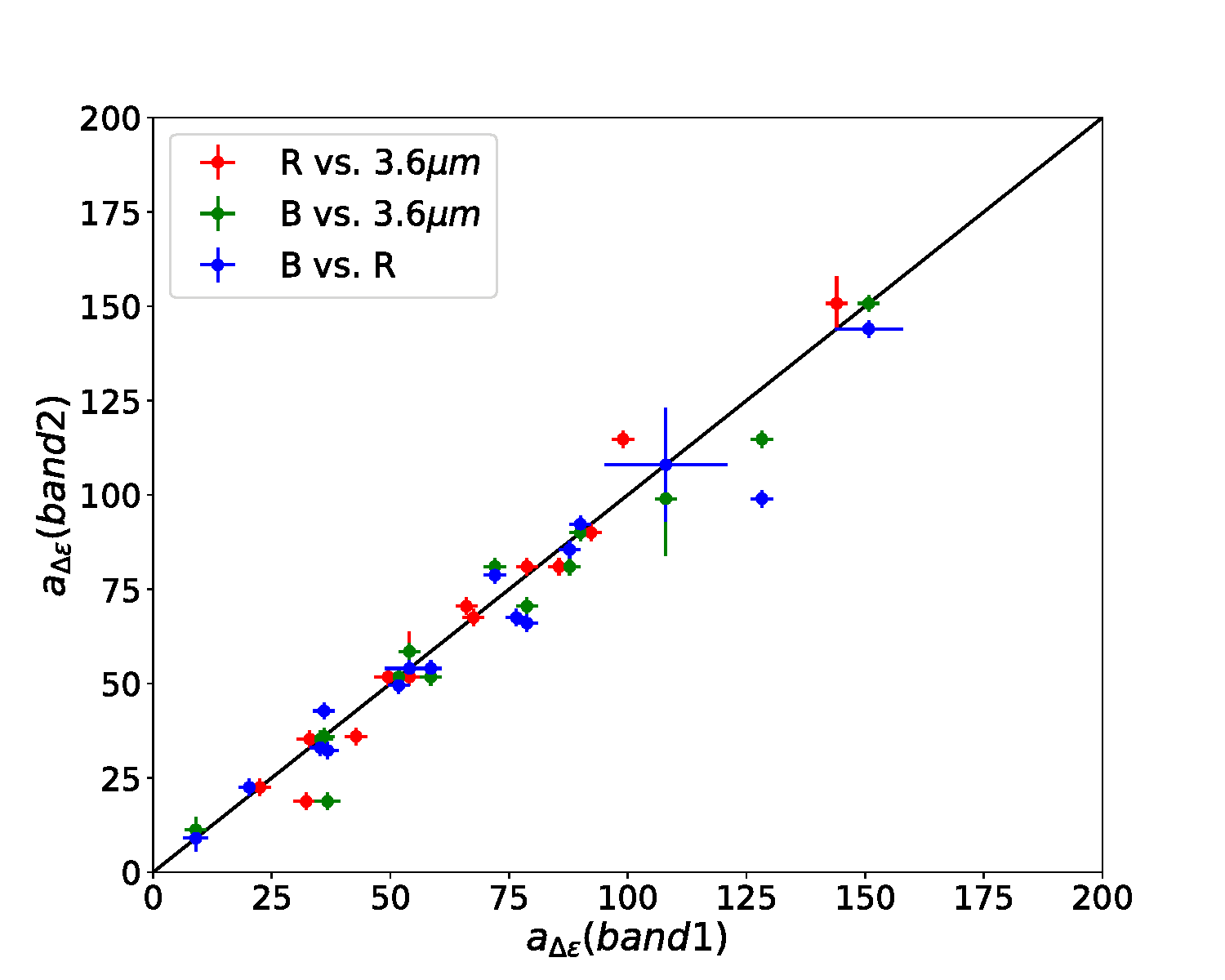}
\includegraphics[scale=0.3]{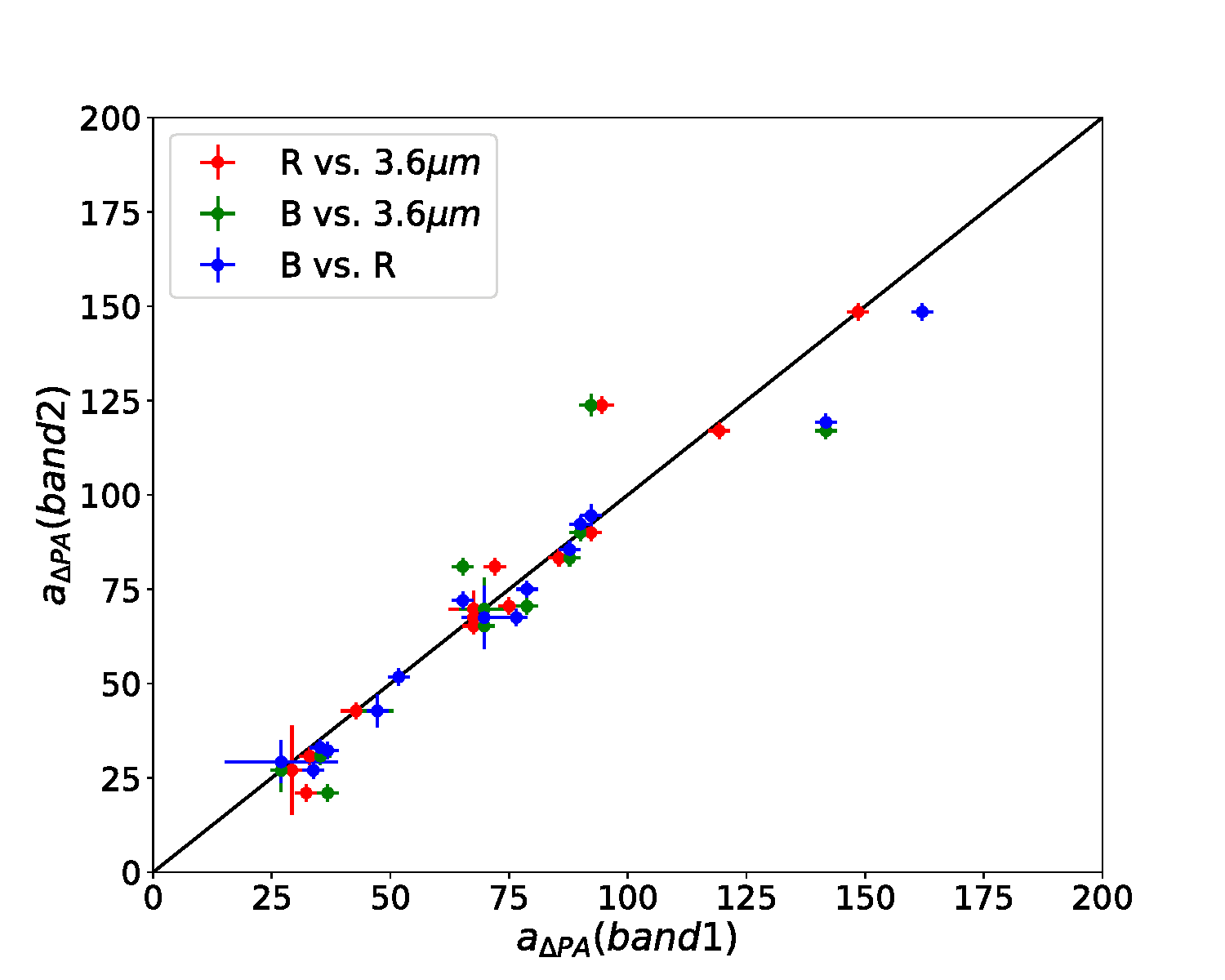}

\caption{Measured bar sizes as a function of wavelength for the galaxies in our sample; each panel corresponds to one of the three bar size definitions adopted for the analysis (see Section \ref{processing} for details). Each galaxy is represented by 3 data points, where the value on the x-axis corresponds to that measured in the bluer band of the waveband pair being considered. We note that the bulk of individual bar size measurements in the B-, R- and 3.6$\mu$m-bands is offset downwards from the identity line (solid line), indicating that bar sizes measured in bluer bands are larger. On average, the bar is measured to be $\sim$9\% longer in the B-band compared to that in the 3.6~$\mu$m.}
\label{length}
\end{figure}

In an effort to investigate the wavelength variation of a more complete set of commonly used measurements, we consider the following {\it three} alternative measures of bar length in our analysis: $a_{\epsilon \rm{max}}$, defined to be the SMA of the isophotal fit with the  maximum ellipticity (e.g., \citealt{wozniak91, regan97});  $a_{\Delta\epsilon}$, the SMA of the isophote where the largest ellipticity drop takes place; and $a_{\Delta \rm{PA}}$, the SMA of the isophote where the sharp change in PA occurs, when present \citep{erwin05b}. To identify the bar length based on all three definitions (a$_{\epsilon \rm{max}}$, a$_{\Delta\epsilon}$ and a$_{\Delta \rm{PA}}$) we analyze not only the ellipticity and PA profiles, but also profiles of the {\it variation} in these parameters, shown in Fig.\,\ref{profiles}.  Following \citet{md07}, to effectively identify a profile as a bar signature we required a bar to display a projected ellipticity, $\epsilon_{\rm{max}}$, greater than 0.2, and the end of the bar to be marked by a change in ellipticity, $\Delta\epsilon>0.1$, with an optional accompanying change in the position angle $\Delta\rm{PA} > 10^\circ$. In this work, both $a_{\Delta\epsilon}$ and $a_{\Delta \rm{PA}}$ are identified within a window of $\sim 30\arcsec$ beyond where the maximum ellipticity is reached. 

Based on a least-squares fitting approach, the {\it ellipse} task provides uncertainties for the geometric parameters of each isophotal fit. These errors, however, only reflect internal uncertainties to convey the amount of deviation from an ellipse in the light distribution at each SMA for the best-fit ellipticity and PA combination. In an effort to include into our reported errors the uncertainty in the assumed SMA for the isophotal fit, we adopt the same approach as \citet{md07} in the case of a$_{\epsilon \rm{max}}$. Considering that the presence of spiral arms may cause the highly-elliptical isophotes that characterize the end of the bar to undergo a soft change in PA with close-to-constant ellipticity (see Fig. 2 for an example), the determination of exactly which SMA corresponds to the maximum ellipticity may become uncertain. \citet{md07} quantify the error in the bar size measurement by considering the SMA range encompassing the tip of the ellipticity peak in the bar signature; that is, the region over which $\epsilon > (\epsilon_{\rm{max}} - \delta \rm{\epsilon})$, where $\delta\rm{\epsilon} =0.01$, a value we adopt as the uncertainty on the ellipticity of the bar. We adopt this value as it corresponds to the maximum error found by the least-squares fit {\it ellipse} task for the ellipticity parameter. For the two alternative measures of bar length -- $a_{\Delta\epsilon}$ and $a_{\Delta \rm{PA}}$ -- we adopt the following approach. Although $a_{\Delta\epsilon}$ is defined as the SMA of the isophote where the largest ellipticity drop takes place, the ellipticity change is not necessarily sharp, extending over a few arcsecs (see ellipticity profile of NGC 3049 as traced by the 3.6 $\mu$m band in Fig.\,\ref{profiles}); we consider the SMA range over which the ellipticity changes around the ellipticity drop at the end of the bar and associate an uncertainty to the bar length equivalent to the FWHM of a Gaussian fit to the ellipticity variation. In the case of extra-sharp drops in ellipticity, we establish a minimum uncertainty equivalent to one resolution element (1.5$\arcsec$ the {\it higher resolution optical-through-MIR} and 6$\arcsec$ in the {\it low-resolution UV-through-MIR}). We define an equivalent approach to the uncertainties in bar length measures based on $a_{\Delta \rm{PA}}$, considering the variation in the PA profile.

We note that in this study we are mainly concerned with the {\it relative} bar properties in the different bands at which each galaxy is studied. For this reason, we present the bar properties prior to any correction for inclination (i.e., deprojection).  For the scope of this work we compare the measured {\it projected} bar sizes and ellipticities for each galaxy from the UV to the MIR as part of the {\it low-resolution UV-through-MIR} study and from the optical to the MIR as part of the {\it higher resolution optical-through-MIR} study.

%
%

\section{Results}\label{results}
 
As part of the {\it low-resolution UV-through-MIR} study, we apply the above method to the entire array of FUV-, NUV-, B-, R- and 3.6$\mu$m-band images at the common resolution of $\sim6\arcsec$ for the 16 barred galaxies in our sample. We repeat this analysis for the same galaxies using the restricted set of B-, R- and 3.6$\mu$m-band images at the finer resolution of $\sim1.5\arcsec$ as part of the {\it higher-resolution} study. The resulting collection of individual bar measurements for the {\it higher resolution optical-through-MIR} and the {\it low-resolution UV-through-MIR} studies are shown in Table\,\ref{resultstab}.

\subsection{Half of all Bars disappear in the UV}\label{diss_uv}

The first result of this analysis is that there are multiple cases where we lose the ability to identify bars in the UV bands. Figure\,\ref{LostBarsInUV} shows the 9 out of the 16 galaxies in our sample with bars that disappear in the NUV images; an additional bar dissapears when we go out to the FUV bands (NGC~4321). This represents 56 and 63\% of the full sample for the NUV and FUV bands, respectively.

In many of these cases (e.g., NGC 1512, NGC 3351, NGC4321, NGC4579, NGC 4725) the ellipticity profiles in the UV bands trace what could be erroneously identified as a bar signature: a maximum in ellipticity, followed by a sharp drop. However, careful examination of the images shows that this ellipticity maximum does not spatially coincide with the end of the bar as seen in the optical or MIR, but to nodes of star formation beyond the bar region, at times significantly offset from the bar PA (e.g., NGC~1512, NGC~3351; see Appendix \ref{appendixB_profiles_lores}).

The images on the FUV and NUV bands often present great challenges for the implementation of our methodology. Compact star-forming regions dominate the brightness distribution, hindering the convergence of the {\it ellipse} routine, particularly within the central regions of the galaxy (e.g., NGC 1512, NGC 1566, NGC 3351, NGC 4321, NGC 4579, NGC 4725). Therefore, these ``fake" bar signatures, although displaying a maximum in ellipticity followed by a sharp drop are never preceded by the monotonic increase that is characteristic of the ellipticity profile in a real bar (see Appendix \ref{appendixB_profiles_lores}).

\subsection{Bar Properties as a function of waveband}\label{bar props}

In Figures \,\ref{length} -\,\ref{ellipticityuv} we present the measured bar length and ellipticity as a function of wavelength for both the {\it higher resolution optical-through-MIR} and the {\it low-resolution UV-through-MIR} studies. The former, based on higher-resolution images, enables a finer evaluation of the variation in bar properties from optical-through-MIR, while the latter allows us to extend our bar characterization to the rest-frame UV. We choose to not elaborate on the optical-through-MIR behavior within the lower-resolution study, as this is done at a finer resolution within the higher-resolution study.


The bar length measurements on Table\,\ref{resultstab} show the following systematics for the bulk of our sample: $a_{\epsilon \rm{max}} \lesssim a_{\Delta\epsilon}  \lesssim a_{\Delta \rm{PA}}$. These trends are a direct consequence of our adopted bar length definitions. The location of maximum ellipticity in a galaxy's ellipticity profile necessarily precedes (or at the closest, coincides) with the ellipticity drop,  which in turn means that the bar length measured at the location of maximum ellipticity ($a_{\epsilon \rm{max}}$) is equal to or lower than that measured at the ellipticity drop ($a_{\Delta\epsilon}$): a$_{\epsilon max} \lesssim a_{\Delta\epsilon}$. Since the PA change -- when present -- is typically found at larger SMA values beyond the ellipticity drop (see Fig.\,\ref{profiles}), the bar length measured at the location of PA variation ($a_{\Delta PA}$) is (in most cases) significantly larger than either the bar length at maximum ellipticity or at the ellipticity drop. For completeness, we note that our ellipse-based bar length measurements a$_{\epsilon max}$ are in good agreement with other visual bar length measurements (e.g., \citealt{herrera-endoqui+15} based on 3.6$\mu$m band S$^4$G images), which we have included for the reader's appreciation in Table\,\ref{resultstab}.

In Fig.\,\ref{length} we compare our bar length measurements from the optical through the MIR, based on our three definitions. The bar length measurements in different bands are set against each other, with the bar length measured in the bluer band displayed in the $x$-axis. Most of our galaxies lie below the line of equality, pointing to bars being longer at bluer bands. The only exception to this trend is when comparing the two reddest bands (R- and 3.6$\mu$m-bands) where the measured lengths are roughly consistent with each other, suggesting that similar stellar populations are traced at these bands.  Based on median values measured as part of our {\it higher resolution optical-through-MIR} study, a bar measured following the ellipticity-maximum method is measured to be $3.6$\% longer in the B-band compared to in the R-band and $8.6$\% longer in the B-band compared to that measured in the 3.6$~\mu$m (see Table \ref{wilcoxon_table_hires}). 

%
%

\begin{figure*}
\centering
\includegraphics[scale=0.3]{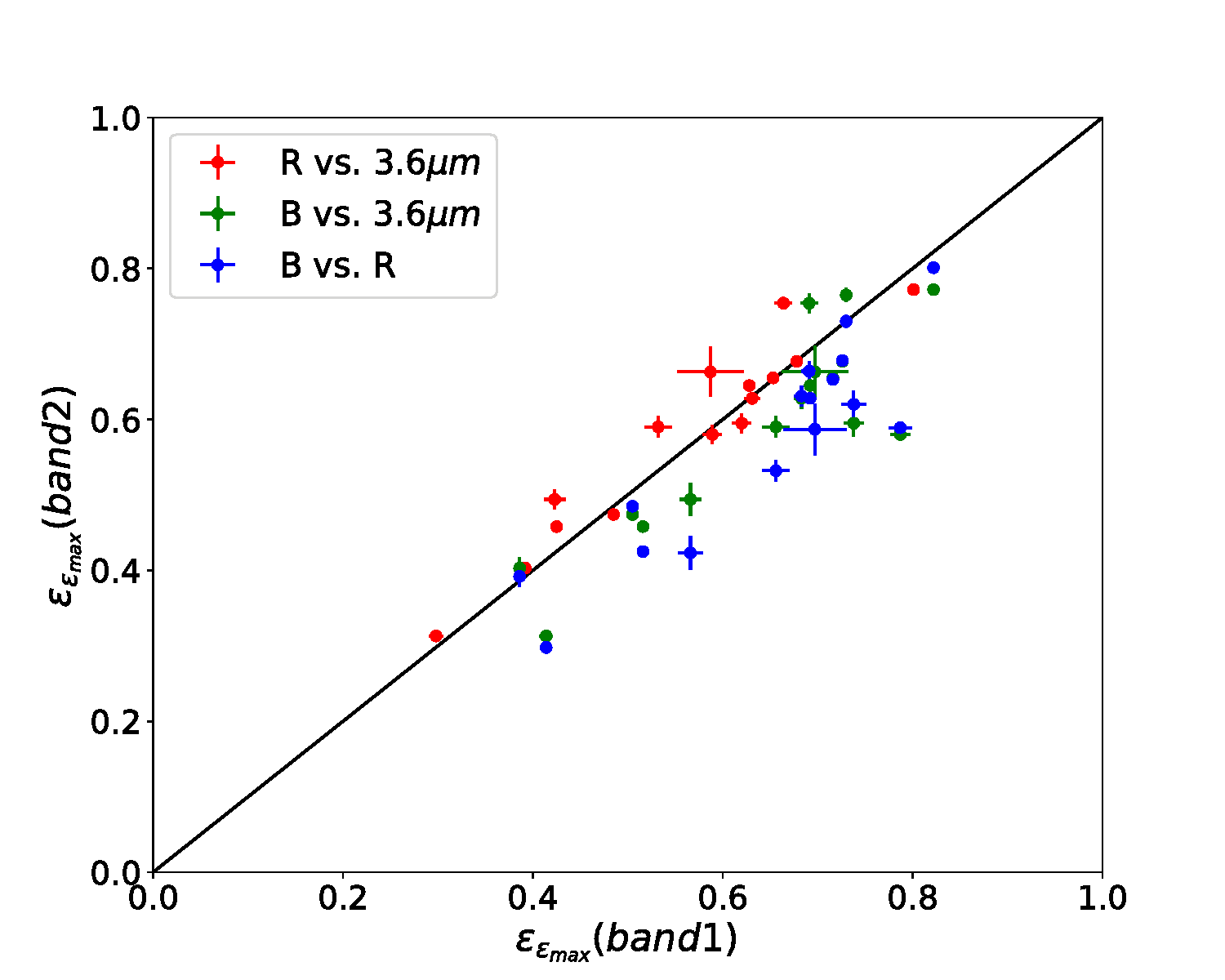}
\caption{Following a similar format to that of Fig.\,\ref{length}, distribution of measured bar ellipticities in our sample. The individual measurements of ellipticities are generally offset downwards from the unity line (solid line), indicating that bar ellipticities measured in the B-band are for the most part higher than those measured in the R- and $3.6~\mu$m bands. }
\label{ellipticity}
\end{figure*}


In Fig.\,\ref{ellipticity} we compare the measured bar ellipticities for each galaxy in the B-, R- and 3.6$\mu$m bands and find that the bar ellipticity is measured to be consistently higher in the B- than in the R- and 3.6$\mu$m bands. We find that $\sim90\%$ ($\sim80\%$) of the sample have measured ellipticity ratios $\epsilon_{\rm{B}}/\epsilon_{\rm{R}}>1$ ($\epsilon_{\rm{B}}/\epsilon_{3.6\mu\rm{m}}>1$). With such skewed distributions towards high ellipticity ratios, this indicates that bar ellipticity is measured to be higher in the B-band than in the redder R and $3.6\mu$m bands, or, equivalently, that bars appear thinner in the B-band. We find that the ellipticity of a bar is typically measured to be $\sim8-10$\% higher in the B-band than in the 3.6$\mu$m- and R-bands. However, in $\sim$30\% of our sample this difference increases to $\gtrsim20-35$\%.


We summarize our results for the {\it higher resolution optical-through-MIR} study -- both for bar length and bar strength multi-band measurements -- in Table\,\ref{wilcoxon_table_hires}, where we also display the results of statistical tests used to evaluate the significance of these results. We perform two different statistical tests: a paired difference test and a Wilcoxon signed-rank test. The two serve a similar purpose, although the latter does not assume the data to be normally distributed. With these tests we essentially seek to verify whether bar length and bar strength measurements in different bands yield distinct means, i.e., if a set of galaxies is measured to have bars with  statistically-significant  longer semi-major axis lengths (or higher ellipticity) in a given band with respect to another. The resulting $p$-values convey the probability of the null-hypothesis that the samples have identical means. We consider a threshold of $p<0.05$ to exclude this hypothesis. In Table\,\ref{wilcoxon_table_hires} we quantify (in terms of \%) the impact that looking at different bands has on the bar length ($a_{\epsilon \rm{max}}$, $a_{\Delta\epsilon}$) and the bar ellipticity for all band pairs, as well as the  $p$-values for the Wilcoxon statistical test. We complement the results shown in Table\,\ref{wilcoxon_table_hires} for both statistical tests in Appendix \ref{appendixC_stats}. Based on this analysis, Table\,\ref{wilcoxon_table_hires} shows that band-shifting effects are statistically significant when comparing bar length and ellipticity (strength) measurements in the B- and in the 3.6$\mu$m-band. To a first order approximation, we may put forward a correction for the band-shifting effects on the measurement of bar properties when considering the rest frame B- and 3.6$\mu$m bands of 9\% and 8\% for the bar length and bar ellipticity, respectively. 

We explored the potential difference in band-shifting effects for stronger vs. weaker bars, as well as for longer vs. smaller bars. However, we find no correlation between either of these bar properties (length or ellipticity) and the shifts caused by making these measurements at different bands.  

%
%

\begin{figure*}
\centering
\includegraphics[scale=0.3]{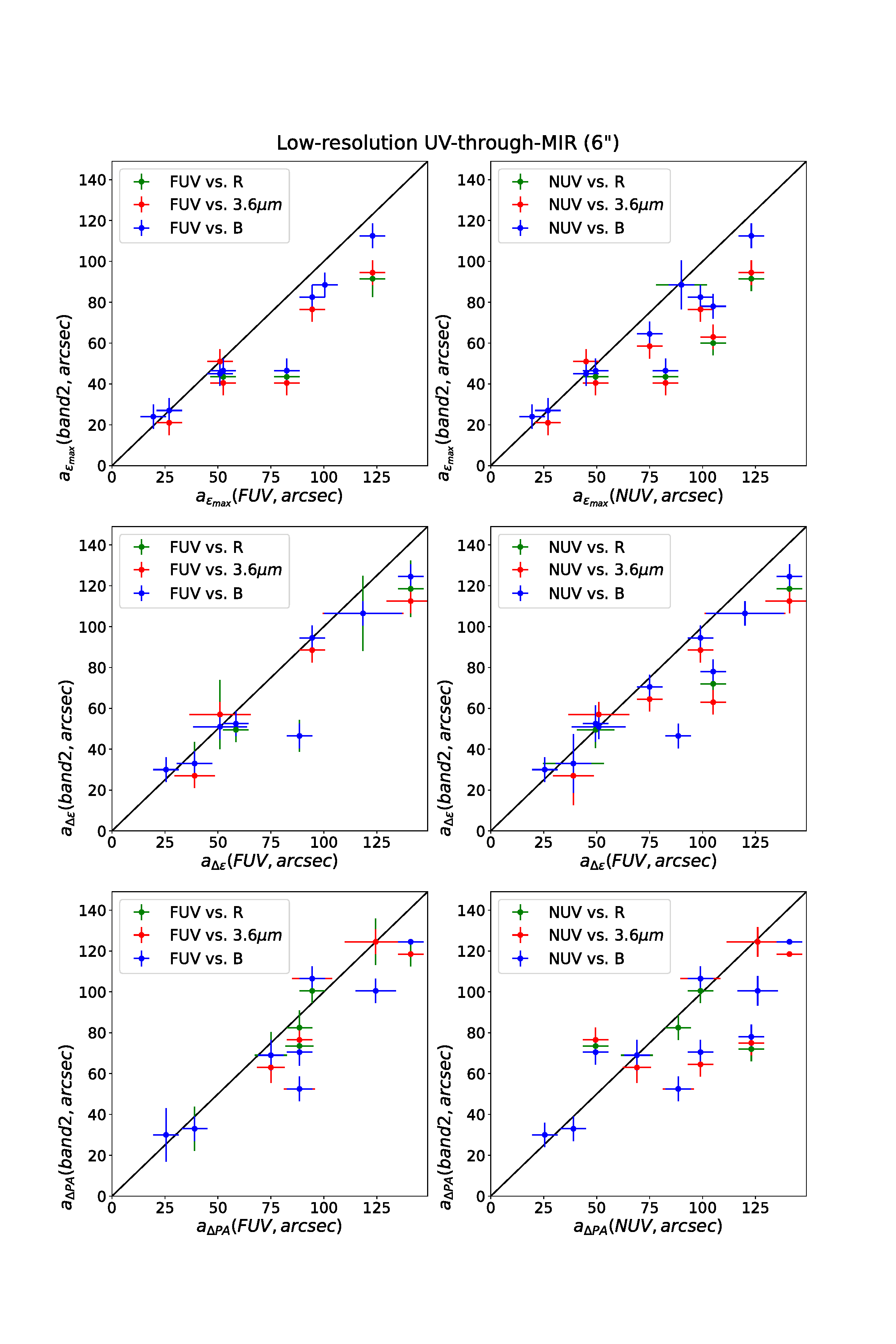}
\caption{Following a similar format to that of Fig.\,\ref{length}, distribution of measured bar lengths in our sample based on the {\it low-resolution UV-through-MIR} study. Bar lengths measured in the B-, R- and 3.6$\mu$m- bands are shown relative to those measured in the NUV (left) and FUV (right) bands, for the 7 (6) galaxies in our sample where bars are still identifiable in the NUV (FUV). These plots show that the results found for the {\it higher-resolution optical-through-MIR} study extend out to UV bands.}
\label{lengthuv}
\end{figure*}


To evaluate whether these results extend to the rest-frame UV, we consider the bar measurements based on the {\it low-resolution UV-through-MIR} study. An immediate consequence of degrading the resolution is that, although bar lengths are not significantly affected (variation typically within $\sim 5$\%; see Table\,\ref{resultstab}) the smaller bars in our sample are washed out (e.g., NGC\,4625; see Appendix \ref{appendixB_profiles_lores}). Furthermore, the measured ellipticities drop by $\sim10$\% at the coarser resolution, compared to the measurements made within the same band for the finer resolution of the {\it optical-through-MIR} study. 

Following the same format of Fig.\,\ref{length} and Fig.\,\ref{ellipticity}, we show in Fig.\,\ref{lengthuv} and Fig.\,\ref{ellipticityuv} the bar length and ellipticity measurements for the {\it low-resolution UV-through-MIR} study. The great majority of bars are measured to be longer in the FUV and NUV, compared to measurements in the optical and MIR, to a degree significantly higher than that in the {\it higher-resolution optical-through-MIR} study. We measure bars to be $\sim$8\% longer in the NUV than in the R band (see Table\,\ref{wilcoxon_table_lores}); however, this increases to $\sim$25\% when comparing to that measured in the 3.6$\mu$m band. In terms of ellipticity, we observe that the bar ellipticity in the NUV band is typically $\sim20$\% higher when comparing to values determined in the R- and 3.6$\mu$m bands. The $p$-values do not exclude the possibility of these difference being just a result of a random selection of the same parent population; however, we note that this trend is merely based on 7 (6) galaxies with bars still identifiable at NUV (FUV) bands, placing this result in the realm of small-number statistics. We provide context and discuss the impact of these results in the next section.

We note that although there is a tendency for bars to appear slightly longer and thinner in the FUV band -- compared to the NUV band, these differences fall within the uncertainties imposed by the $6\arcsec$ resolution we adopted for the {\it low-resolution UV-through-MIR} study.

%
%

\section{DISCUSSION}\label{discussion}

\subsection{Bars in the UV}\label{uv}

Half of all bars in our sample disappear when viewed in the rest-frame UV.  At $\lambda_{\rm{FUV, NUV}}=1516$~\AA, 2267~\AA, the light traced by GALEX FUV, NUV bands is dominated by young OB star complexes along spiral arms, inner and outer starforming rings and nuclear emission from an AGN or nuclear starburst (see Fig.\,\ref{LostBarsInUV}). Although various works have pointed to star formation found associated with the bar structure (e.g., \citealt{fraser-mckelvie+20, diaz-garcia+20}), bar stellar orbits are mainly occupied by old stellar populations that contribute to the emission redward of the Balmer discontinuity (3646~\AA) and 4000~\AA~break. Hence our ability to recognize the presence of a bar in a galaxy is compromised when our analysis is based on rest-frame UV imaging, even considering the relatively deep coverage provided for the galaxies in our sample by the GALEX NGS (\citealt{gildepaz+04, bianchi+03a, bianchi+03b}), down to a surface brightness of  $\mu_{\rm{FUV, NUV}}=27.25$ mag/arcsec$^{2}$, 27.35mag/arcsec$^{2}$, respectively \citep{thilker+07}. According to \cite{erwin08}, typical surface brightness values at the outer extent of galaxy bars are approximately $\mu_R \sim 21-22$ mag arcsec$^{-2}$. Considering that NUV-r colours of galaxies can vary between 1-2~mag for bluer systems dominated by young stellar populations to 5-6~mag for early-type galaxies dominated by older stellar populations \citep{wyder+07}, bars hosting young stellar populations should be easily detectable in the UV at our detection limit of $\mu_{\rm NUV} = 27.4$ mag arcsec$^{-2}$. However,  this would be close to the noise level for older stellar populations. We conclude that a significant fraction of bars are bound to disappear, since they are stellar structures traditionally hosted by older stellar populations. Indeed, \citet{diaz-garcia+20} have shown, for similar detection limits, that UV emission is detected in bars which host star-forming regions. The UV emission along these structures will not be detected in passive bars unless surface brightness limits are considerably deeper (i.e. ~5 mag) than the optical data at hand. In order to identify the stars that typically support the bar structure, we must probe emission redwards of the Balmer/4000~\AA~breaks. 

The impaired ability to trace bars in the UV had been previously emphasized by \citet{sheth03}, where the authors identify an acute drop in the local bar fraction based on the visual identification of bars in 139 local spirals using SDSS u-band ($\lambda_{\rm{SDSS\,u}}=3453$~\AA) imaging.  Considering that the evolution of the bar fraction  with redshift has been the focus of numerous studies \citep{abraham99, sheth03, elmegreen04, jogee04, sheth08, melvin14}, the artificial paucity of bars in the UV acquires critical importance. At a fixed photometric band the morphology of a hypothetical barred galaxy at increasingly higher redshifts is imaged at incrementally bluer rest-frame wavebands. Therefore, it is imperative that we restrict the analysis to redshifts where we probe emission redward of the Balmer/4000~\AA~breaks.

\subsection{Bars {\it appear} thinner and longer in bluer bands}\label{bluer}

Our results show that at bluer wavebands both the measured bar length and the bar ellipticity increase. Although we find that ~50-60\% of the bars disappear in the UV, the results on bar ellipticity and length extend to those cases in which the bar is still visible in the UV.  The increase in maximum ellipticity and the location of the bar end further away from the galaxy center can be clearly seen in the ellipticity profiles for NGC 1097 in Fig.\,\ref{profiles}. 

At face value we attribute the increase in bar length towards bluer bands to the prominent star forming knots that are frequently present at the end of bars. While the contribution from bar stellar populations diminishes as we probe bluer emission, these star-forming regions at both ends of the bar become significantly brighter. The fitted isophotes stretch further out into the disc, resulting in the ``artificial lengthening" of the bar signature. The trend found for the bar ellipticity may be driven by the apparent bulge size at the different wavebands, since the semi-minor axis of the measured bar isophotes are limited by the size of the central bulge. In this manner, a larger, more prominent bulge reduces the bar ellipticity (see also \citealt{diaz-garcia+16}). Bulges are composed primarily of older, redder stars and are thus significantly more prominent -- with a larger apparent size -- at longer-wavelengths (e.g., \citealt{mollenhoff04}). For a given barred galaxy, at shorter wavelengths the bulge appears smaller than at longer wavelengths and thus the bar isophotes may reach higher ellipticities in the bluer bands. Interestingly, \citet{speltincx08} find an increase of $\sim25$\% in bar strength from H-band to B-band for 152 galaxies from the Ohio State University Bright Spiral Galaxy Survey, based on the gravitational bar torque method (Q$_b$, the maximum tangential force normalized by the radial force). These authors also attribute this increase to the diminished prominence of the bulge in optical bands, as the radial forces that it introduces may be underestimated, leading to an increase in the measured bar strength. The impact of an increased contribution from the bulge, with radially-oriented stellar orbits, has also been the center of discussion when considering the reduced ellipticities associated with bars in lenticular galaxies (e.g., \citealt{speltincx08, buta+10}). Work by \citet{hilmi+20}, based on simulated Milky Way (MW)-like galaxies formed in a cosmological context, also present findings pointing to longer and stronger bars when the bar ends are connected to the spiral structure.

We note that boxy/peanut bulges (see \citealt{athanassoula16} for a review), which have been shown to reside in the majority of massive disk galaxies \citep{erwin+debattista17}, have been identified in NGC\,1097, NGC\,1291, NGC\,1512, NGC\,3627 and NGC\,4725 (\citealt{erwin+debattista13, erwin+debattista17,laurikainen+14}). These structures, identified based on the isophotes in moderately inclined galaxies \citep{erwin+debattista13}, may in principle affect bar ellipticity measurements, due to the projection of a vertically-extended bar structure. Although the expectation would be that these boxy/peanut bulges -- which typically host older stellar populations -- present rounder isophotes, with lower ellipticities, the galaxies in our sample with boxy/peanut bulges do not appear to present a particular trend within our general results. We consider that both the presence of boxy/peanut bulges, as well as the variation in bulge prominence associated with a diverse range of morphological Hubble types, may also present interesting trends. Such an analysis, however, calls for a more thorough analysis based on a broader sample and goes beyond the scope of this paper.

%
%

\begin{figure*}
\centering
\includegraphics[scale=0.4]{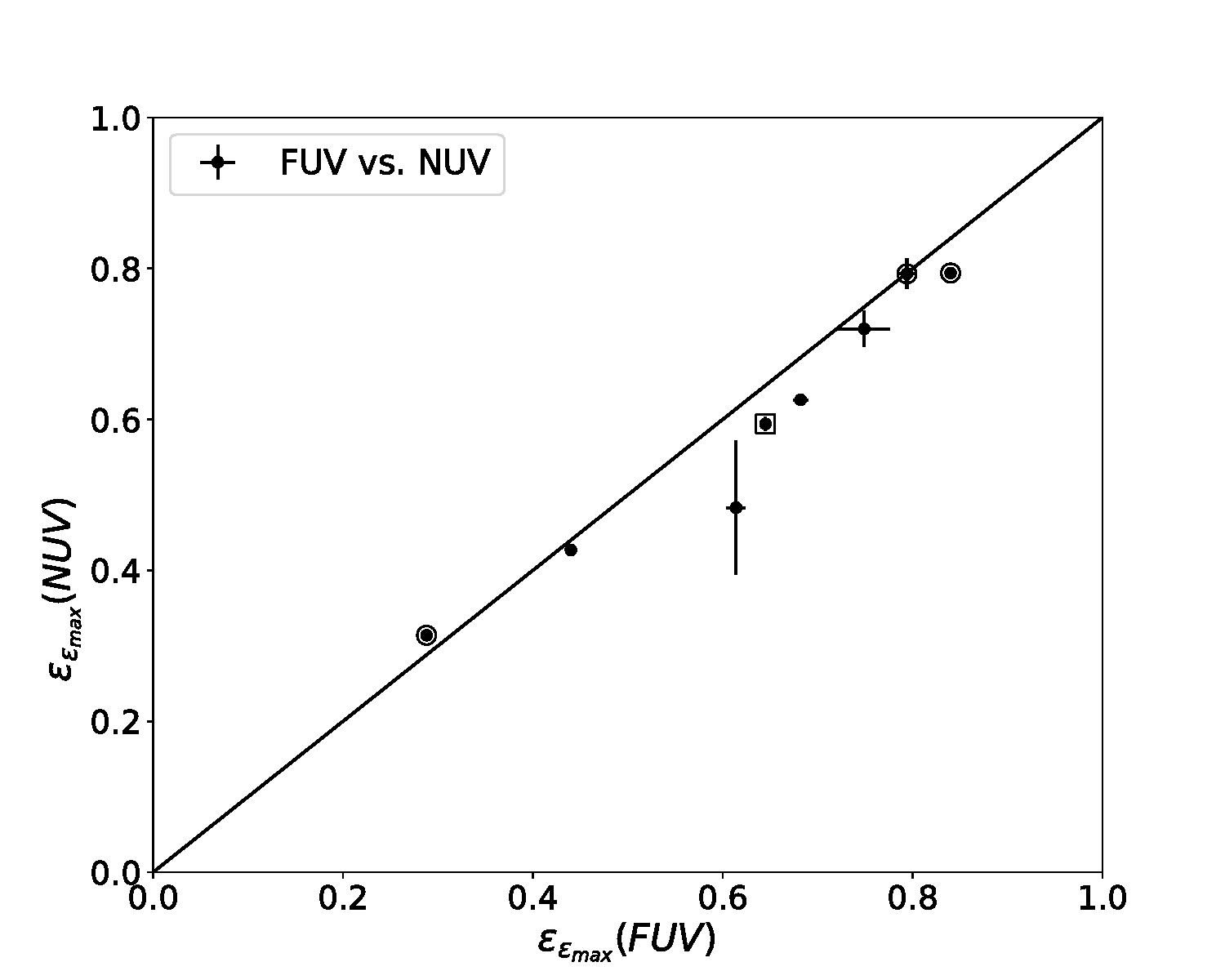}
\includegraphics[scale=0.4]{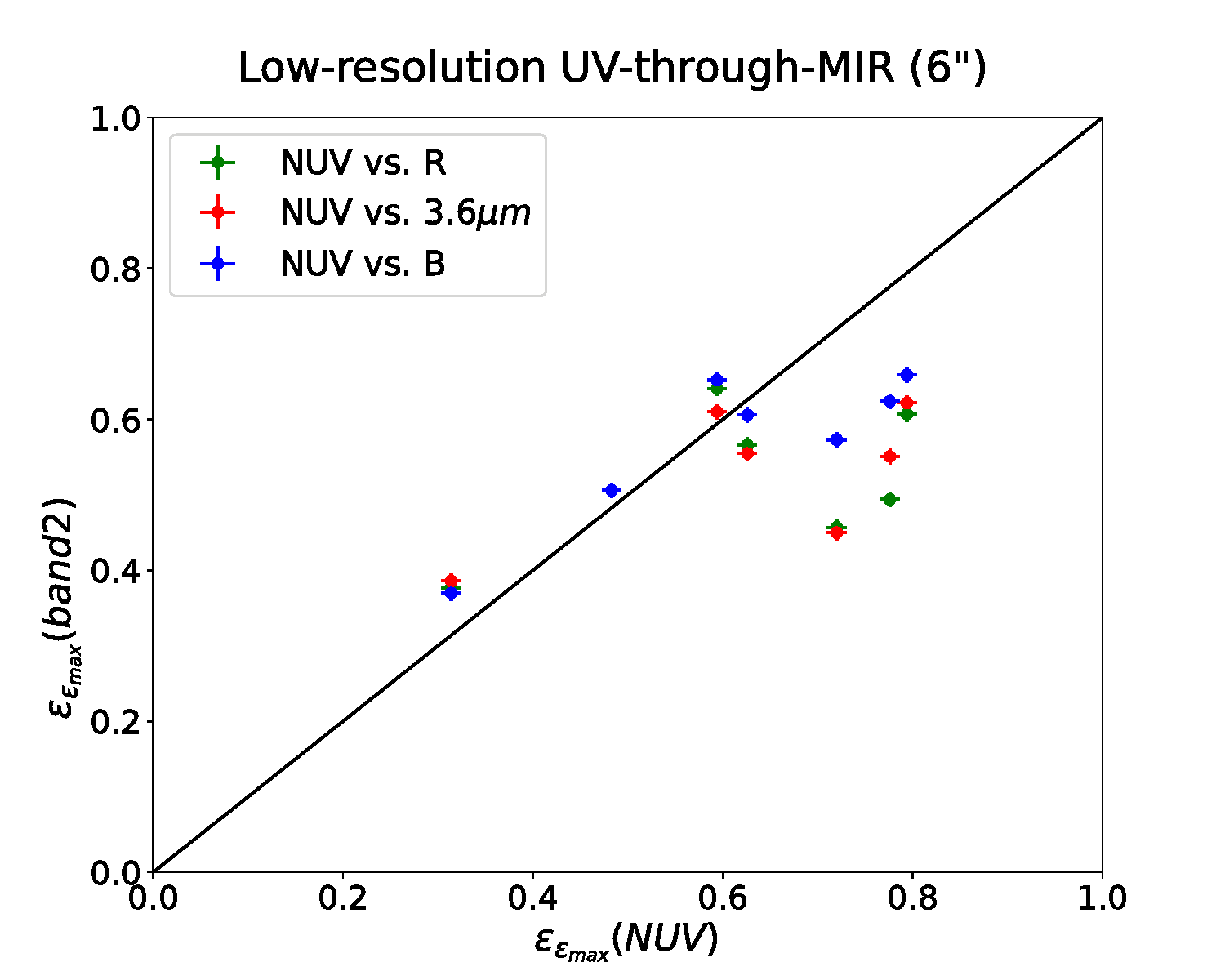}
\caption{Following a similar format to that of Fig.\,\ref{lengthuv}, distribution of measured bar ellipticities in our sample based on the {\it low-resolution UV-through-MIR} study.}
\label{ellipticityuv}
\end{figure*}

Results from recent works by \citet{neumann+20} and \citet{bittner+21} show that the stellar populations in bars are on average older and more metal-rich than the surrounding regions of the galaxy -- both of which contribute to redder colours. They also show that along the major axis of the bar one finds, on average, younger stellar populations compared to the sides/edges of the bar, due to  kinematic fractionation \citep{fragkoudi+17, athanassoula+17, debattista+17}. This leads to the ends of the bar showing younger stars because they will have, on average, more elongated orbits than older stars. So even in the absence of recent star formation we may expect to see longer and more eccentric bars in the bluer wavelengths. These findings suggest that the higher values of bar length and ellipticity we find in bluer bands may not necessarily be ``artificial", as the young stars may indeed be in bar orbits.  

We note that the increase in bar ellipticity (and bar strength) at bluer bands is to some extent related with the {\it ellipse-fit} method (and gravitational bar torque method) used. A full-2D decomposition of the bar and bulge components would allow for the bar's ellipticity and strength to be measured with no contamination from the bulge. This has been in fact performed by \citet{gadotti08} using the BUDDA decomposition code \citep{souza04} on {\it V} and {\it R}-band images for a similarly-sized sample of 17 barred galaxies, where the authors conclude that the bar ellipticity is typically underestimated by $\sim20$\%  when based on the peak in the ellipticity profile. This supports the use of complex methods to identify and characterize bars that are based on modeling and breakdown of galaxy components to obtain enriching results for individual galaxies and small samples; e.g., BUDDA \citep{desouza+04}, GALFIT \citep{peng+02}, GASP2D \citep{mendez-abreu+08, mendez-abreu+14}. Although these complex 2D decompositions of barred galaxies are hard to implement for large samples, several works have analyzed significant samples: e.g., $\sim$3500 galaxies in \citet{kruk+18}, $\sim$800 galaxies in S$^{4}$G \citep{salo+15}, 291 galaxies in \citet{gadotti09}, 162 galaxies with GASP2D \citep{mendez-abreu+17}. However, these decomposition approaches are not necessarily practical nor efficient means of identifying and characterizing bars at high redshift, where samples are plagued with low signal-to-noise imaging and surface-brightness dimming. Precisely the relative simplicity of the ellipse-fit methodology -- the analysis of one-dimensional profiles of ellipticity and position angle adjustment based on the elliptical isophotes in two-dimensional brightness distribution of barred galaxies -- allows it to be extended efficiently to large samples out to the distant universe. For this very reason, studies have in the past years extended the use of the ellipse-fit method towards large samples (e.g., \citealt{reganelmegreen97, zheng05, knapen00, laine02, laurikainen02, sheth02, sheth03, marinova07, md07, aguerri09, cheung13}) and at increasingly earlier cosmological times (e.g., \citealt{sheth08, barazza09}). We must therefore consider with great care the caveats that the widely-applied ellipse-fit method entail.

Our results show that measuring bar properties at different bands introduces potentially artificial effects on the measured ellipticity and length. In going from the 3.6$\mu$m- to the B-band, bars appear longer and thinner by $\sim$9\% and $\sim$8\%, respectively; extending the waveband range further to the blue, bars increase their median length and ellipticity at NUV/FUV bands by $\sim$25\% and $\sim$20\% with respect to their canonical ``true values" at 3.6$~\mu$m. These variations are significant in the light of the reported differences in bar properties with respect to: (1) Hubble type -- with early-type bars displaying higher ellipticities than late-types by $\sim10-20$\% (e.g., \citealt{md07, aguerri09, hoyle11}); (2) environment -- e.g., {\citet{barazza09}} find that bars in cluster environments are longer than in the field; and (3) active vs. normal galaxy type -- where an increase in bar ellipticity of $\sim30$\% and bar strength Q$_b$ of $\sim10$\% has been measured in non-active galaxies, with respect to active ones (Seyferts, starbursts) by \citet{chapelon99} and \citet{laurikainen04}, respectively. The fact that measuring bar properties in different bands may lead to similar variations that are found when analyzing the impact that other galaxy properties may have on bar properties, alerts us on the cautions that must be taken when discussing these intrinsic changes in bar structural parameters at different cosmological times, free of band-shifting effects.

The bar length measurement is also important in the context of the bar pattern speed, where the latter is often parametrised by the bar rotation rate, defined as the ratio between the corotation resonance (CR) radius R$_{CR}$ and the bar semi-major axis R$_{bar}$, R$_{CR}$/R$_{bar}$ \citep{debattista+sellwood2000}. Considering that the bar interacts with other galaxy components, exchanging energy, mass and angular momentum, N-body simulations show that as bars grow, they also slow down (e.g., \citealt{weinberg85, debattista+sellwood2000, martinez-valpuesta+06, athanassoula13}). Although this result seemed observationally supported at first (e.g., \citealt{aguerri+98, rautiainen+08, garma-oehmichen+20}), more recent work has pointed to bars staying fast through their lifetimes (e.g., \citealt{aguerri+15, guo+19}), bringing tension into the picture with cosmological simulations (e.g., \citealt{algorry+17} based on EAGLE and \citealt{peschken+19} based on Illustris). Recent work by \citet{fragkoudi+21} adds to this context, suggesting that simulations reproduce the observed tendency for fast bars all the way to z$\sim$0 when considering less prominent dark matter halos that in turn act less as an angular momentum sink, limiting significantly the bar slow-down predicted by earlier simulations. Within this context, although the 9\% difference in bar length that we report here, when measuring in the B and 3.6~$\mu$m bands, is unlikely to make a difference between studies reporting different populations of slow and fast bars, our results suggests that future studies calculating bar pattern speeds should do so based on consistent measurement of bar lengths in the same band.

\subsection{Corrections for high-redshift bar studies}\label{highz}

Theoretical work shows that a bar grows more elliptical and longer with time (e.g., \citealt{sellwood81, athanassoula03}), which in turn reflects changes within the underlying disk dynamical structure. Hence, an evolution in the median structural parameters of bars with cosmological time may be interpreted as an indication of a transformation in the dynamical maturity of disks. \citet{sheth08} shows that bars in the most massive galaxies ($>10^{10.5}$M$_{\rm{Sun}}$) are already in place at $z\sim0.8$, while active bar growth (via an increase of the bar fraction with decreasing redshift) is still today taking place in the less massive systems ($<10^{10.5}$M$_{\rm{Sun}}$). Within the context of bars being long-lived structures (e.g., \citealt{kraljic+12}), this suggests that the median bar grows -- both in size and strength -- towards lower redshifts. 

Considering that at a fixed photometric band the morphology of a hypothetical barred galaxy at increasingly higher redshifts is imaged at incrementally bluer rest-frame wavebands, our results indicate that the bar size and the bar ellipticity will suffer a systematic shift towards higher values. That is, unless we make careful corrections for band-shifting effects, we may perceive a shift in the bar length and bar ellipticity distributions towards {\it longer} and {\it thinner} -- thus {\it stronger} -- bars in the high-redshift population. These potentially  ``artificial" lengthening and thinning of the bar structure with redshift could mask any intrinsic evolution of the bar structural parameters.  \citet{barazza09}, based on the ellipse-fitting technique applied to HST/ACS i-band imaging of 63 barred galaxies, undertook the pioneering investigation of bar properties out to intermediate redshifts; they found no variation in the median bar size and ellipticity in three redshift bins: $0.4< z < 0.55$, $0.55< z < 0.7$, $0.7< z < 0.8$. Although within this redshift range the i-band merely shifts from rest frame V- to B-band, the potential impact of band-shifting effects in studies comparing bar properties in different redshift bins such as this one has not yet been well discussed. The results of our study caution high-redshift bar studies not merely against relying on rest frame UV imaging when identifying the presence of a bar, but also against tracing the evolution in bar properties -- even in the optical and near-IR -- at face value, with no consideration of band-shifting effects. This is particularly important in the light of the current availability of large datasets based on deep, high-resolution optical imaging. 

The Wide-Field Camera 3 (WFC3) on the HST provides near-IR imaging with the F160W filter that allows us to probe rest-frame B-band out to $z\lesssim2.5$. Our results indicate that high-redshift studies will be able to put interesting constraints on any detected evolution in the bar size distribution. With a spatial angular resolution of 0.13$\arcsec$/pix for the WFC3 IR channel corresponding to $\sim1$~kpc at $0.8 \lesssim z \lesssim2.5$, WFC3 allows us to trace evolution in bar properties for what corresponds to a typical local bar, with $a\sim4.2$~kpc \citep{md07}. Recent results by \citet{kim+21} are pioneering the exploration of bar properties in large samples out to $z\sim0.8$, exploiting restframe optical data based on HST. With the advent of the JWST, the significantly finer angular resolution of the MIRI and NIRcam imagers (MIRI: 0.11$\arcsec$/pix at $5.6~\mu$m; NIRcam: 0.03, 0.6$\arcsec$/pix at $2~\mu$m and $4~\mu$m, respectively) and their wavelength coverage out to MIR wavelengths ($\lambda\sim5-25~\mu$m) will allow investigations of bar properties down to even sub-typical bar size values at higher redshifts \citep{guo+23, costantin+23}.

We emphasize that the ability of reliably quantifying any change in the bar size distribution in light of band-shifting effects does not preclude the bias against smaller bars imposed by a decrease in spatial resolution. As studies probe larger distances, these become increasingly biased against the smaller bars that fall below the instrument's physical spatial resolution. A number of studies have clearly established the need for careful consideration of bar size limit in sample selection when investigating the presence of bars at higher redshifts \citep{sheth03, erwin05b, md07, erwin18}. Cosmological surface brightness dimming also irremediably impacts our ability to study structures in the distant universe, as has been discussed by previous works (e.g., \citealt{sheth08, melvin14, kruk+19}). \citet{sheth08} explored in detail the impact that surface brightness dimming due to cosmologically-significant distances had in their ability to detect underlying disks and showed that their measured bar fractions did not have a dependence on the surface brightness of their galaxies, especially at the low surface brightness end of their sample. \citet{kruk+19}, with a focus on the redshift evolution of boxy/peanut-shaped bulges, explicitly explore the impact of cosmological surface brightness dimming by artificially redshifting their SDSS galaxy images out the redshifts covered by their COSMOS barred population at redshifts up to z$\sim1$. Taking these works into consideration and our bar characterization approach (based on the ellipse fit methodology, which inherently demands the detection of an underlying disk), we note that the ellipticity signature on which our bar length and ellipticity measures are based on will continue to be an effective approach to measure bar properties at higher redshifts. Our work suggests that, once bar size limit considerations and band-shifting effects are taken into account, high-redshift studies are in a position to reliably trace intrinsic changes in the distribution of bar sizes as a function of redshift.

Interestingly, the increase in measured bar ellipticity as bluer rest-frame wavebands are probed due to band-shifting effects may have an unexpected consequence that it may facilitate the identification of weak bars at high redshift. Although intrinsically-weaker bars -- and their shallow ellipticity signature -- are more difficult to pick out in general, the ``artificial" enhancement of the bar ellipticity due to band-shifting effects may make them increasingly easier to recognize at higher redshifts. In this way, studies are in the advantageous position of being able to probe a wider range in bar strengths at increasingly higher redshifts.

\section{Conclusions}\label{conclusions} 

We present a detailed study of bar properties as a function of wavelength for a sample of 16 large nearby barred galaxies, spanning the wavelength range from the ultraviolet (GALEX FUV/NUV) through the optical (SINGS B, R) out to the mid-IR (MIR; Spitzer/IRAC/3.6~$\mu$m). While the MIR provides the optimal window to  probe the details of stellar structure in galaxies, including bars, we choose to include the UV bands in our study to extend the validity of our calibration to redshifts beyond $z\sim0.8$, when HST/optical bands start tracing restframe UV.

Based on the ellipticity and position angle profiles resulting from fitting elliptical isophotes to the full two-dimensional light distribution of each galaxy, we determine the bar length and bar ellipticity for each galaxy, in each band. Our main results are the following:

(1) We find that at bluer wavebands both the bar length and the bar ellipticity increase. We attribute the increase in bar length to the frequent presence of star forming knots at the end of bars: these regions become more prominent in bluer bands, resulting in an apparent lengthening of the bar. Recent studies have also pointed to distinct stellar populations in bars, where kinematic fractionation may lead to younger stars pertaining to an otherwise older bar; this suggests that the bluer bands may perhaps be allowing us to trace these younger stars in bar orbits. The increase in bar ellipticity, on the other hand, we interpret as a result of the fact that the bulge, composed primarily of old and red stars, is less prominent at bluer bands, allowing for thinner ellipses to be fit within the bar region. Part of the ellipticity increase, however, could be due to the differences between the properties of the bar as outlined by the young stars, compared to the older ones, in good agreement with what is seen in simulations. The resulting effect is that bars appear longer and thinner when traced in bluer wavebands. 

(2) We find that, to first order, bars are measured to have a size $\sim9$\% longer in the B-band, compared to their ``canonical" size at 3.6~$\mu$m and that their ellipticity increases by $\sim8$\%. This result mainly places constraints on the impact that band-shifting effects may have for bar studies at high redshift.  

(3) Although we find that $\gtrsim50\%$ of the bars disappear in the GALEX NUV/FUV bands, the results on bar ellipticity and length extend to those cases in which the bar is still visible in the UV. Our results can be used as a reference to implement band-shifting corrections to reliably gauge any intrinsic redshift evolution of bar properties beyond $z\sim0.8$, when optical filters start tracing rest-frame UV bands.

These results are pioneering in providing a reference for band-shifting corrections in order to reliably gauge any intrinsic redshift evolution of bar properties beyond $z\sim0.8$. This opens the door for current and future studies to exploit current high-resolution imaging surveys and extend detailed morphological studies of bars out to the high-redshift universe.

\section*{Acknowledgements}

The authors wish to thank the referee for comments and suggestions that improved this paper, as well as the the entire S4G team for their efforts with this project. This work is based on observations made with the Spitzer Space Telescope, which is operated by the Jet Propulsion Laboratory, California Institute of Technology under a contract with NASA. Support for this work was provided by NASA through an award issued by JPL/Caltech. This publication makes use of data products from the Two Micron All Sky Survey, which is a joint project of the University of Massachusetts and the Infrared Processing and Analysis Center/California Institute of Technology, funded by the National Aero- nautics and Space Administration and the National Science Foundation. KMD thanks the support of the Serrapilheira Institute (grant Serra-1709-17357) as well as that of the Brazilian National Research Council (CNPq grant 308584/2022-8) and of the Rio de Janeiro State Research Foundation (FAPERJ grant E-26/200.952/2022), Brazil. TSG would also like to thank the support of CNPq (Productivity in Research grant 314747/2020-6) and the FAPERJ (Young Scientist of Our State grant E-26/201.309/2021). TK acknowledges support from the Basic Science Research Program through the National Research Foundation of Korea (NRF) funded by the Ministry of Education (RS-2023-00240212 and No. 2019R1I1A3A02062242) and the grant funded by the Korean government (MSIT) (No. 2022R1A4A3031306 and WISET 2022-804). DG acknowledges support from STFC grants ST/T000244/1 and ST/X001075/1. EA and AB acknowledge support from the Centre National d’Etudes Spatiales (CNES), France. JHK acknowledges financial support from the State Research Agency (AEI-MCINN) of the Spanish Ministry of Science and Innovation under the grant "The structure and evolution of galaxies and their central regions" with reference PID2019-105602GB-I00/10.13039/501100011033, from the ACIISI, Consejería de Economía, Conocimiento y Empleo del Gobierno de Canarias and the European Regional Development Fund (ERDF) under grant with reference PROID2021010044, and from IAC project P/300724, financed by the Ministry of Science and Innovation, through the State Budget and by the Canary Islands Department of Economy, Knowledge and Employment, through the Regional Budget of the Autonomous Community. REGM acknowledges support from CNPq through grants 303426/2018-7 and 406908/2018-4, as well as the support from Funda\c{c}\~{a}o de Apoio \`{a} Ci\^{e}ncia, Tecnologia e Inova\c{c}\~{a}o do Paran\'{a} through grant 18.148.096- 3 – NAPI Fen\^{o}menos Extremos do Universo.

%
%

\begin{table*}
\caption{Bar Properties}
\begin{threeparttable}
\centering
\begin{tabular}{lllllrrrr}
\hline
\hline
Galaxy &  & \multicolumn{3}{c}{S$^4$G [RC3] Classification\textsuperscript{a}} & & Inclination\textsuperscript{b} [deg] & Distance\textsuperscript{b} [Mpc] & Visual bar length\textsuperscript{c} [\arcsec] \\ \\
{\it Band} & $\epsilon _{max}$ & a$_{\epsilon max}$ & a$_{\Delta\epsilon}$ & a$_{\Delta PA}$ & $\epsilon _{max}$ & a$_{\epsilon max}$ & a$_{\Delta\epsilon}$ & a$_{\Delta PA}$ \\
 & $\delta\epsilon = 0.01$ & [$\arcsec$] & [$\arcsec$] & [$\arcsec$] & $\delta\epsilon = 0.01$ & [$\arcsec$] & [$\arcsec$] & [$\arcsec$]\\
\hline
& \multicolumn{4}{c}{\it Optical-to-MIR high-resolution study} & \multicolumn{4}{c}{\it UV-through-MIR low-resolution study} \\
{\bf NGC0337}	&				& \multicolumn{3}{c}{	SAB(s)cd:  pec [SB(s)d]} 	&				&	50.6	&	22.49	&	19.4\\
3.6$\mu$m	&	0.66	&	26	$\pm$	2	&	35	$\pm$	3	&	31	$\pm$	2	&	0.45	&	21	$\pm$	6	&	27	$\pm$	10	&	33	$\pm$	6	\\
R	&	0.59	&	26	$\pm$	2	&	33	$\pm$	2	&	33	$\pm$	2	&	0.46	&	27	$\pm$	6	&	33	$\pm$	11	&	33	$\pm$	11	\\
B	&	0.70	&	31	$\pm$	2	&	35	$\pm$	2	&	35	$\pm$	2	&	0.57	&	27	$\pm$	6	&	33	$\pm$	8	&	33	$\pm$	6	\\
NUV	&	$-$	$-$		&	$-$	$-$	&	$-$	$-$	&	$-$	$-$	&	0.72	&	27	$\pm$	6	&	39	$\pm$	14	&	39	$\pm$	14	\\
FUV	&	$-$	$-$		&	$-$	$-$	&	$-$	$-$	&	$-$	$-$	&	0.75	&	27	$\pm$	6	&	39	$\pm$	6	&	39	$\pm$	6	\\
\\
{\bf NGC1097}	&				& \multicolumn{3}{c}{	(R)SB(rs,nr)ab pec [SB(s)b]} 	&				&	37.0	&	15.21	&	94.0\\
3.6$\mu$m	&	0.65	&	94	$\pm$	4	&	115	$\pm$	2	&	117	$\pm$	2	&	0.62	&	94	$\pm$	6	&	112	$\pm$	11	&	118	$\pm$	6	\\
R	&	0.63	&	99	$\pm$	9	&	99	$\pm$	2	&	119	$\pm$	2	&	0.61	&	92	$\pm$	9	&	118	$\pm$	14	&	118	$\pm$	6	\\
B	&	0.69	&	112	$\pm$	2	&	128	$\pm$	2	&	142	$\pm$	2	&	0.66	&	112	$\pm$	6	&	124	$\pm$	6	&	124	$\pm$	6	\\
NUV	&	$-$	$-$		&	$-$	$-$	&	$-$	$-$	&	$-$	$-$	&	0.79	&	120	$\pm$	6	&	141	$\pm$	6	&	141	$\pm$	6	\\
FUV	&	$-$	$-$		&	$-$	$-$	&	$-$	$-$	&	$-$	$-$	&	0.84	&	123	$\pm$	6	&	141	$\pm$	6	&	141	$\pm$	6	\\
\\
{\bf NGC1291}	&				& \multicolumn{3}{c}{	(R)SAB(l,nb)0$^+$ [(R)SB(s)0$/$a]} 	&				&	29.4	&	8.47	&	97.2\\
3.6$\mu$m	&	0.40	&	94	$\pm$	6	&	99	$\pm$	2	&	124	$\pm$	3	&	0.39	&	88	$\pm$	6	&	106	$\pm$	19	&	124	$\pm$	15	\\
R	&	0.39	&	94	$\pm$	6	&	108	$\pm$	13	&	94	$\pm$	2	&	0.38	&	88	$\pm$	6	&	106	$\pm$	18	&	124	$\pm$	11	\\
B	&	0.39	&	90	$\pm$	4	&	108	$\pm$	15	&	92	$\pm$	3	&	0.37	&	88	$\pm$	6	&	106	$\pm$	18	&	100	$\pm$	9	\\
NUV	&	$-$	$-$		&	$-$	$-$	&	$-$	$-$	&	$-$	$-$	&	0.31	&	90	$\pm$	12	&	120	$\pm$	6	&	126	$\pm$	6	\\
FUV	&	$-$	$-$		&	$-$	$-$	&	$-$	$-$	&	$-$	$-$	&	0.29	&	100	$\pm$	6	&	118	$\pm$	6	&	124	$\pm$	6	\\
\\
{\bf NGC1512}	&				& \multicolumn{3}{c}{	(RL)SB(r,nr)a [SB(r)a]} 	&				&	68.3	&	9.16	&	71.4\\
3.6$\mu$m	&	0.66	&	74	$\pm$	6	&	90	$\pm$	2	&	90	$\pm$	2	&	0.62	&	76	$\pm$	6	&	88	$\pm$	6	&	106	$\pm$	9	\\
R	&	0.65	&	73	$\pm$	2	&	92	$\pm$	2	&	92	$\pm$	2	&	0.62	&	82	$\pm$	6	&	88	$\pm$	6	&	100	$\pm$	6	\\
B	&	0.72	&	90	$\pm$	2	&	90	$\pm$	2	&	90	$\pm$	2	&	0.67	&	82	$\pm$	6	&	94	$\pm$	6	&	106	$\pm$	6	\\
NUV	&	$-$	$-$		&	$-$	$-$	&	$-$	$-$	&	$-$	$-$	&	$-$	$-$	&	$-$	$-$	&	$-$	$-$	&	$-$	$-$	\\
FUV	&	$-$	$-$		&	$-$	$-$	&	$-$	$-$	&	$-$	$-$	&	$-$	$-$	&	$-$	$-$	&	$-$	$-$	&	$-$	$-$	\\
\\
{\bf NGC1566}	&				& \multicolumn{3}{c}{	(R1')SAB(s)b [SAB(s)bc]} 	&				&	47.9	&	17.46	&	35.8\\
3.6$\mu$m	&	0.49	&	32	$\pm$	2	&	36	$\pm$	2	&	43	$\pm$	3	&	0.34	&	34	$\pm$	6	&	34	$\pm$	6	&	40	$\pm$	6	\\
R	&	0.42	&	32	$\pm$	2	&	43	$\pm$	2	&	43	$\pm$	2	&	0.30	&	34	$\pm$	6	&	34	$\pm$	6	&	40	$\pm$	6	\\
B	&	0.57	&	34	$\pm$	2	&	36	$\pm$	2	&	47	$\pm$	4	&	0.30	&	34	$\pm$	6	&	34	$\pm$	6	&	40	$\pm$	6	\\
NUV	&	$-$	$-$		&	$-$	$-$	&	$-$	$-$	&	$-$	$-$	&	$-$	$-$	&	$-$	$-$	&	$-$	$-$	&	$-$	$-$	\\
FUV	&	$-$	$-$		&	$-$	$-$	&	$-$	$-$	&	$-$	$-$	&	$-$	$-$	&	$-$	$-$	&	$-$	$-$	&	$-$	$-$	\\
\\
{\bf NGC3049}	&				& \multicolumn{3}{c}{	SB(s)ab:  [SB(rs)ab]} 	&				&	58.1	&	21.28	&	36.2\\
3.6$\mu$m	&	0.77	&	38	$\pm$	3	&	52	$\pm$	2	&	65	$\pm$	2	&	0.61	&	40	$\pm$	6	&	52	$\pm$	6	&	76	$\pm$	6	\\
R	&	0.80	&	38	$\pm$	6	&	54	$\pm$	2	&	68	$\pm$	2	&	0.64	&	44	$\pm$	6	&	50	$\pm$	6	&	74	$\pm$	6	\\
B	&	0.82	&	43	$\pm$	8	&	58	$\pm$	2	&	70	$\pm$	2	&	0.65	&	46	$\pm$	6	&	52	$\pm$	6	&	70	$\pm$	6	\\
NUV	&	$-$	$-$		&	$-$	$-$	&	$-$	$-$	&	$-$	$-$	&	0.59	&	50	$\pm$	6	&	50	$\pm$	9	&	50	$\pm$	9	\\
FUV	&	$-$	$-$		&	$-$	$-$	&	$-$	$-$	&	$-$	$-$	&	0.65	&	52	$\pm$	6	&	58	$\pm$	6	&	88	$\pm$	6	\\
\\
{\bf NGC3198}	&				& \multicolumn{3}{c}{	SAB(rs)bc [SB(rs)c]} 	&				&	70.0	&	12.13	&	28.0\\
3.6$\mu$m	&	0.77	&	68	$\pm$	3	&	81	$\pm$	2	&	83	$\pm$	2	&	0.71	&	70	$\pm$	6	&	88	$\pm$	6	&	82	$\pm$	7	\\
R	&	0.73	&	76	$\pm$	9	&	86	$\pm$	2	&	86	$\pm$	2	&	0.69	&	76	$\pm$	6	&	88	$\pm$	9	&	82	$\pm$	17	\\
B	&	0.73	&	74	$\pm$	10	&	88	$\pm$	2	&	88	$\pm$	2	&	0.70	&	76	$\pm$	6	&	88	$\pm$	8	&	82	$\pm$	6	\\
NUV	&	$-$	$-$		&	$-$	$-$	&	$-$	$-$	&	$-$	$-$	&	$-$	$-$	&	$-$	$-$	&	$-$	$-$	&	$-$	$-$	\\
FUV	&	$-$	$-$		&	$-$	$-$	&	$-$	$-$	&	$-$	$-$	&	$-$	$-$	&	$-$	$-$	&	$-$	$-$	&	$-$	$-$	\\
\\
{\bf NGC3351}	&				& \multicolumn{3}{c}{	(R')SB(r,nr)a [SB(r)b]} 	&				&	41.5	&	8.79	&	51.4\\
3.6$\mu$m	&	0.46	&	60	$\pm$	6	&	68	$\pm$	2	&	68	$\pm$	2	&	0.42	&	58	$\pm$	6	&	64	$\pm$	6	&	64	$\pm$	6	\\
R	&	0.42	&	54	$\pm$	3	&	68	$\pm$	2	&	68	$\pm$	2	&	0.38	&	58	$\pm$	6	&	64	$\pm$	6	&	64	$\pm$	6	\\
B	&	0.52	&	68	$\pm$	2	&	76	$\pm$	2	&	76	$\pm$	2	&	0.45	&	64	$\pm$	6	&	70	$\pm$	6	&	70	$\pm$	6	\\
NUV	&	$-$	$-$		&	$-$	$-$	&	$-$	$-$	&	$-$	$-$	&	$-$	$-$	&	$-$	$-$	&	$-$	$-$	&	$-$	$-$	\\
FUV	&	$-$	$-$		&	$-$	$-$	&	$-$	$-$	&	$-$	$-$	&	$-$	$-$	&	$-$	$-$	&	$-$	$-$	&	$-$	$-$	\\
\\
\hline
\end{tabular}
\label{resultstab}
\begin{tablenotes}
\small
\item (a)~S$^4$G classification from \citet{buta+10} (b)~from Hyperleda (c)~S$^4$G classification from \citet{herrera-endoqui+15}
\end{tablenotes}
\end{threeparttable}
\end{table*}

\setcounter{table}{0}
\begin{table*}
\caption{Continuation}
\begin{threeparttable}
\centering
\begin{tabular}{lllllrrrr}
\hline
\hline
Galaxy &  & \multicolumn{3}{c}{S$^4$G [RC3] Classification\textsuperscript{a}} & & Inclination\textsuperscript{b} [deg] & Distance\textsuperscript{b} [Mpc] & Visual bar length\textsuperscript{c} [\arcsec] \\ \\
{\it Band} & $\epsilon _{max}$ & a$_{\epsilon max}$ & a$_{\Delta\epsilon}$ & a$_{\Delta PA}$ & $\epsilon _{max}$ & a$_{\epsilon max}$ & a$_{\Delta\epsilon}$ & a$_{\Delta PA}$ \\
 & $\delta\epsilon = 0.01$ & [$\arcsec$] & [$\arcsec$] & [$\arcsec$] & $\delta\epsilon = 0.01$ & [$\arcsec$] & [$\arcsec$] & [$\arcsec$]\\
\hline
& \multicolumn{4}{c}{\it Optical-to-MIR high-resolution study} & \multicolumn{4}{c}{\it UV-through-MIR low-resolution study} \\
{\bf NGC3627}	&				& \multicolumn{3}{c}{	SB(s)b pec [SAB(s)b]} 	&				&	57.3	&	11.27	&	66.4\\
3.6$\mu$m	&	0.75	&	61	$\pm$	3	&	81	$\pm$	2	&	81	$\pm$	2	&	0.69	&	64	$\pm$	6	&	76	$\pm$	6	&	82	$\pm$	6	\\
R	&	0.66	&	63	$\pm$	2	&	79	$\pm$	2	&	72	$\pm$	2	&	0.60	&	64	$\pm$	6	&	76	$\pm$	9	&	88	$\pm$	9	\\
B	&	0.69	&	63	$\pm$	2	&	72	$\pm$	2	&	65	$\pm$	2	&	0.62	&	64	$\pm$	6	&	82	$\pm$	6	&	82	$\pm$	6	\\
NUV	&	$-$	$-$		&	$-$	$-$	&	$-$	$-$	&	$-$	$-$	&	$-$	$-$	&	$-$	$-$	&	$-$	$-$	&	$-$	$-$	\\
FUV	&	$-$	$-$		&	$-$	$-$	&	$-$	$-$	&	$-$	$-$	&	$-$	$-$	&	$-$	$-$	&	$-$	$-$	&	$-$	$-$	\\
\\
{\bf NGC4321}	&				& \multicolumn{3}{c}{	SAB(rs,nr)bc [SAB(s)bc]} 	&				&	30.0	&	23.99	&	59.2\\
3.6$\mu$m	&	0.59	&	59	$\pm$	4	&	70	$\pm$	2	&	70	$\pm$	2	&	0.55	&	63	$\pm$	6	&	63	$\pm$	6	&	75	$\pm$	6	\\
R	&	0.53	&	66	$\pm$	4	&	66	$\pm$	2	&	75	$\pm$	2	&	0.49	&	60	$\pm$	6	&	72	$\pm$	6	&	72	$\pm$	6	\\
B	&	0.66	&	76	$\pm$	2	&	79	$\pm$	2	&	79	$\pm$	2	&	0.62	&	78	$\pm$	6	&	78	$\pm$	6	&	78	$\pm$	6	\\
NUV	&	$-$	$-$		&	$-$	$-$	&	$-$	$-$	&	$-$	$-$	&	0.78	&	105	$\pm$	6	&	105	$\pm$	6	&	105	$\pm$	6	\\
FUV	&	$-$	$-$		&	$-$	$-$	&	$-$	$-$	&	$-$	$-$	&	$-$	$-$	&	$-$	$-$	&	$-$	$-$	&	$-$	$-$	\\
\\
{\bf NGC4559}	&				& \multicolumn{3}{c}{	SB(s)cd [SAB(rs)cd]} 	&				&	64.8	&	13.93	&	12.6\\
3.6$\mu$m	&	0.58	&	14	$\pm$	2	&	19	$\pm$	3	&	21	$\pm$	2	&	$-$	$-$	&	$-$	$-$	&	$-$	$-$	&	$-$	$-$	\\
R	&	0.59	&	28	$\pm$	2	&	32	$\pm$	2	&	32	$\pm$	2	&	$-$	$-$	&	$-$	$-$	&	$-$	$-$	&	$-$	$-$	\\
B	&	0.79	&	30	$\pm$	2	&	37	$\pm$	2	&	37	$\pm$	2	&	0.51	&	24	$\pm$	6	&	30	$\pm$	6	&	30	$\pm$	6	\\
NUV	&	$-$	$-$		&	$-$	$-$	&	$-$	$-$	&	$-$	$-$	&	0.48	&	20	$\pm$	6	&	26	$\pm$	6	&	26	$\pm$	6	\\
FUV	&	$-$	$-$		&	$-$	$-$	&	$-$	$-$	&	$-$	$-$	&	0.61	&	20	$\pm$	6	&	26	$\pm$	6	&	26	$\pm$	6	\\
\\
{\bf NGC4579}	&				& \multicolumn{3}{c}{	(R)SB(rs)a [SAB(rs)b]} 	&				&	39.0	&	22.91	&	40.7\\
3.6$\mu$m	&	0.47	&	43	$\pm$	2	&	52	$\pm$	3	&	52	$\pm$	2	&	0.40	&	40	$\pm$	6	&	46	$\pm$	6	&	52	$\pm$	6	\\
R	&	0.48	&	45	$\pm$	2	&	50	$\pm$	2	&	52	$\pm$	2	&	0.40	&	44	$\pm$	6	&	46	$\pm$	8	&	52	$\pm$	8	\\
B	&	0.51	&	45	$\pm$	2	&	52	$\pm$	2	&	52	$\pm$	2	&	0.42	&	46	$\pm$	6	&	46	$\pm$	6	&	52	$\pm$	6	\\
NUV	&	$-$	$-$		&	$-$	$-$	&	$-$	$-$	&	$-$	$-$	&	$-$	$-$	&	$-$	$-$	&	$-$ $-$	&	$-$	$-$	\\
FUV	&	$-$	$-$		&	$-$	$-$	&	$-$	$-$	&	$-$	$-$	&	$-$	$-$	&	$-$	$-$	&	$-$	$-$	&	$-$	$-$	\\
\\
{\bf NGC4625}	&				& \multicolumn{3}{c}{	(R)SAB(rs)m [SAB(rs)m pec]} 	&				&	46.1	&	11.75	&	7.7\\
3.6$\mu$m	&	0.31	&	9	$\pm$	2	&	11	$\pm$	2	&	27	$\pm$	2	&	$-$	$-$	&	$-$	$-$	&	$-$	$-$	&	$-$	$-$	\\
R	&	0.30	&	7	$\pm$	2	&	9	$\pm$	3	&	27	$\pm$	2	&	$-$	$-$	&	$-$	$-$	&	$-$	$-$	&	$-$	$-$	\\
B	&	0.41	&	7	$\pm$	2	&	9	$\pm$	3	&	34	$\pm$	2	&	$-$	$-$	&	$-$	$-$	&	$-$	$-$	&	$-$	$-$	\\
NUV	&	$-$	$-$		&	$-$	$-$	&	$-$	$-$	&	$-$	$-$	&	$-$	$-$	&	$-$	$-$	&	$-$	$-$	&	$-$	$-$	\\
FUV	&	$-$	$-$		&	$-$	$-$	&	$-$	$-$	&	$-$	$-$	&	$-$	$-$	&	$-$	$-$	&	$-$	$-$	&	$-$	$-$	\\
\\
{\bf NGC4725}	&				& \multicolumn{3}{c}{	SAB(r,nb)a [SAB(r)ab pec]} 	&				&	54.4	&	19.50	&	123.3\\
3.6$\mu$m	&	0.68	&	137	$\pm$	8	&	151	$\pm$	2	&	148	$\pm$	2	&	0.67	&	130	$\pm$	6	&	148	$\pm$	10	&	148	$\pm$	10	\\
R	&	0.68	&	137	$\pm$	7	&	144	$\pm$	7	&	148	$\pm$	2	&	0.66	&	130	$\pm$	6	&	148	$\pm$	12	&	142	$\pm$	12	\\
B	&	0.73	&	137	$\pm$	2	&	151	$\pm$	2	&	162	$\pm$	2	&	0.71	&	136	$\pm$	6	&	160	$\pm$	10	&	154	$\pm$	10	\\
NUV	&	$-$	$-$		&	$-$	$-$	&	$-$	$-$	&	$-$	$-$	&	$-$	$-$	&	$-$	$-$	&	$-$	$-$	&	$-$	$-$	\\
FUV	&	$-$	$-$		&	$-$	$-$	&	$-$	$-$	&	$-$	$-$	&	$-$	$-$	&	$-$	$-$	&	$-$	$-$	&	$-$	$-$	\\
\\
{\bf NGC5713}	&				& \multicolumn{3}{c}{	(R)SB(rs)ab: pec [SAB(rs)bc pec]} 	&				&	48.2	&	28.05	&	17.2\\
3.6$\mu$m	&	0.59	&	20	$\pm$	2	&	22	$\pm$	2	&	27	$\pm$	2	&	0.29	&	20	$\pm$	6	&	22	$\pm$	6	&	28	$\pm$	6	\\
R	&	0.62	&	20	$\pm$	2	&	22	$\pm$	2	&	29	$\pm$	12	&	0.31	&	16	$\pm$	6	&	22	$\pm$	6	&	34	$\pm$	6	\\
B	&	0.74	&	20	$\pm$	2	&	20	$\pm$	2	&	27	$\pm$	6	&	0.33	&	16	$\pm$	6	&	22	$\pm$	8	&	40	$\pm$	8	\\
NUV	&	$-$	$-$		&	$-$	$-$	&	$-$	$-$	&	$-$	$-$	&	$-$	$-$	&	$-$	$-$	&	$-$	$-$	&	$-$	$-$	\\
FUV	&	$-$	$-$		&	$-$	$-$	&	$-$	$-$	&	$-$	$-$	&	$-$	$-$	&	$-$	$-$	&	$-$	$-$	&	$-$	$-$	\\
\\
{\bf NGC7552}	&				& \multicolumn{3}{c}{	(R)SB(r\underline{s},nr)a [(R)SB(s)ab]} 	&				&	23.6	&	20.23	&	52.2\\
3.6$\mu$m	&	0.63	&	46	$\pm$	8	&	58	$\pm$	2	&	70	$\pm$	5	&	0.56	&	51	$\pm$	6	&	57	$\pm$	14	&	63	$\pm$	14	\\
R	&	0.63	&	45	$\pm$	4	&	54	$\pm$	5	&	68	$\pm$	5	&	0.57	&	45	$\pm$	6	&	57	$\pm$	17	&	69	$\pm$	17	\\
B	&	0.68	&	40	$\pm$	2	&	54	$\pm$	2	&	70	$\pm$	8	&	0.61	&	45	$\pm$	6	&	51	$\pm$	13	&	69	$\pm$	13	\\
NUV	&	$-$	$-$		&	$-$	$-$	&	$-$	$-$	&	$-$	$-$	&	0.63	&	45	$\pm$	6	&	51	$\pm$	6	&	69	$\pm$	6	\\
FUV	&	$-$	$-$		&	$-$	$-$	&	$-$	$-$	&	$-$	$-$	&	0.68	&	51	$\pm$	6	&	51	$\pm$	6	&	75	$\pm$	6	\\
\\
\hline
\end{tabular}
\begin{tablenotes}
\small
\item (a)~S$^4$G classification from \citet{buta+10} (b)~from Hyperleda (c)~S$^4$G classification from \citet{herrera-endoqui+15}
\end{tablenotes}
\end{threeparttable}
\end{table*}

\begin{table*}
\caption{Impact of measuring bar length and ellipticity in different bands for the {\it higher resolution optical-through-MIR study} -- sample average of relative bar measurements and statistical significance.}
\begin{threeparttable}
\centering
\label{wilcoxon_table_hires}
\begin{tabular}{l|cc}
\hline
\hline
Band & R & 3.6 \\
\hline
B & a$_{\epsilon{\rm max}}$: 3.6\% (0.073) & a$_{\epsilon{\rm max}}$: 8.6\% (0.023)\\ 
 & a$_{\Delta\epsilon}$: 3.6\% (0.130) & a$_{\Delta\epsilon}$: 0.0\% (0.090)\\ 
 & $\epsilon_{\rm max}$: 9.9\% (0.001) & $\epsilon_{\rm max}$: 8.0\% (0.005)\\ 
\\ 
R &  & a$_{\epsilon{\rm max}}$: 0.0\% (0.167)\\ 
 &  & a$_{\Delta\epsilon}$: -1.4\% (0.899)\\ 
 &  & $\epsilon_{\rm max}$: -2.7\% (0.065)\\ 
\\ 
\hline\end{tabular}
\begin{tablenotes}
\footnotesize
\item Percentages associated to the mean ratios between measurements for blue and red bands, where the bluer band is in the numerator (e.g., $a_{\epsilon \rm{max}}$ is measured to be $3.6$\% longer in the B-band compared to in the R-band and $8.6$\% longer in the B-band compared to that measured in the 3.6$\mu$m. Numbers in parentheses indicate the p-value for a Wilcoxon test of two related samples, investigating the difference between both measurements. Smaller values reinforce the probability that one sample has a larger values than the other. See Section \ref{bar props} for details.
\end{tablenotes}
\end{threeparttable}
\end{table*}

\begin{table*}
\caption{Impact of measuring bar length and ellipticity in different bands for the {\it low resolution UV-through-MIR study} -- sample average of relative bar measurements and statistical significance.}
\begin{threeparttable}
\centering
\label{wilcoxon_table_lores}
\begin{tabular}{l|cccc}
\hline
\hline
Band & NUV & B & R & 3.6 \\
\hline
FUV & a$_{\epsilon{\rm max}}$: 3.0\% (0.109) & a$_{\epsilon{\rm max}}$: 11.1\% (0.078) & a$_{\epsilon{\rm max}}$: 13.6\% (0.068) & a$_{\epsilon{\rm max}}$: 28.6\% (0.066)\\ 
 & a$_{\Delta\epsilon}$: 0.0\% (0.655) & a$_{\Delta\epsilon}$: 11.3\% (0.078) & a$_{\Delta\epsilon}$: 18.2\% (0.188) & a$_{\Delta\epsilon}$: 11.4\% (0.188)\\ 
 & $\epsilon_{\rm max}$: 7.2\% (0.062) & $\epsilon_{\rm max}$: 16.9\% (0.219) & $\epsilon_{\rm max}$: 20.5\% (0.188) & $\epsilon_{\rm max}$: 22.9\% (0.188)\\ 
\\ 
NUV &  & a$_{\epsilon{\rm max}}$: 1.7\% (0.225) & a$_{\epsilon{\rm max}}$: 7.7\% (0.068) & a$_{\epsilon{\rm max}}$: 25.4\% (0.156)\\ 
 &  & a$_{\Delta\epsilon}$: 12.7\% (0.116) & a$_{\Delta\epsilon}$: 15.4\% (0.104) & a$_{\Delta\epsilon}$: 19.0\% (0.156)\\ 
 &  & $\epsilon_{\rm max}$: 3.3\% (0.469) & $\epsilon_{\rm max}$: 20.7\% (0.219) & $\epsilon_{\rm max}$: 20.2\% (0.219)\\ 
\\ 
B &  &  & a$_{\epsilon{\rm max}}$: 0.0\% (0.027) & a$_{\epsilon{\rm max}}$: 8.2\% (0.014)\\ 
 &  &  & a$_{\Delta\epsilon}$: 2.5\% (0.048) & a$_{\Delta\epsilon}$: 3.4\% (0.031)\\ 
 &  &  & $\epsilon_{\rm max}$: 6.7\% (0.000) & $\epsilon_{\rm max}$: 6.8\% (0.049)\\ 
\\ 
R &  &  &  & a$_{\epsilon{\rm max}}$: 0.0\% (0.465)\\ 
 &  &  &  & a$_{\Delta\epsilon}$: 0.0\% (0.141)\\ 
 &  &  &  & $\epsilon_{\rm max}$: -1.6\% (0.194)\\ 
\\ 
\hline
\end{tabular}
\begin{tablenotes}
\footnotesize
\item Same as Table \ref{wilcoxon_table_hires}, for the low-resolution images.
\end{tablenotes}
\end{threeparttable}
\end{table*}

\section*{Data Availability}

The data were derived from images in the public domain: http://irsa.ipac.caltech.edu/data/SPITZER/S4G. The measurements produced as part of our work are available in the article. Additional access will be shared on reasonable request to the corresponding author.

\bibliographystyle{mnras}
\bibliography{bars}

\clearpage
\appendix

%
%

\section{Radial Profiles for the high-resolution study in the B-, R-, 3.6$\mu$m bands}
\label{appendixA_profiles_hires}

\begin{figure*}
\centering
\includegraphics[scale=0.2]{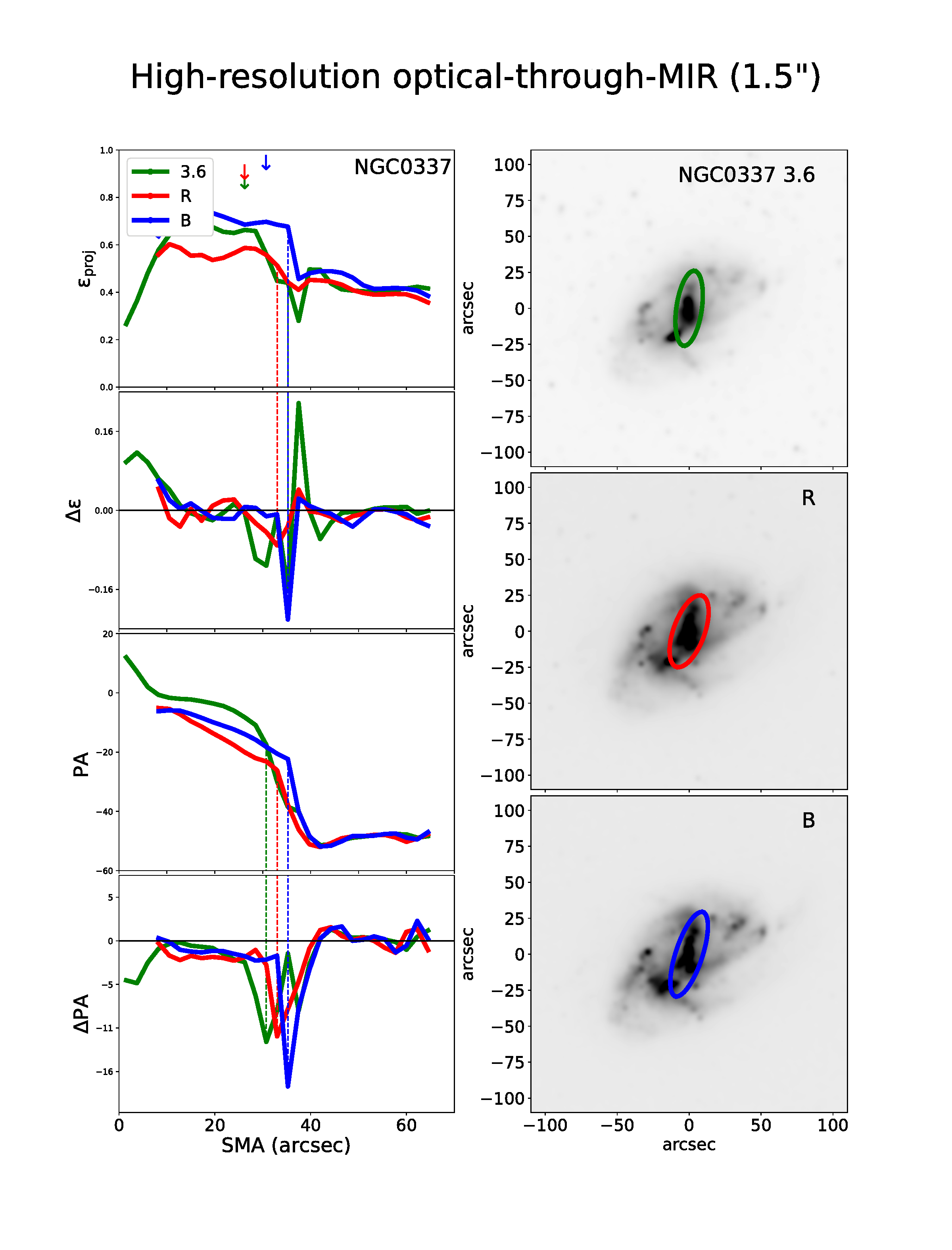}
\includegraphics[scale=0.2]{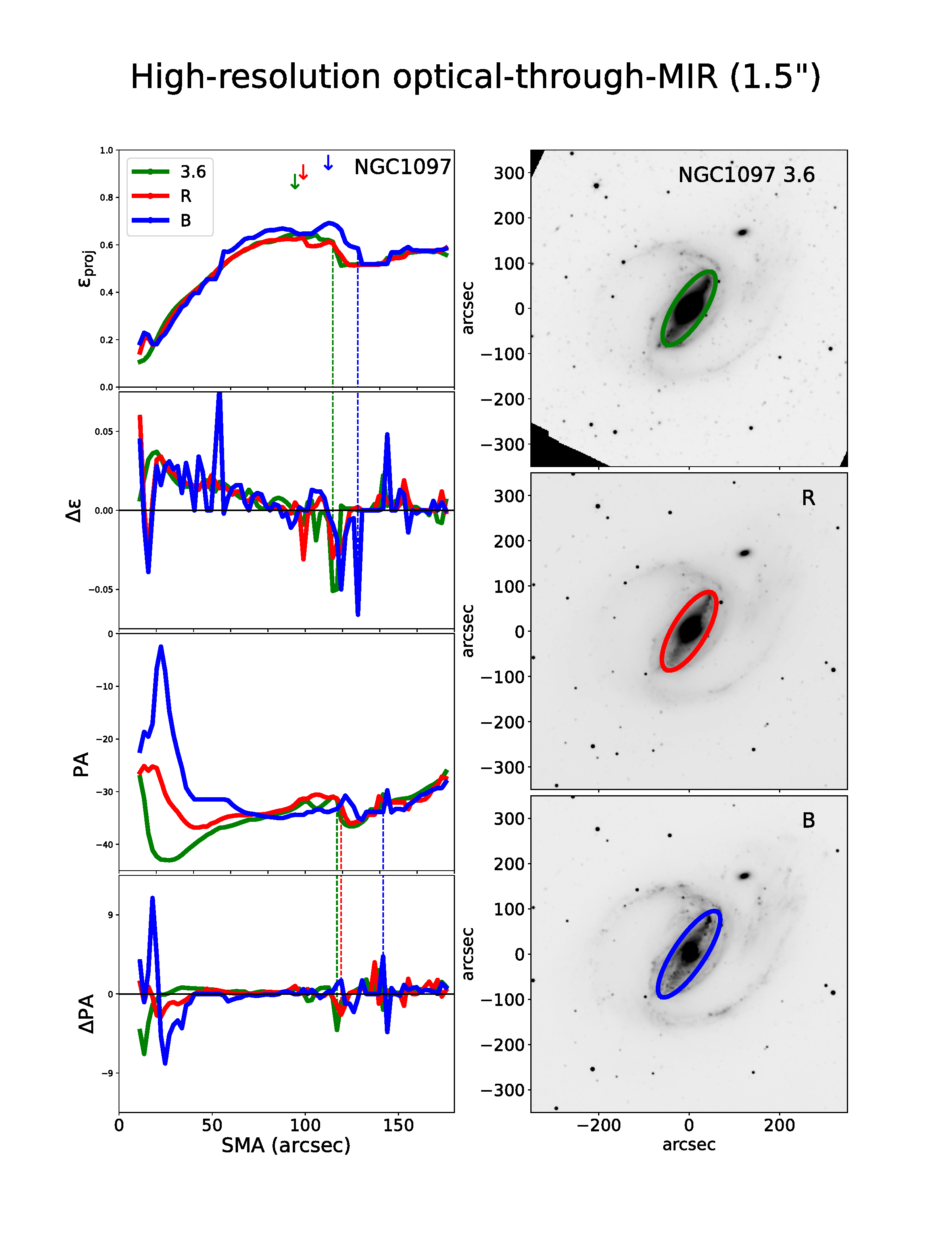}
\includegraphics[scale=0.2]{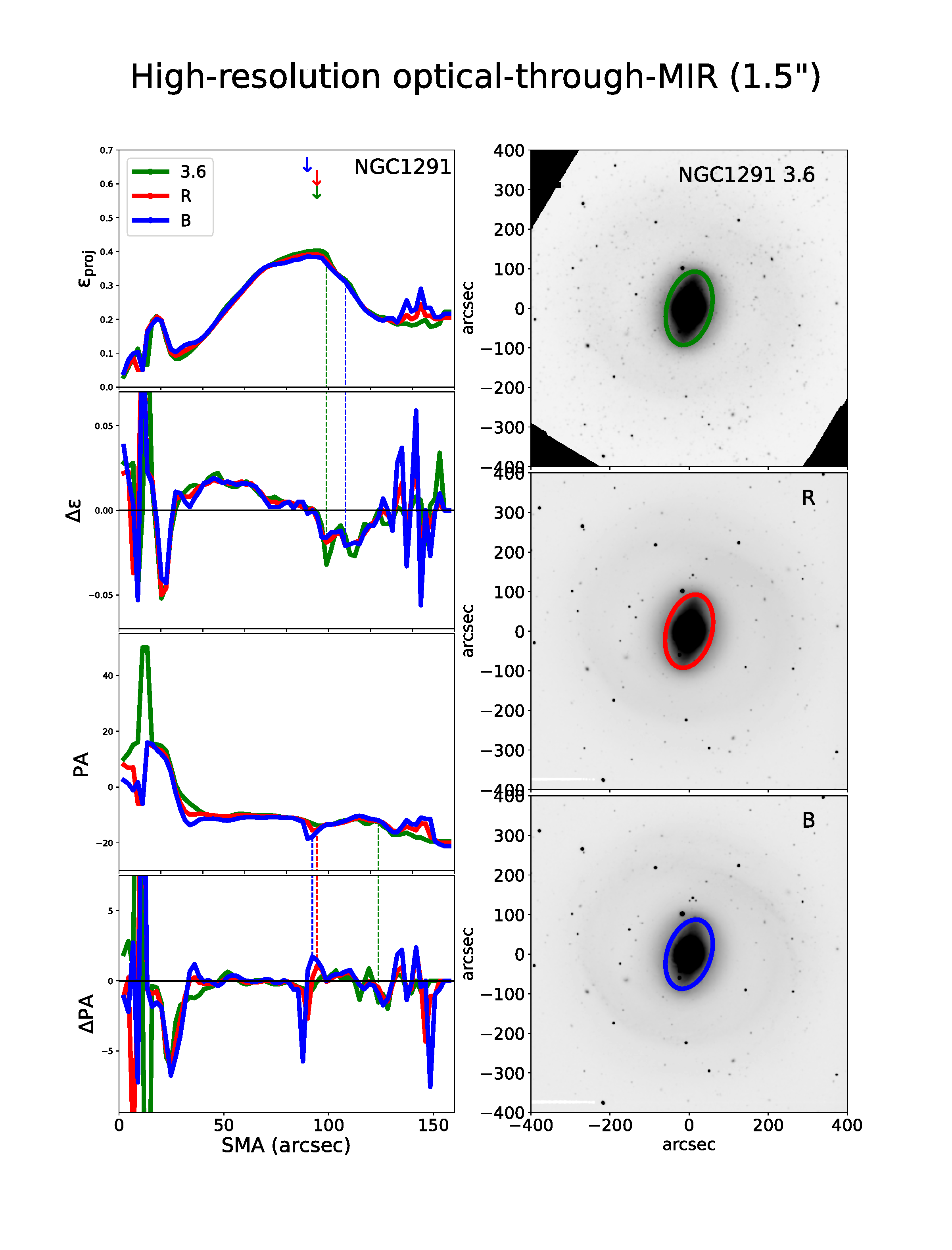}
\includegraphics[scale=0.2]{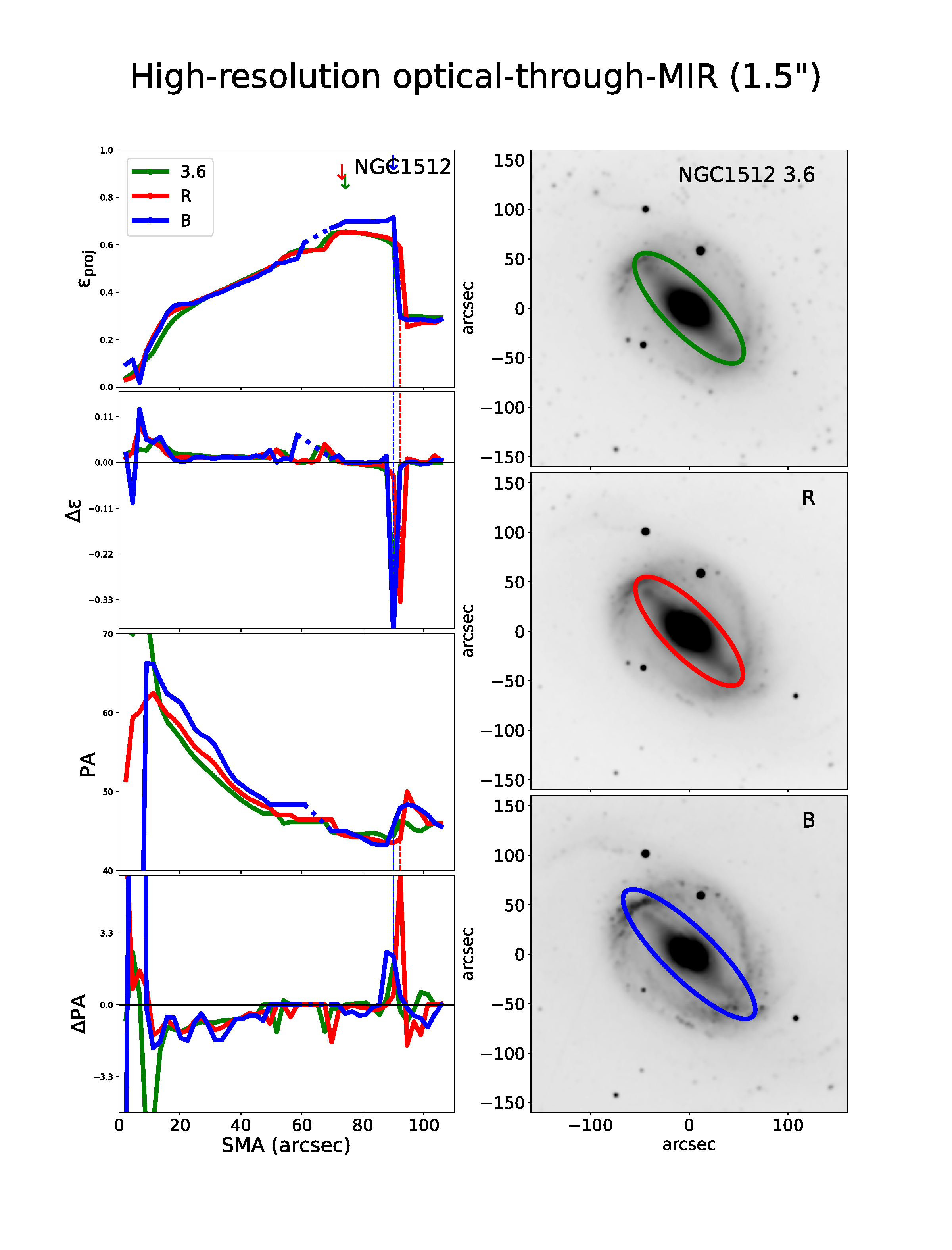}
\caption{Ellipticity and PA profiles for all galaxies in our sample as part of the {\it higher-resolution opt-through-MIR} study, following the format of Fig.\,\ref{profiles}.}
\label{appendixA_profiles_hires_fig}
\end{figure*}

We show in Fig.\,\ref{appendixA_profiles_hires_fig} the ellipse-fit results for all the galaxies in our sample. Following the same format as in Fig.\,\ref{profiles}, for each galaxy we show profiles for ellipticity and its variation, as well as position angle and its variation in the B-, R-, and 3.6$\mu$m-band. The vertical down-pointing arrows indicate the semi-major axis (SMA) of maximum ellipticity, i.e., a$_{\epsilon max}$ for each band.  Postage stamp images in the B-, R-, and 3.6$\mu$m-bands, from top to bottom, are shown for each galaxy with an overlaid ellipse showing the isophote of maximum ellipticity.



%
%

%
%
\begin{figure*}
\centering
\includegraphics[scale=0.2]{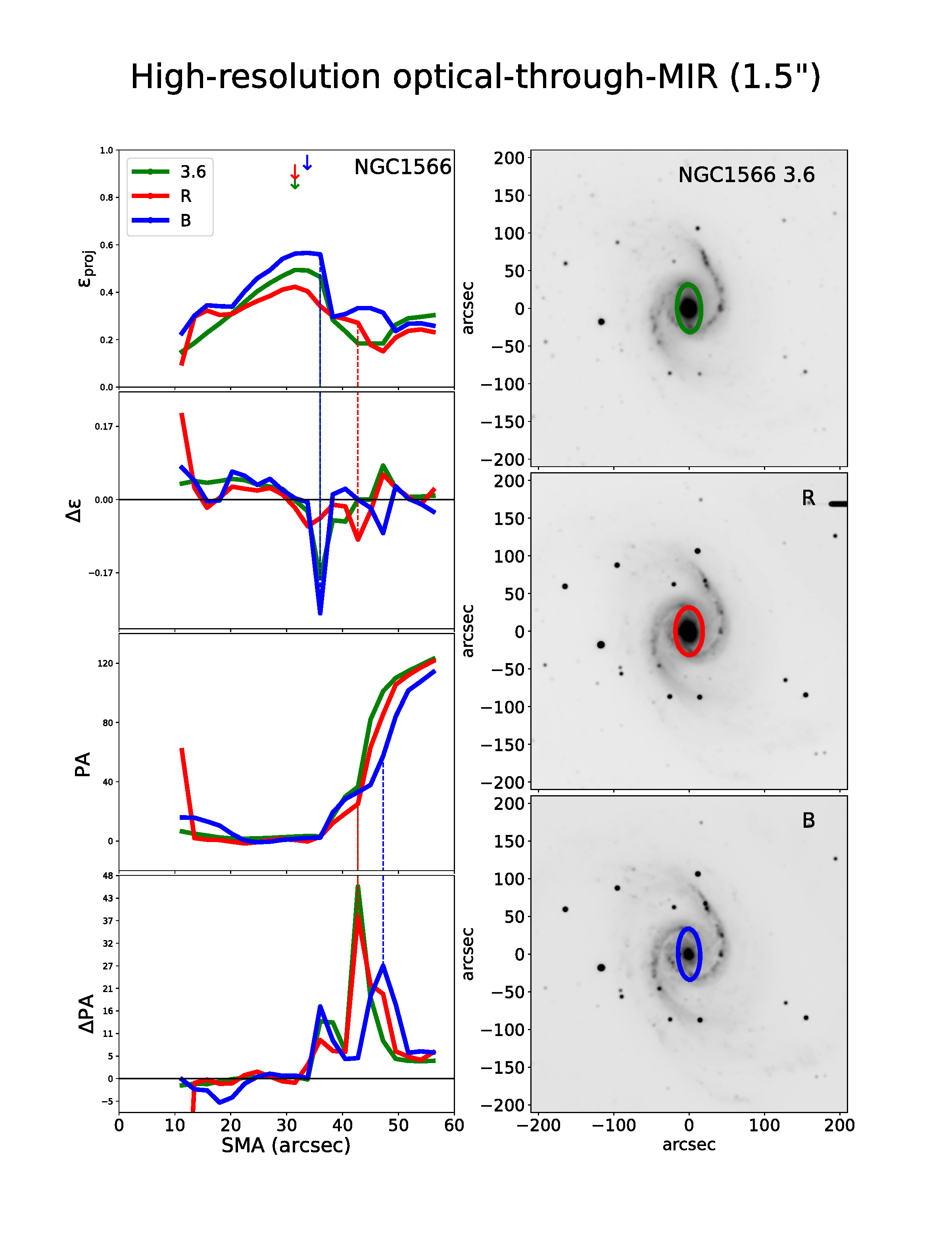}
\includegraphics[scale=0.2]{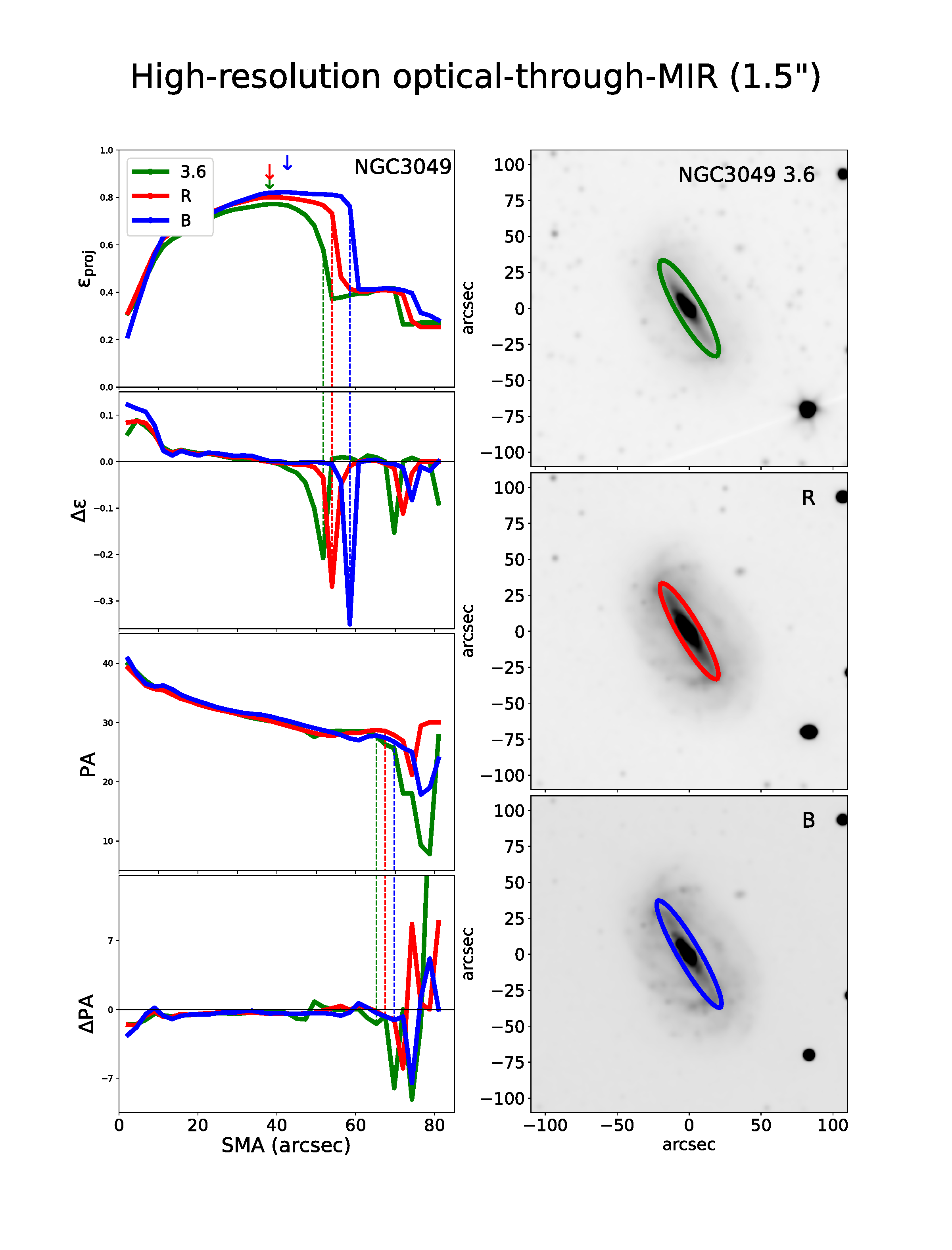}
\includegraphics[scale=0.2]{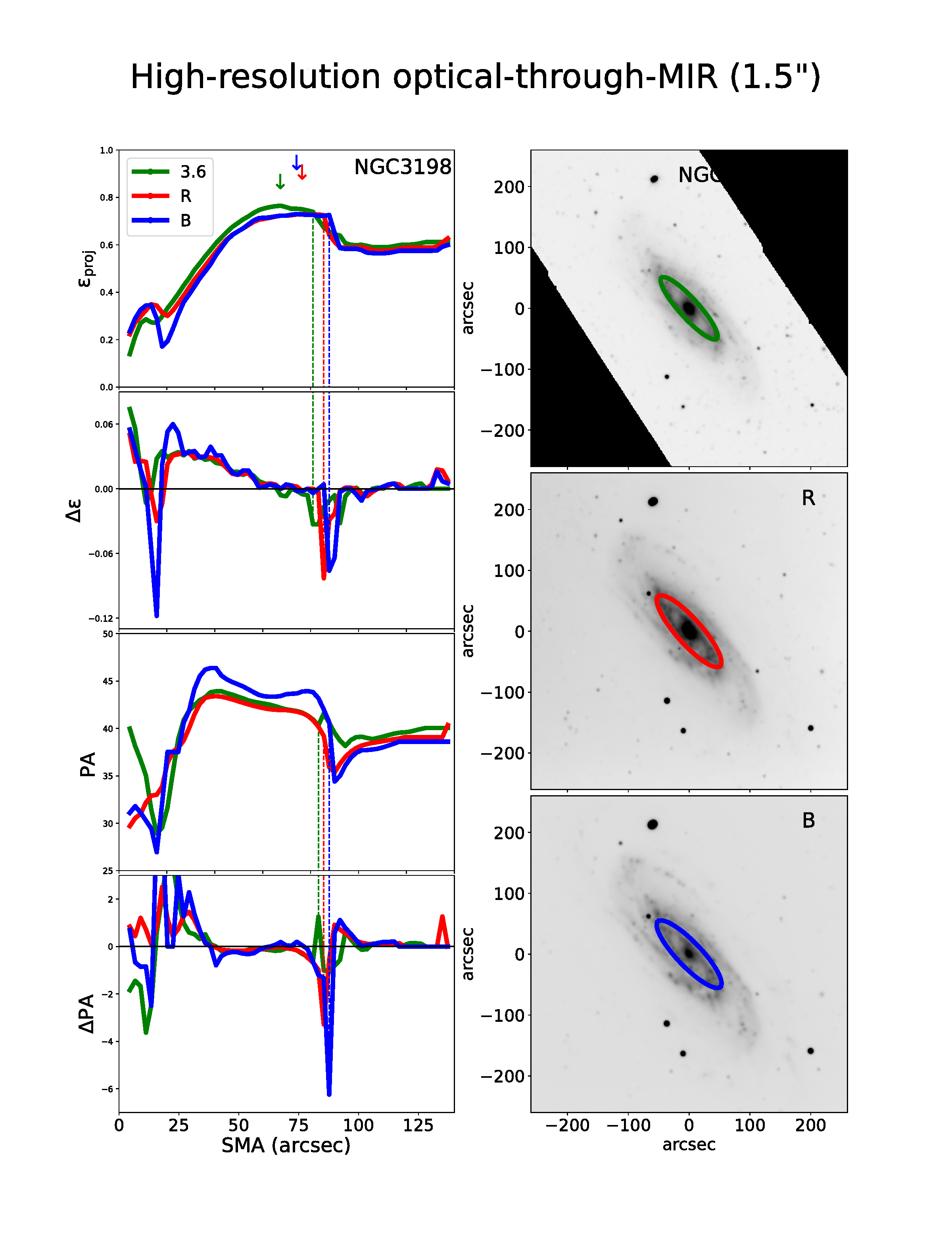}
\includegraphics[scale=0.2]{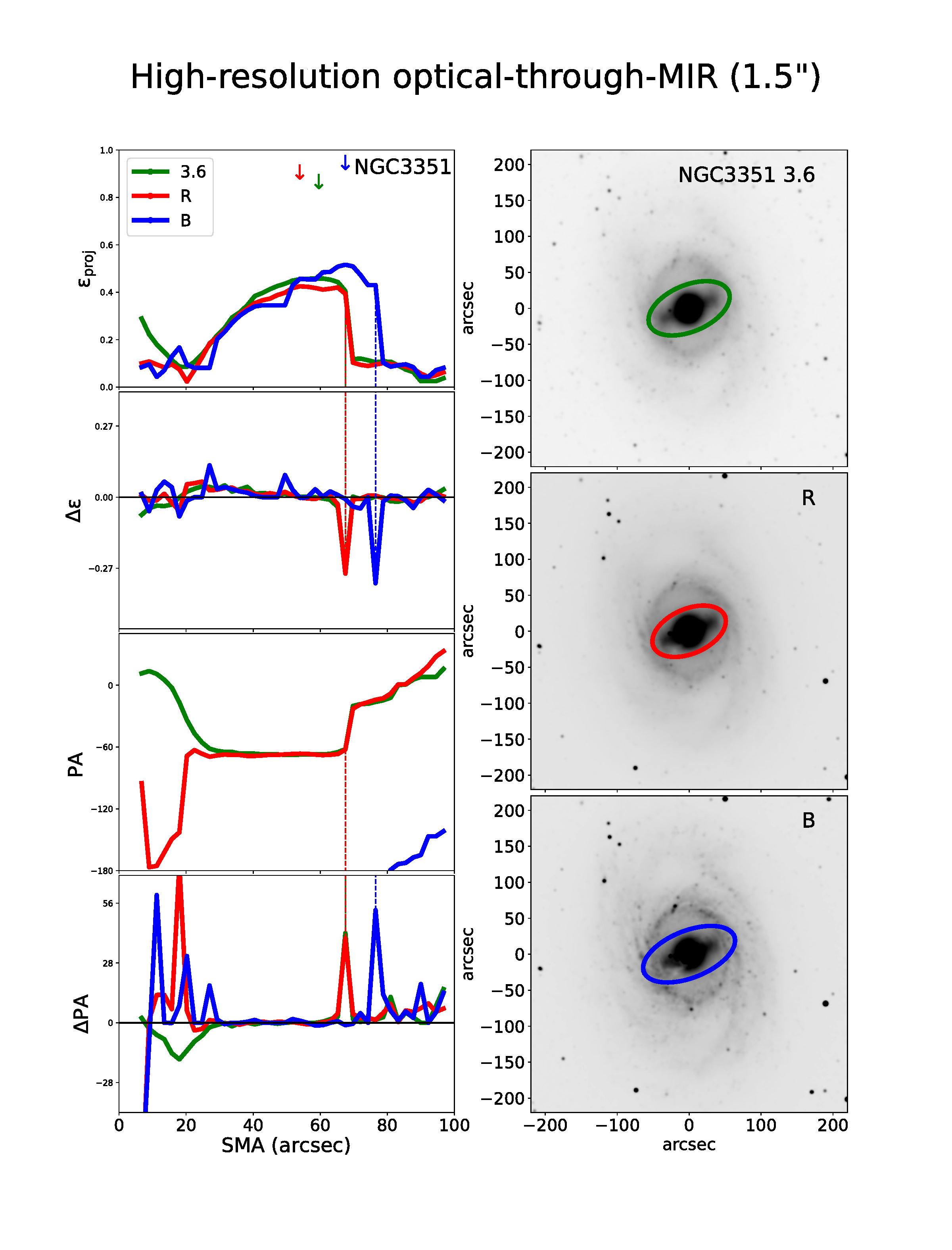}
\centerline{Fig. \ref{appendixA_profiles_hires_fig}.\ --- Continued. }
\end{figure*}

\begin{figure*}
\centering
\includegraphics[scale=0.2]{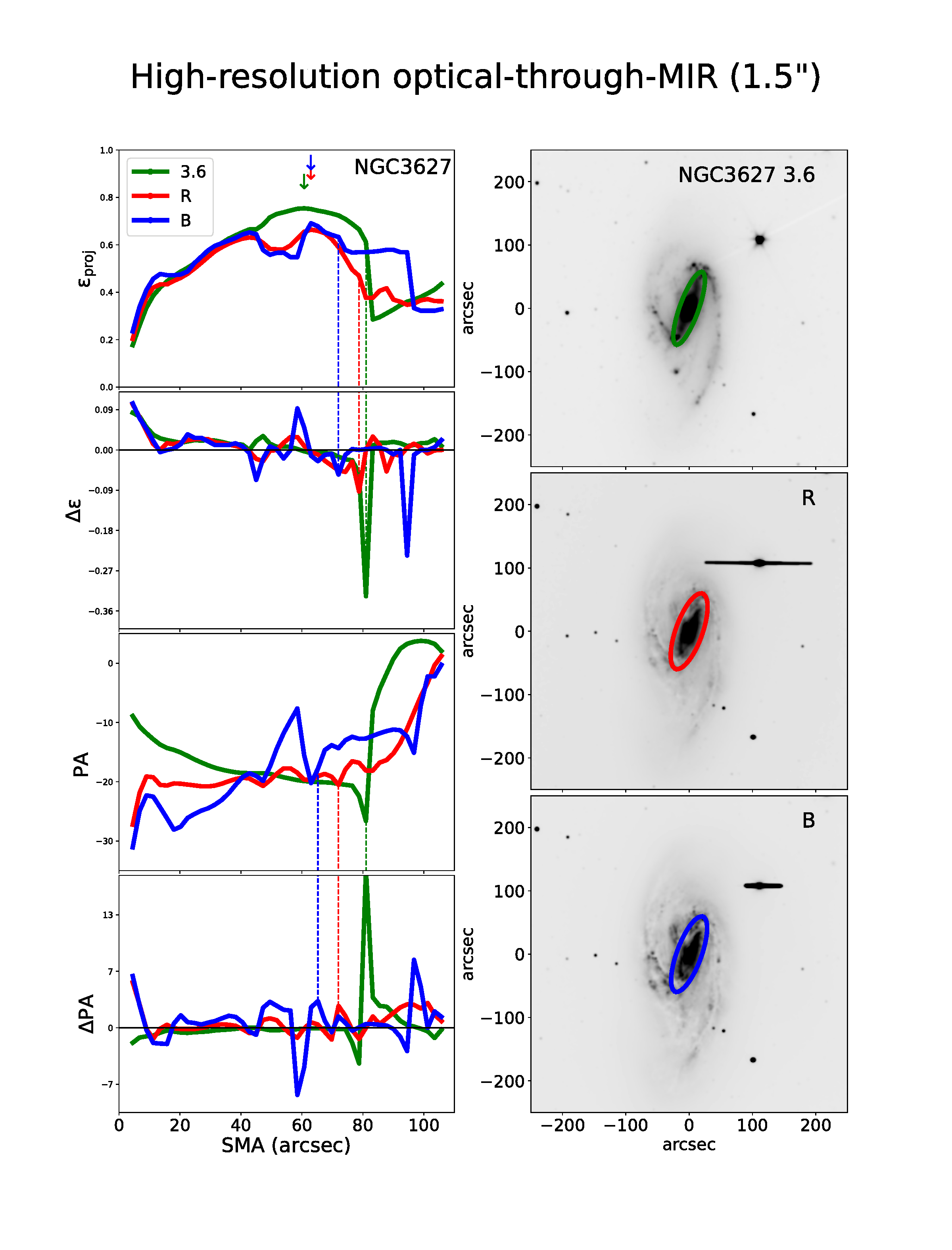}
\includegraphics[scale=0.2]{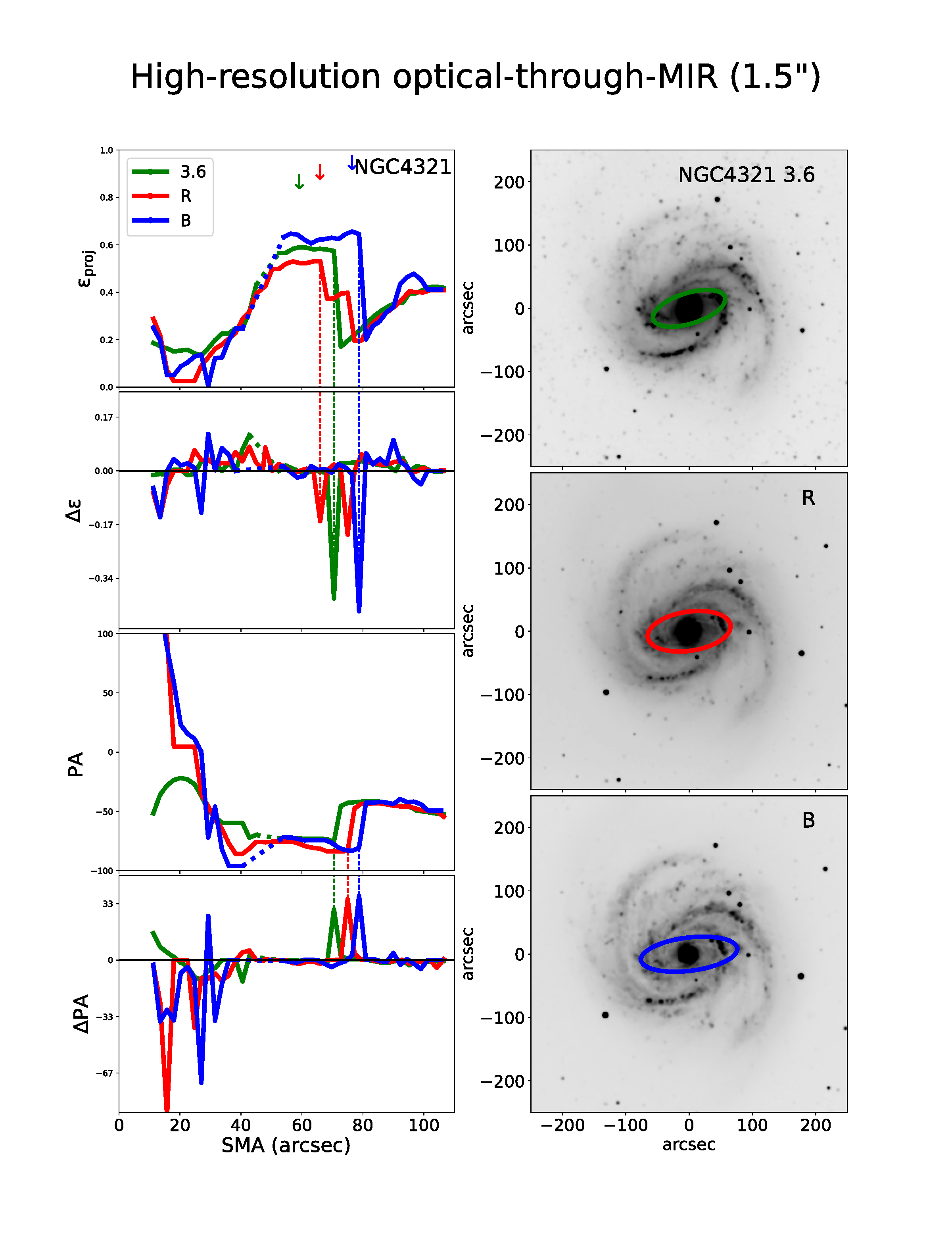}
\includegraphics[scale=0.2]{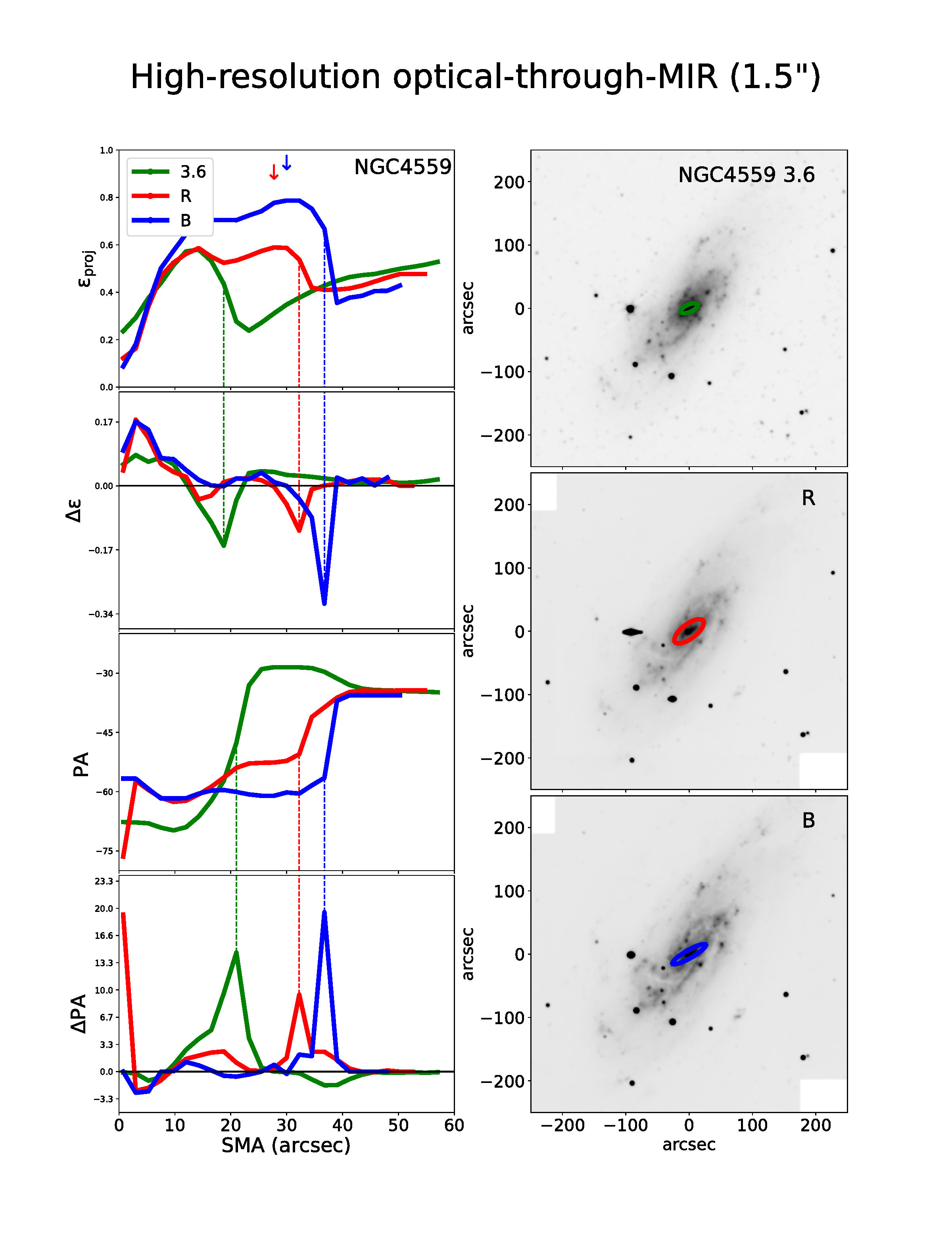}
\includegraphics[scale=0.2]{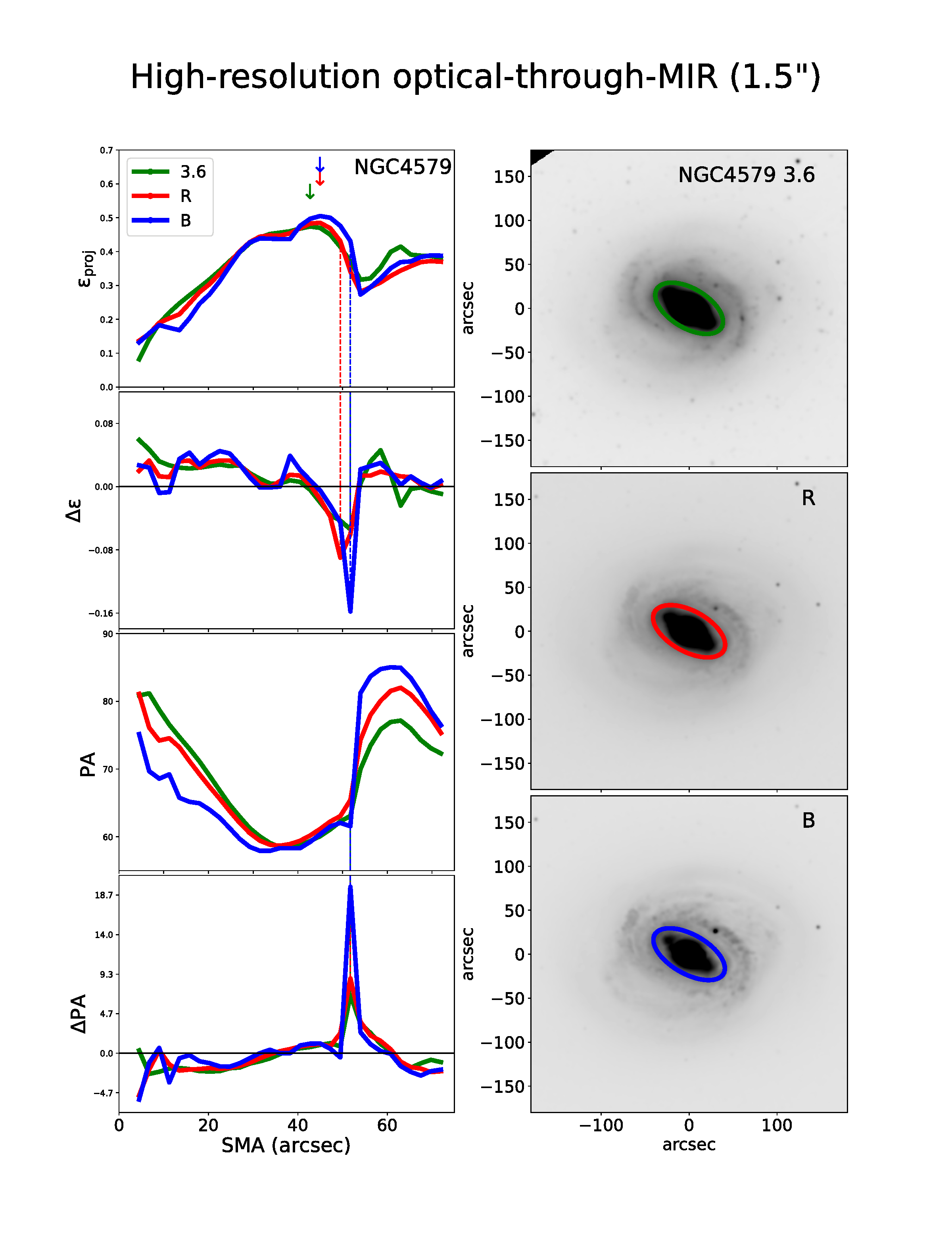}
\centerline{Fig. \ref{appendixA_profiles_hires_fig}.\ --- Continued. }
\end{figure*}

\begin{figure*}
\centering

\includegraphics[scale=0.2]{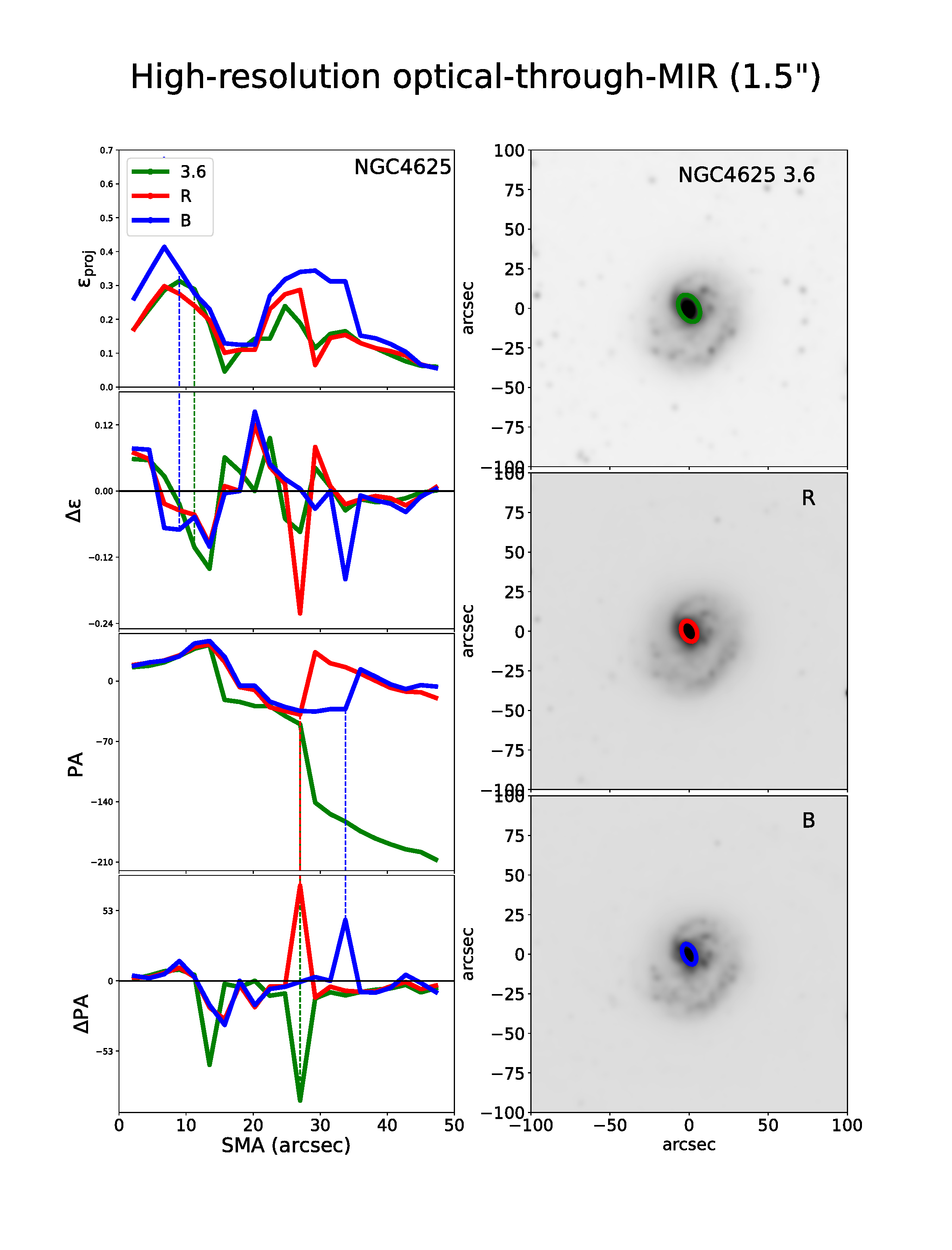}
\includegraphics[scale=0.2]{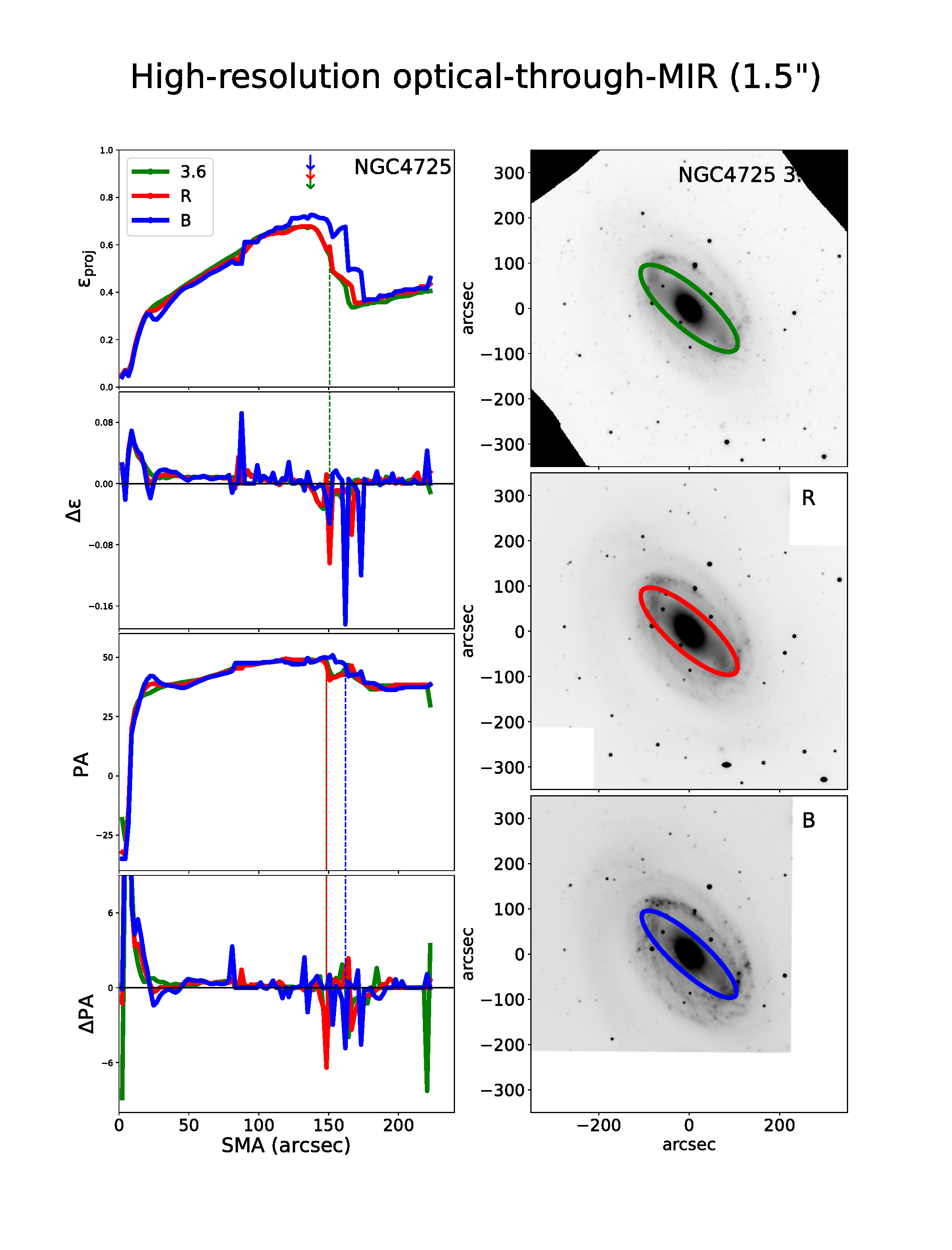}
\includegraphics[scale=0.2]{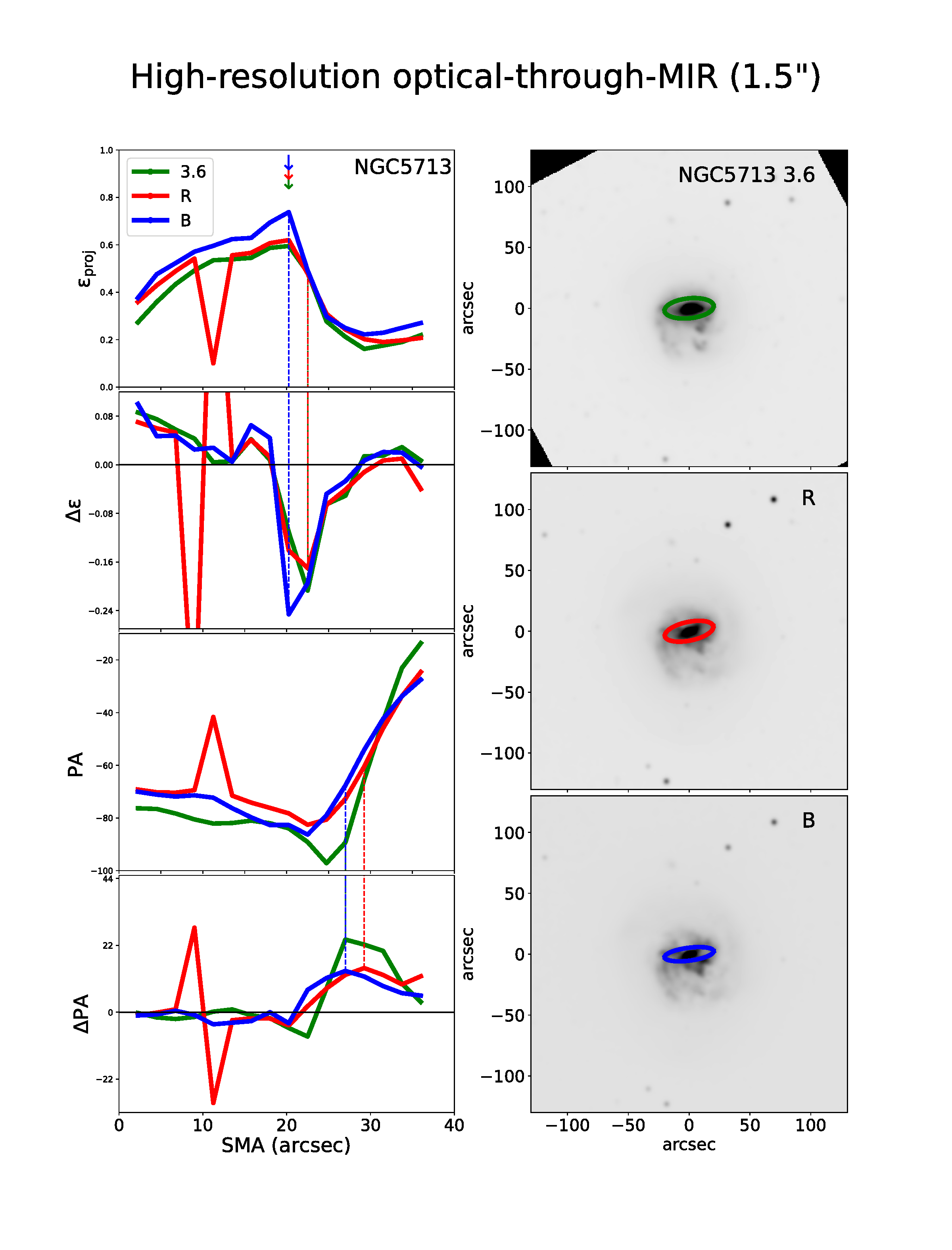}
\includegraphics[scale=0.2]{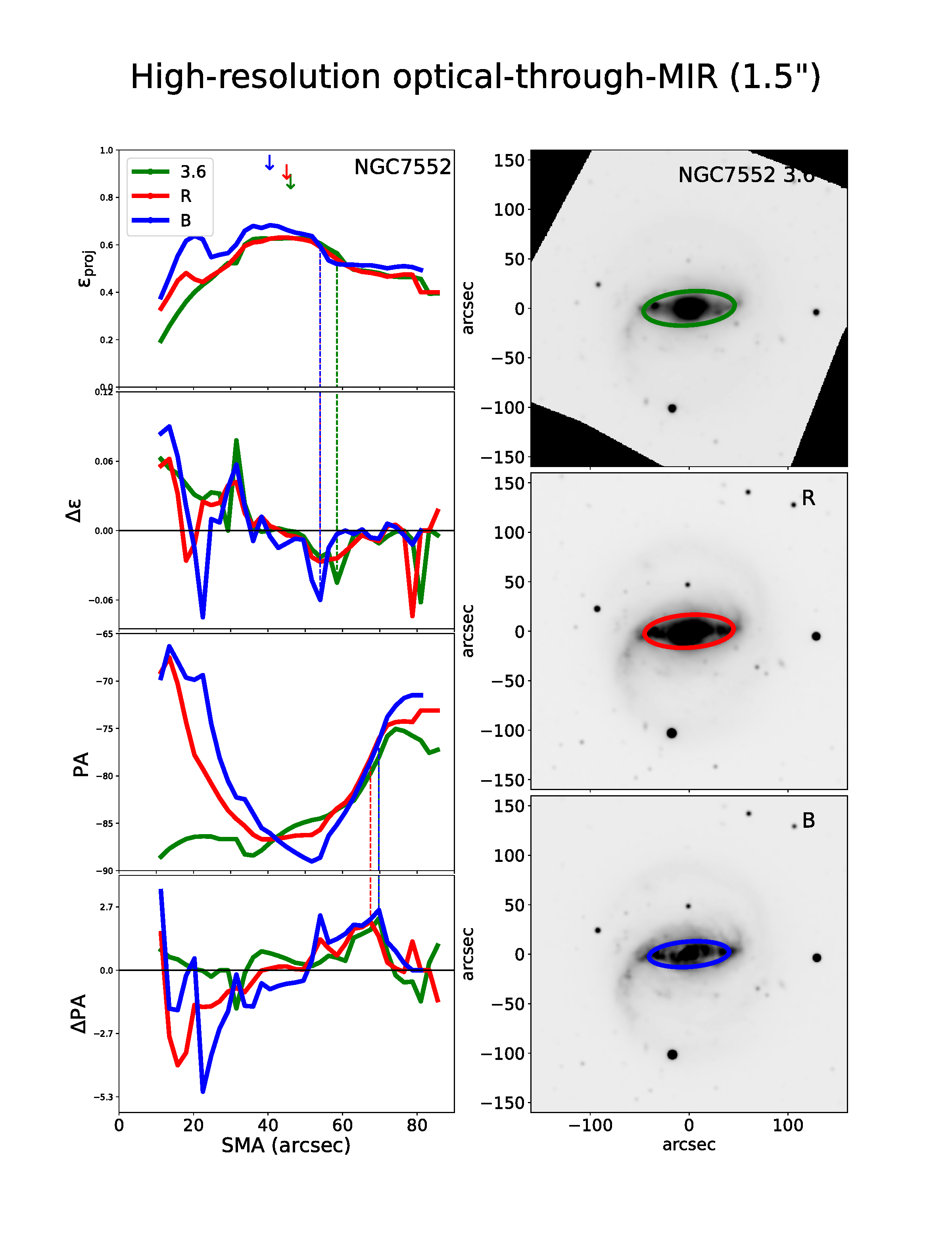}
\centerline{Fig. \ref{appendixA_profiles_hires_fig}.\ --- Continued. }
\end{figure*}

\newpage
\newpage 

%
%

\section{Radial Profiles for the low-resolution study in the FUV-, NUV-, B-, R-, 3.6$\mu$m bands}
\label{appendixB_profiles_lores}

\begin{figure*}[b]
\centering
\includegraphics[scale=0.2]{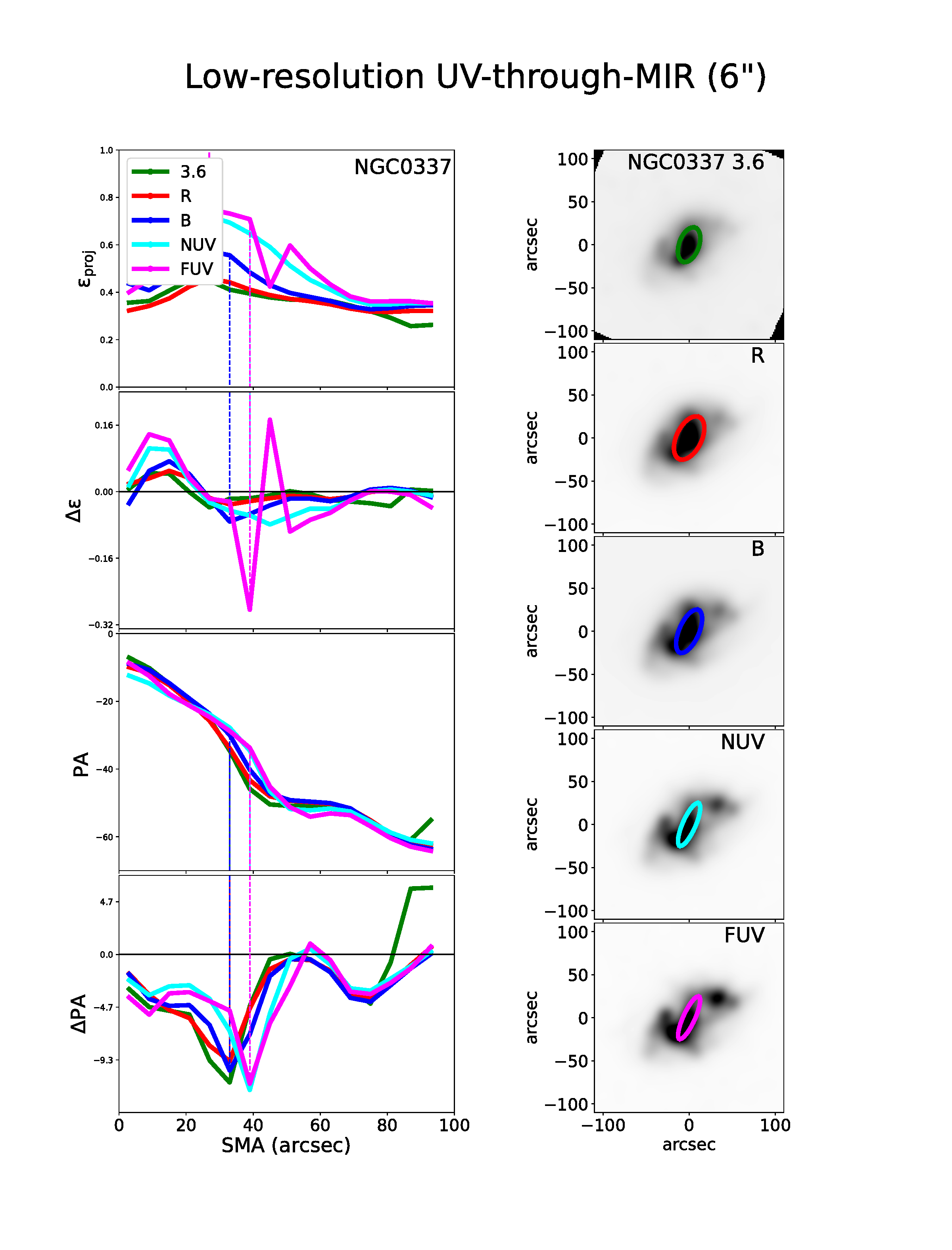}
\includegraphics[scale=0.2]{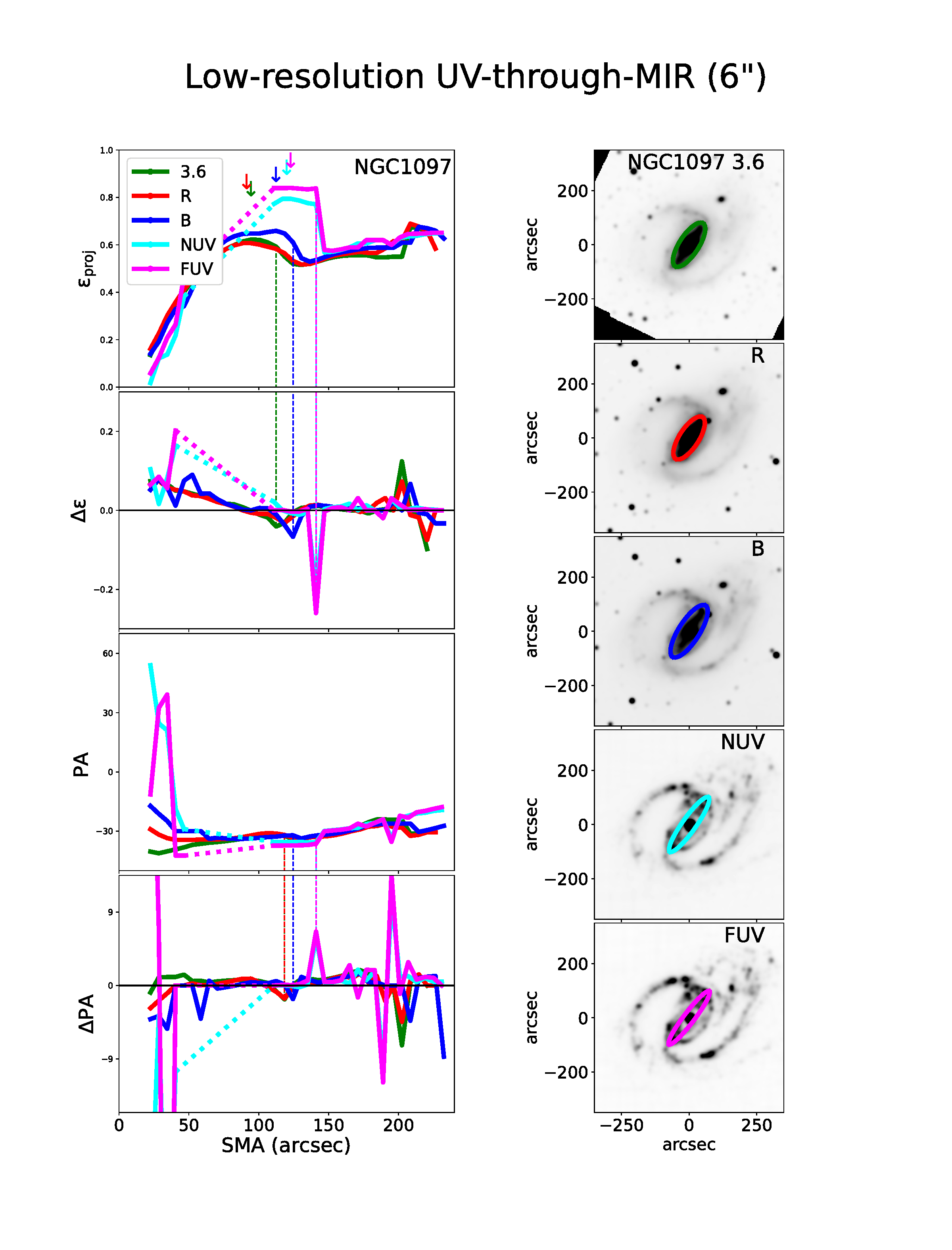}
\includegraphics[scale=0.2]{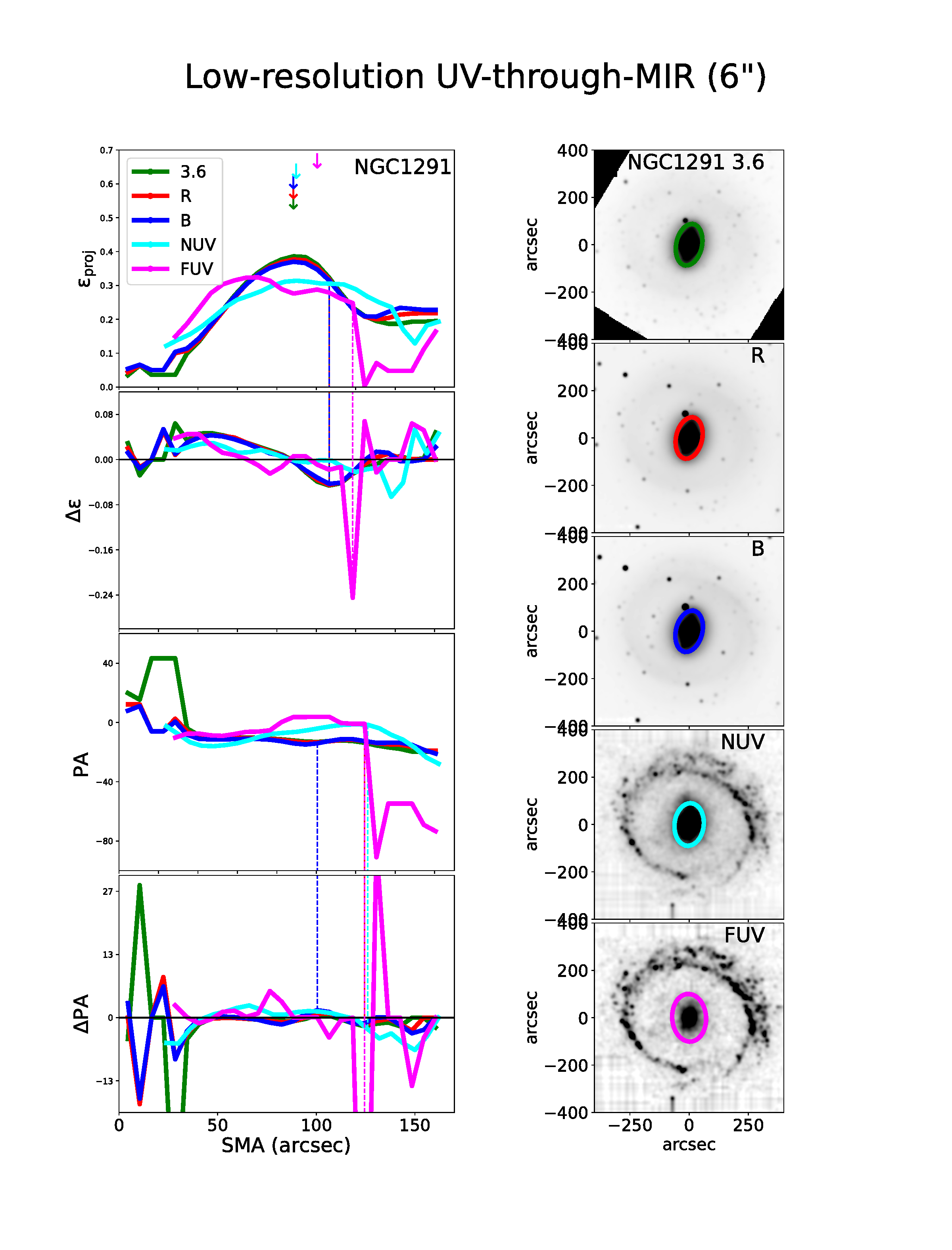}
\includegraphics[scale=0.2]{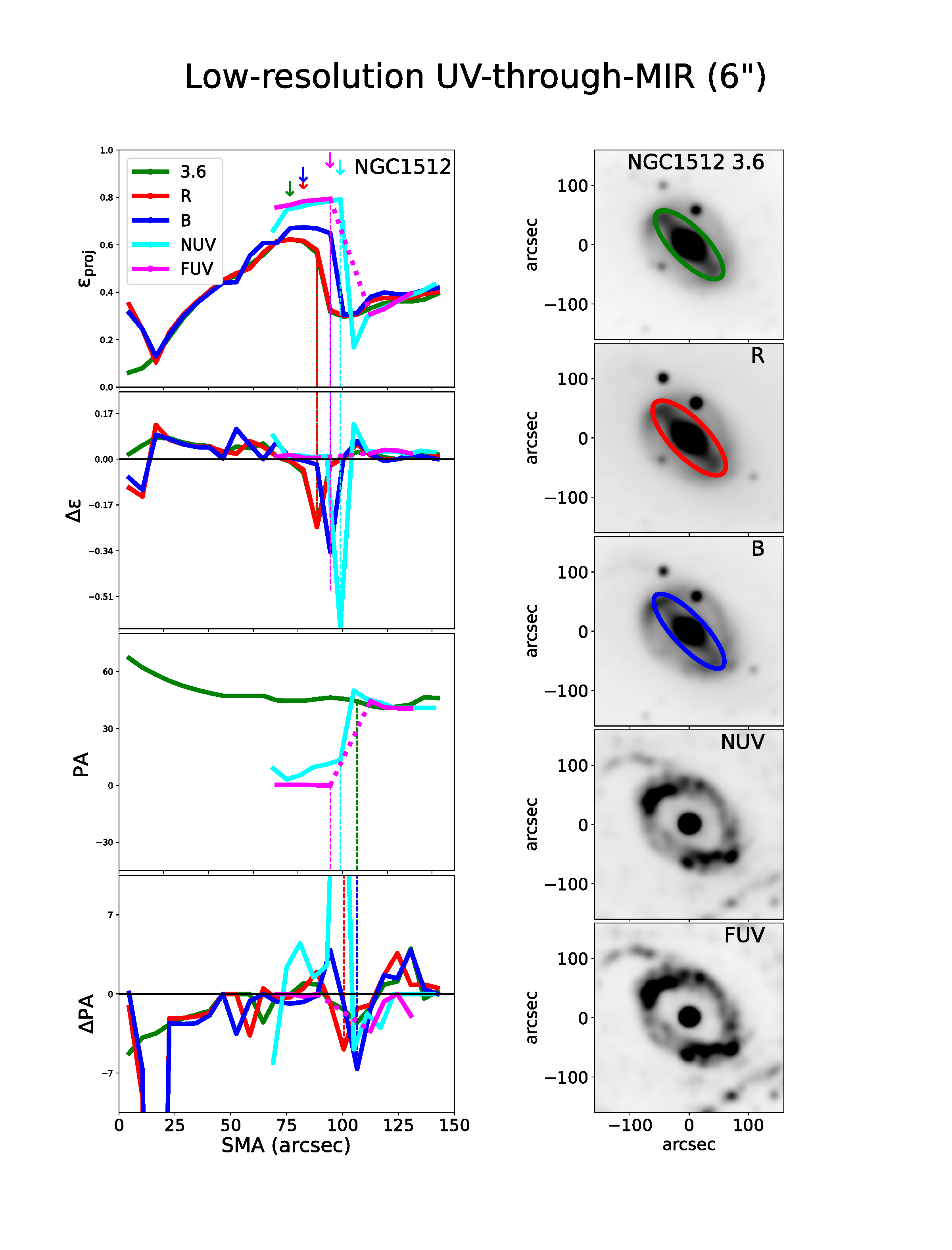}
\caption{Ellipse-fit results for the UV-through-MIR analysis.}
\label{appendixB_profiles_lores_fig}
\end{figure*}

Fig.\,\ref{appendixB_profiles_lores_fig} shows the ellipticity and PA profiles for all galaxies in our sample as part of the {\it low resolution UV-through-MIR} study, following the format of Fig.\,\ref{profiles}.


%
%

%
%
\begin{figure*}
\centering
\includegraphics[scale=0.2]{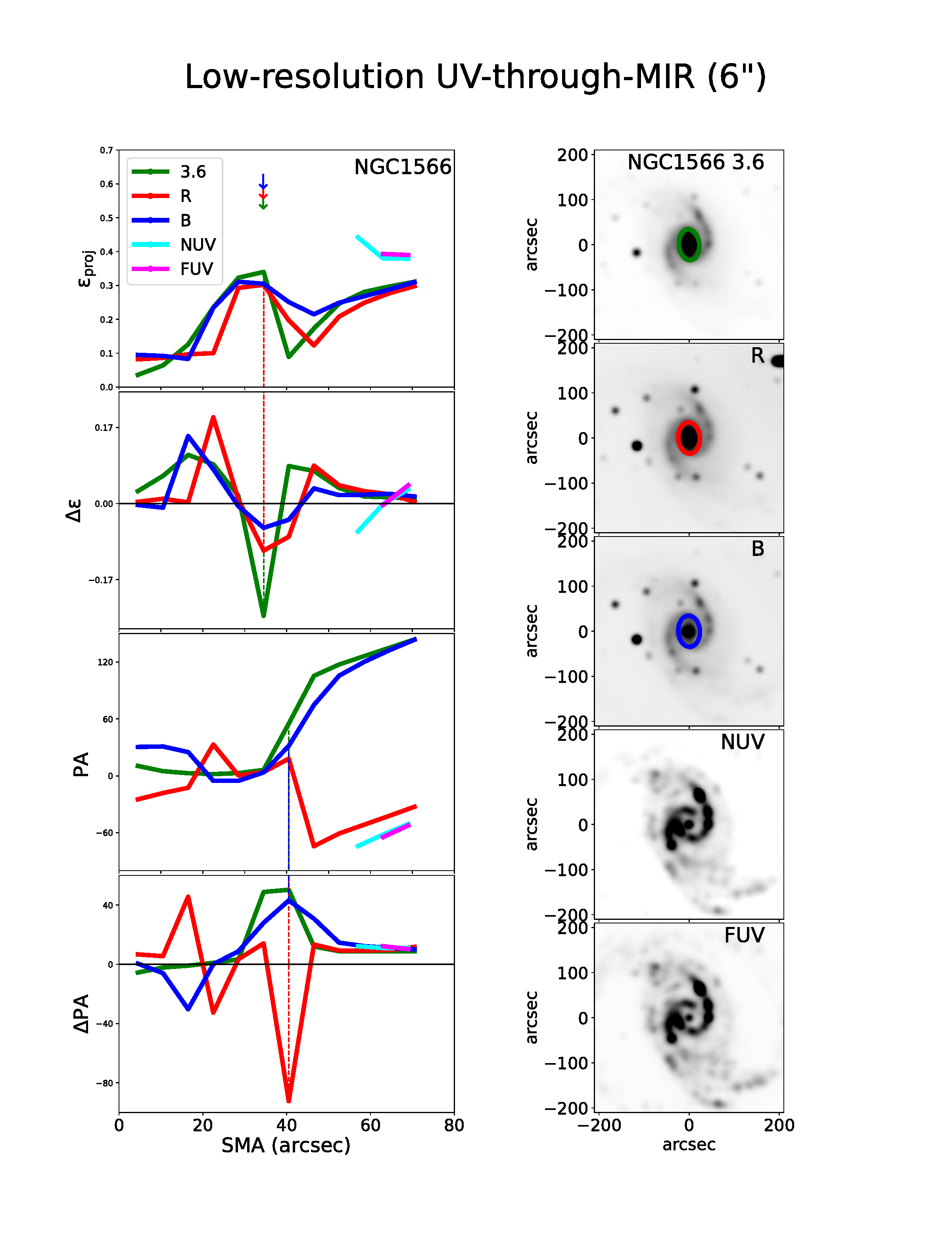}
\includegraphics[scale=0.2]{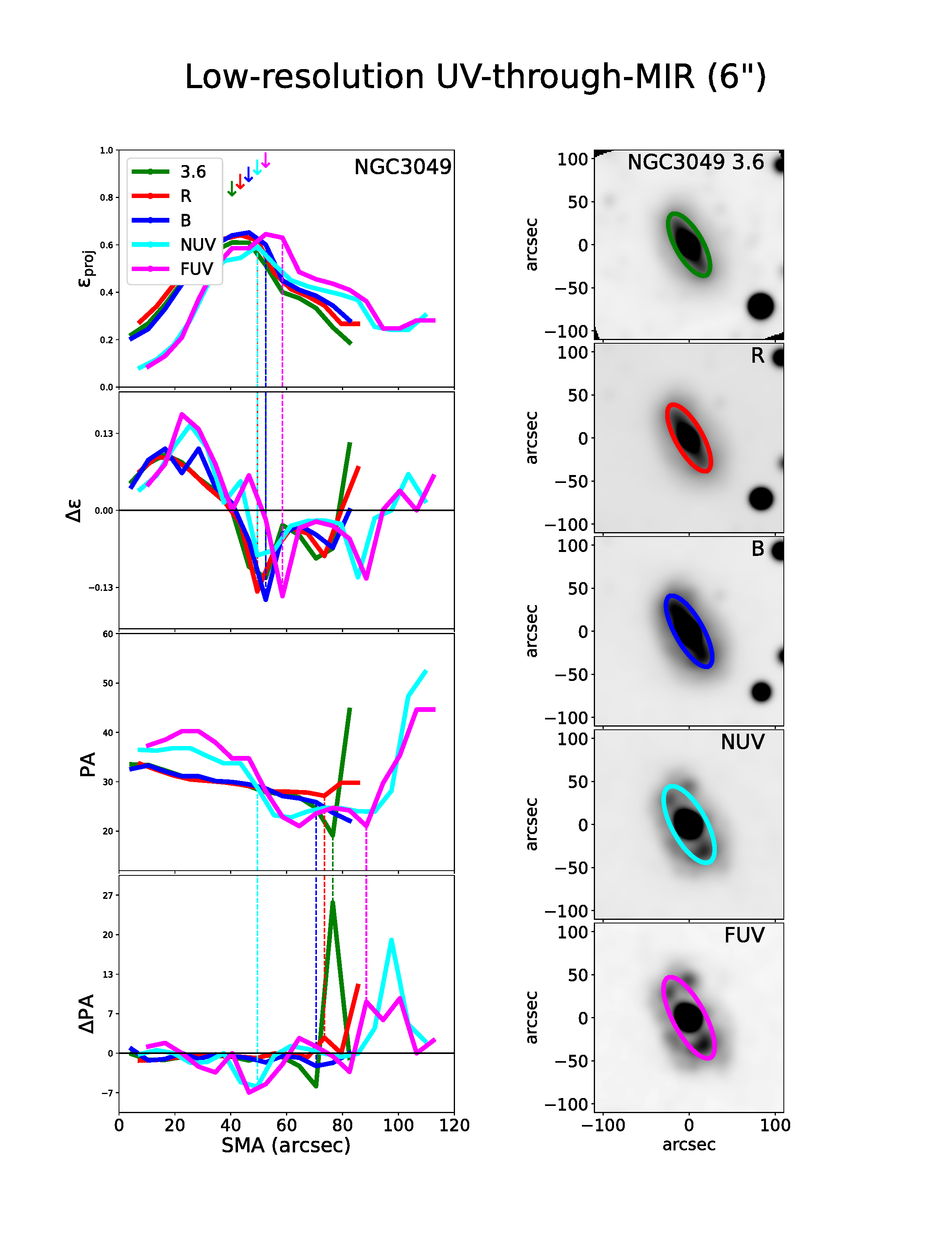}
\includegraphics[scale=0.2]{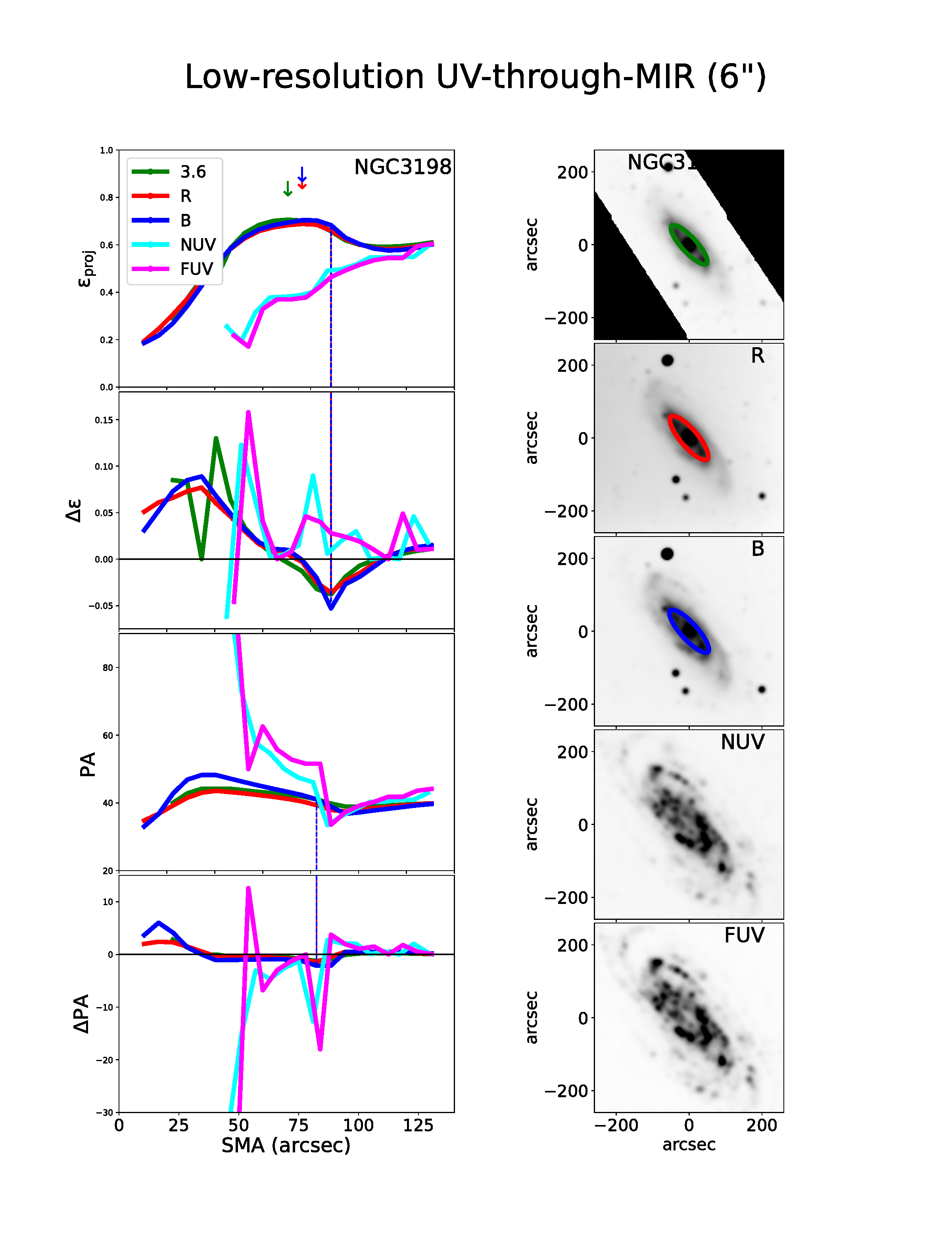}
\includegraphics[scale=0.2]{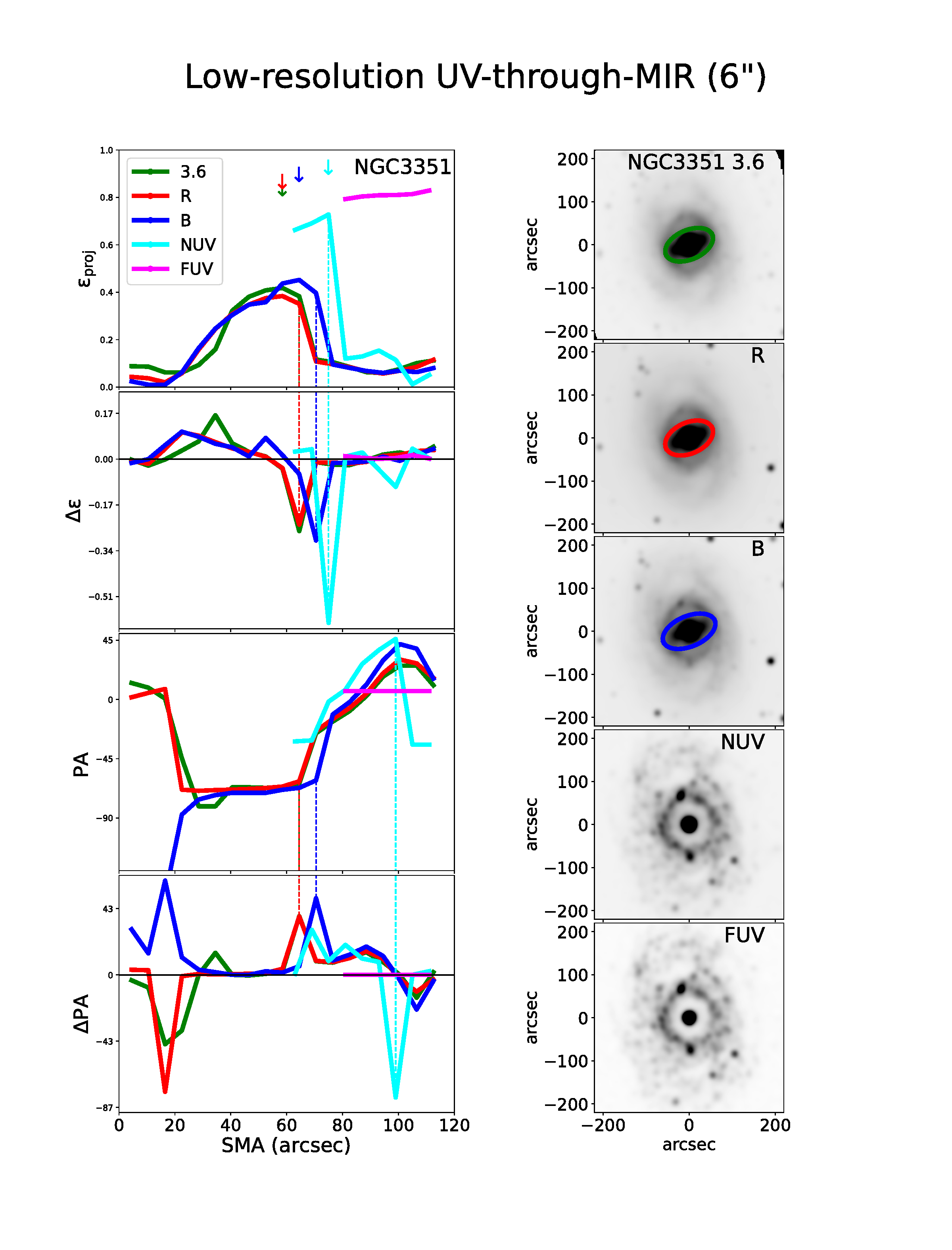}
\centerline{Fig. \ref{appendixB_profiles_lores_fig}.\ --- Continued. }
\end{figure*}

\begin{figure*}
\centering
\includegraphics[scale=0.2]{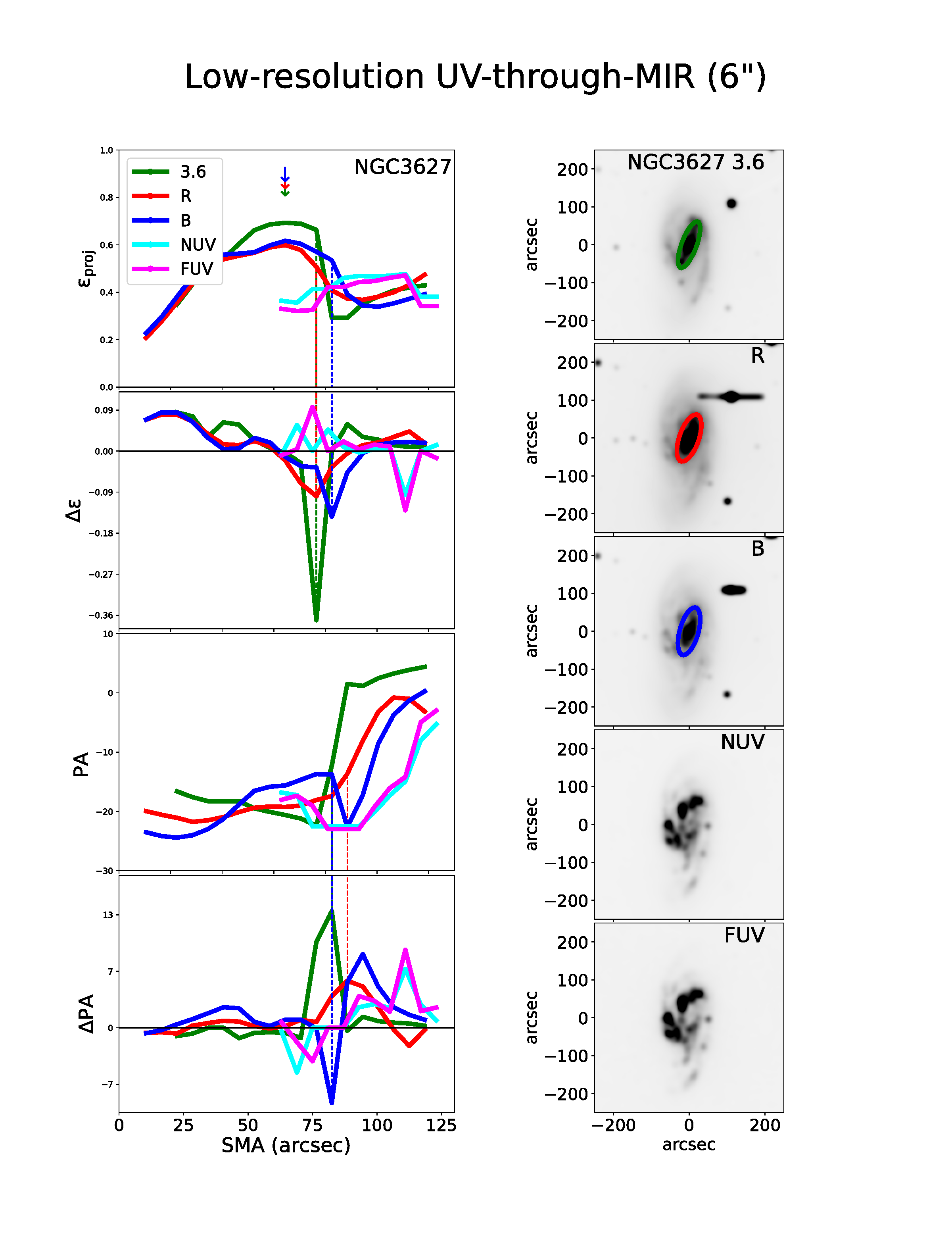}
\includegraphics[scale=0.2]{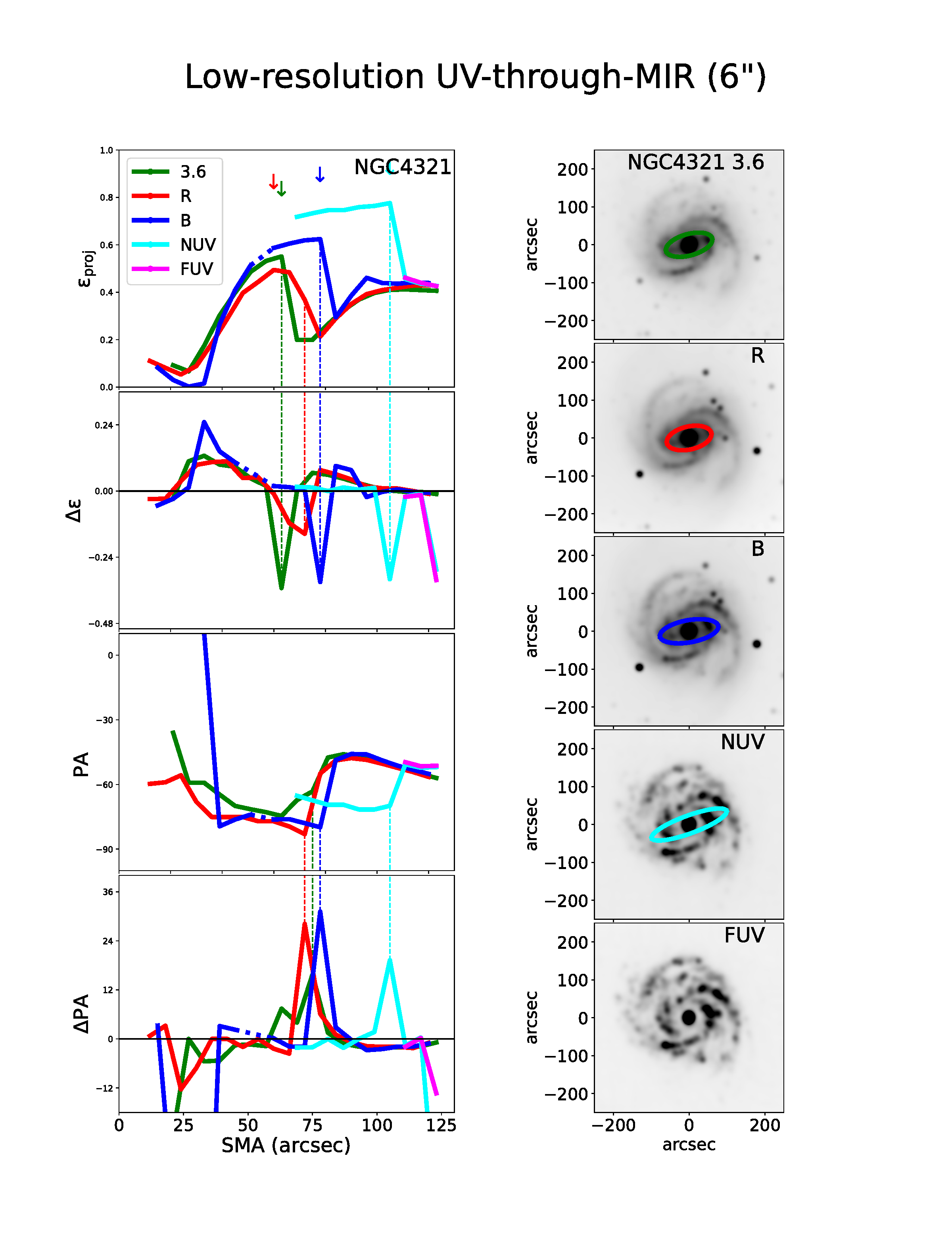}
\includegraphics[scale=0.2]{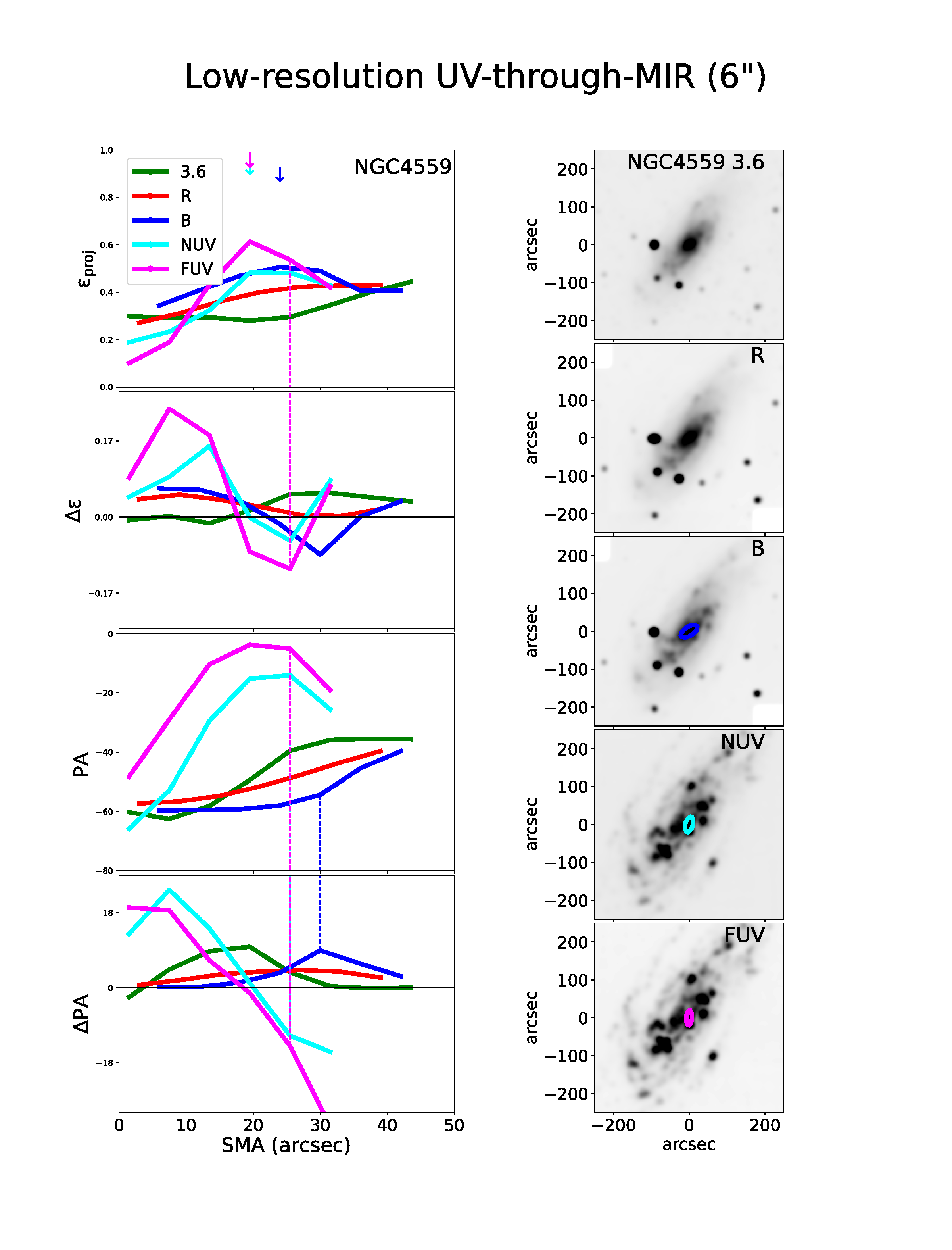}
\includegraphics[scale=0.2]{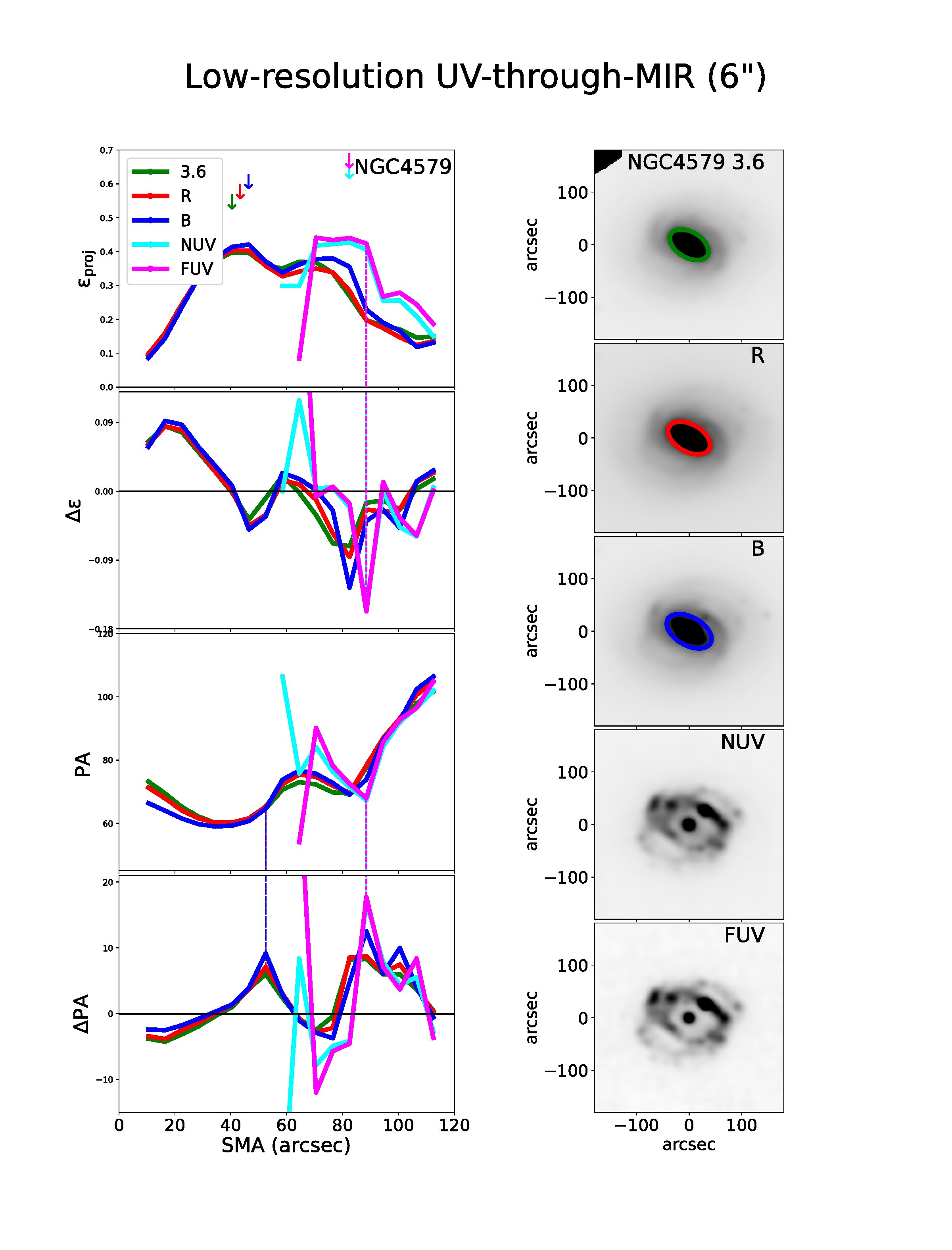}
\centerline{Fig. \ref{appendixB_profiles_lores_fig}.\ --- Continued. }
\end{figure*}

\begin{figure*}
\centering
\includegraphics[scale=0.2]{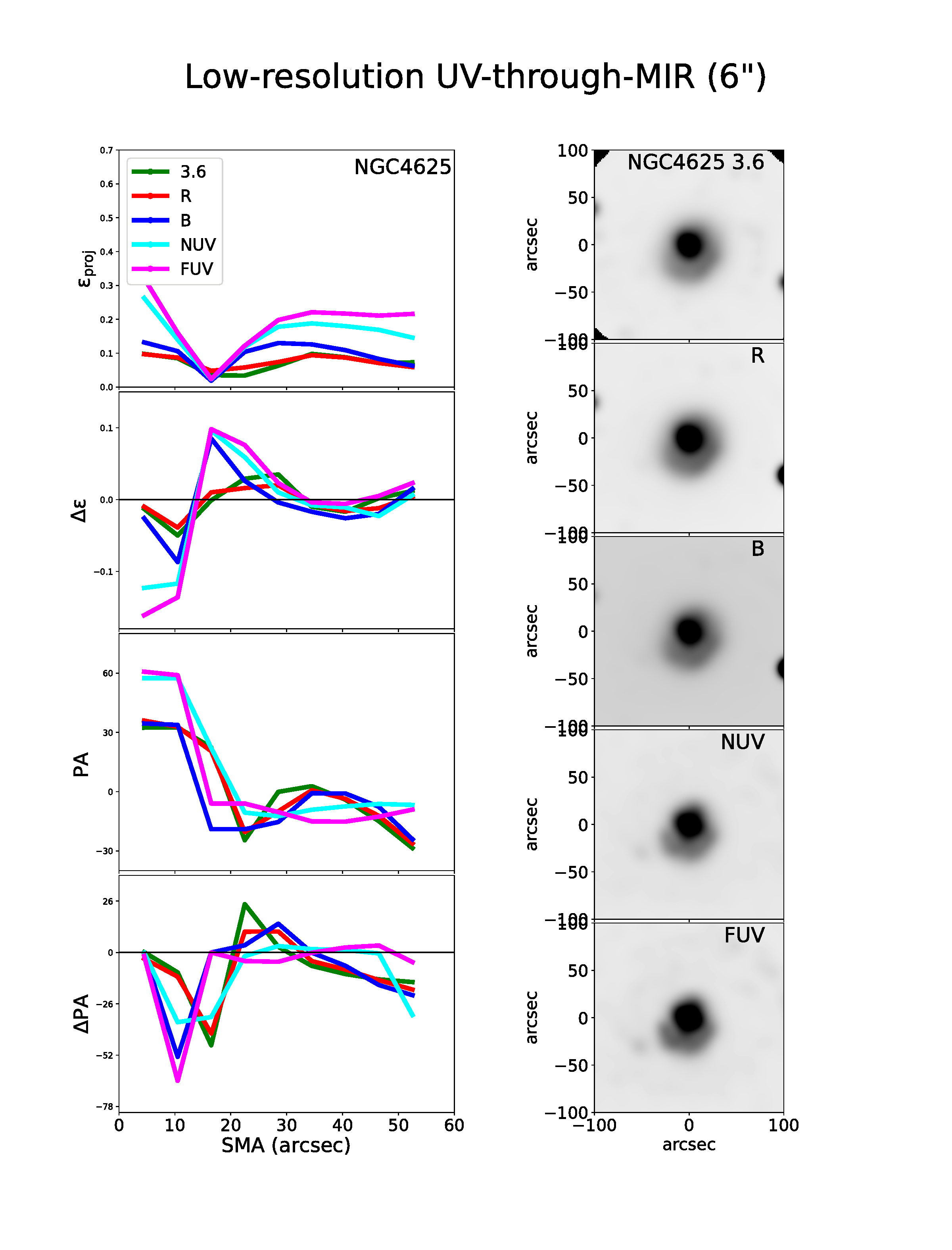}
\includegraphics[scale=0.2]{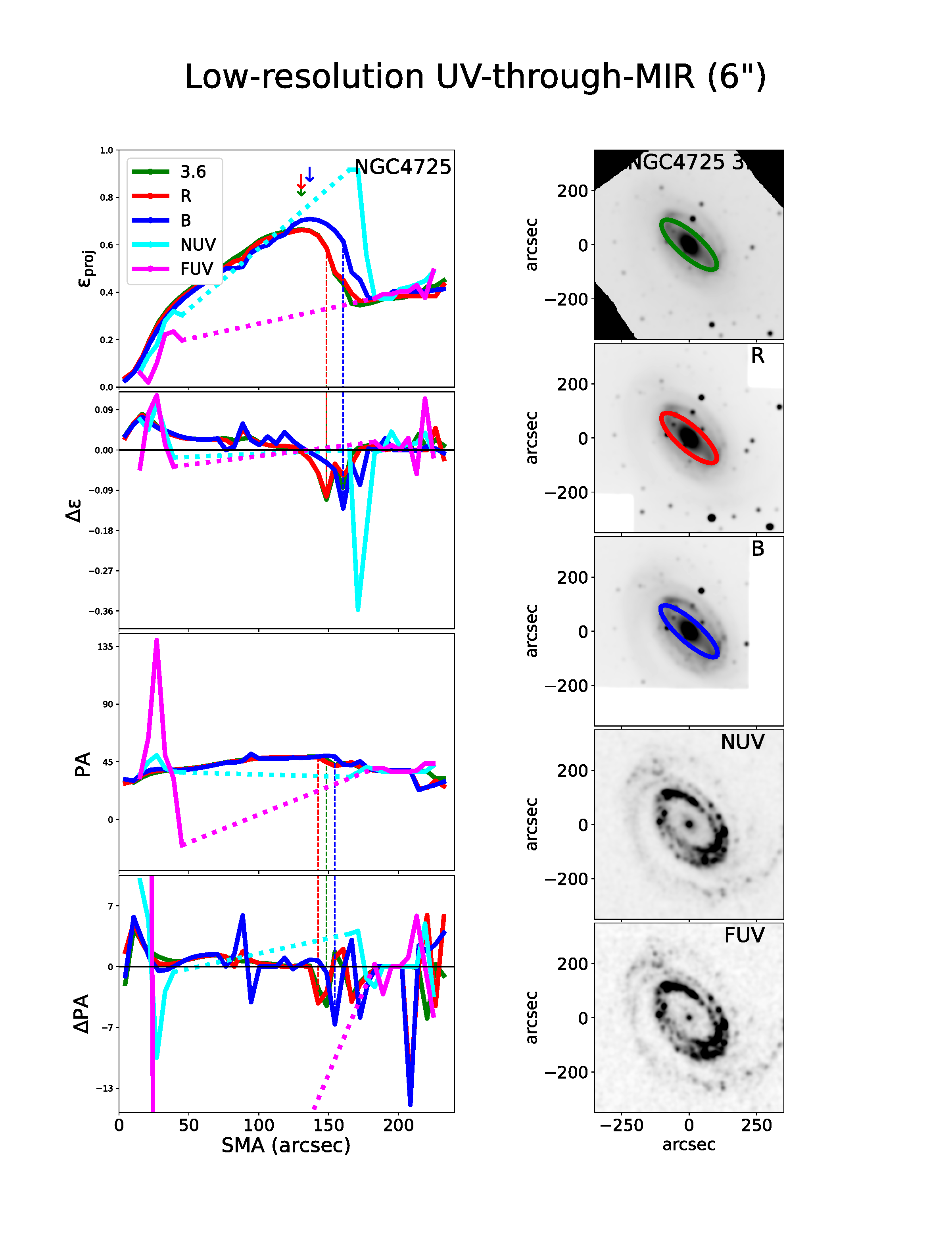}
\includegraphics[scale=0.2]{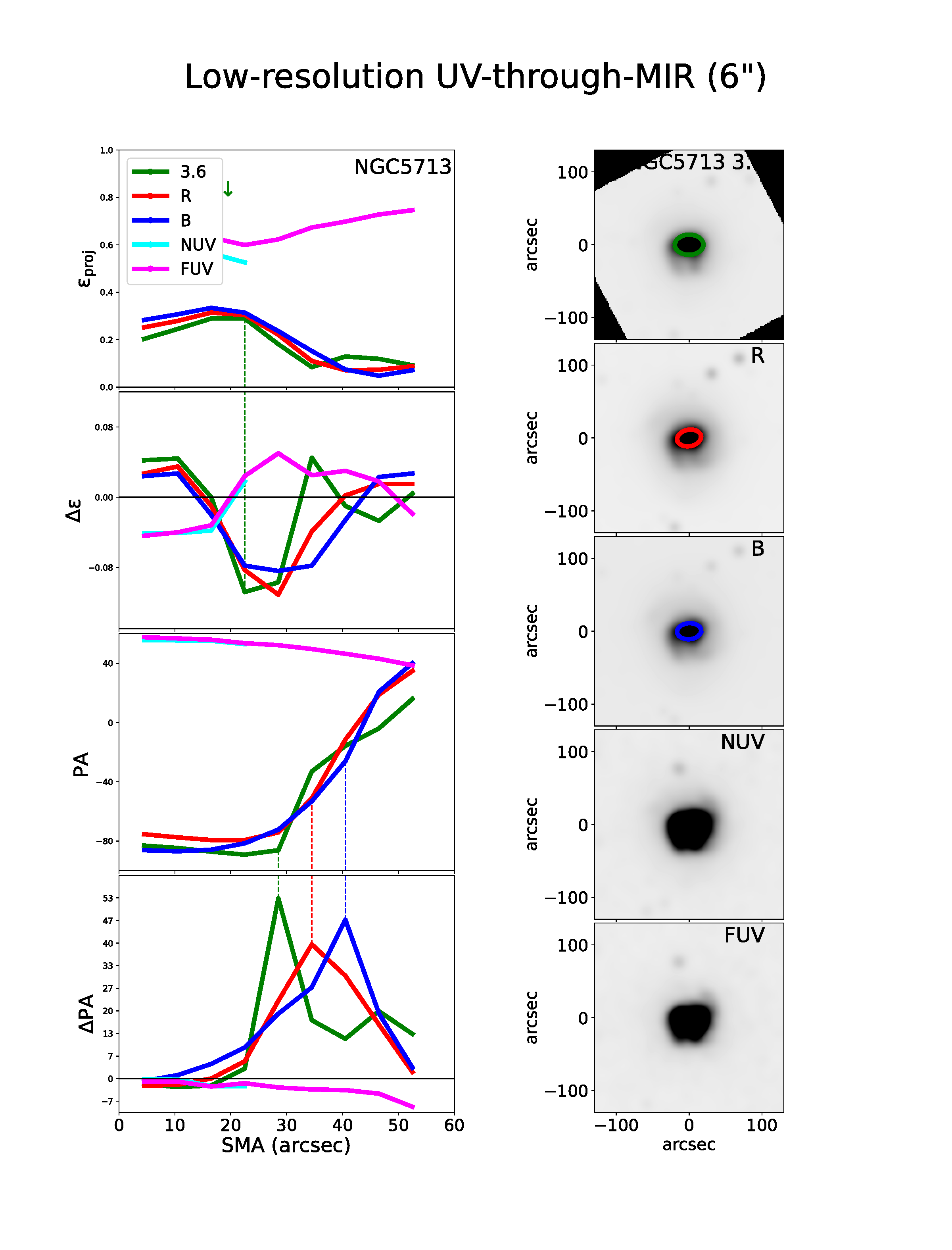}
\includegraphics[scale=0.2]{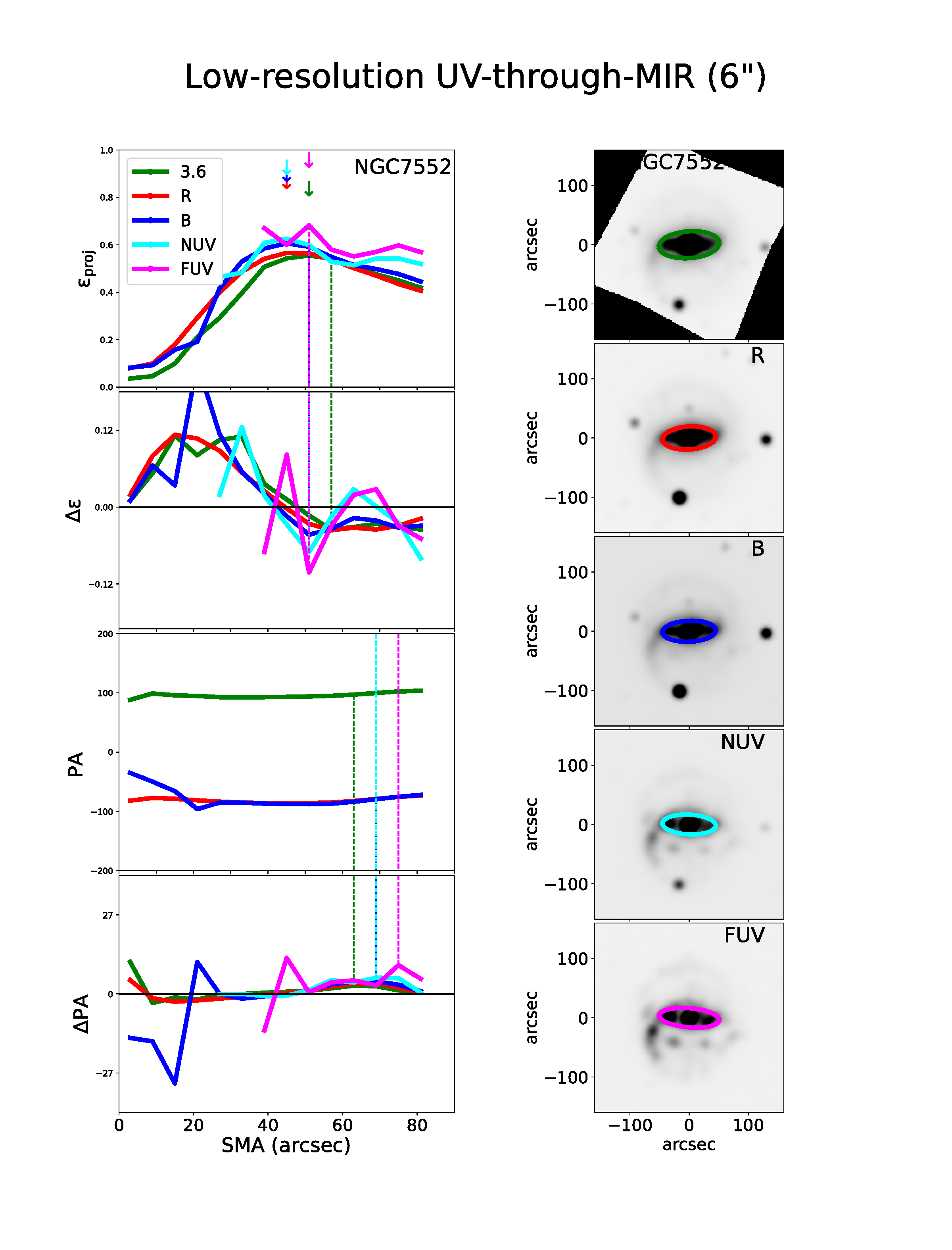}
\centerline{Fig. \ref{appendixB_profiles_lores_fig}.\ --- Continued. }
\end{figure*}

\newpage
\newpage

%
%

\section{Appendix C: Results on the Statistical analysis}
\label{appendixC_stats}

\begin{figure*}
\centering
\includegraphics[scale=0.24]{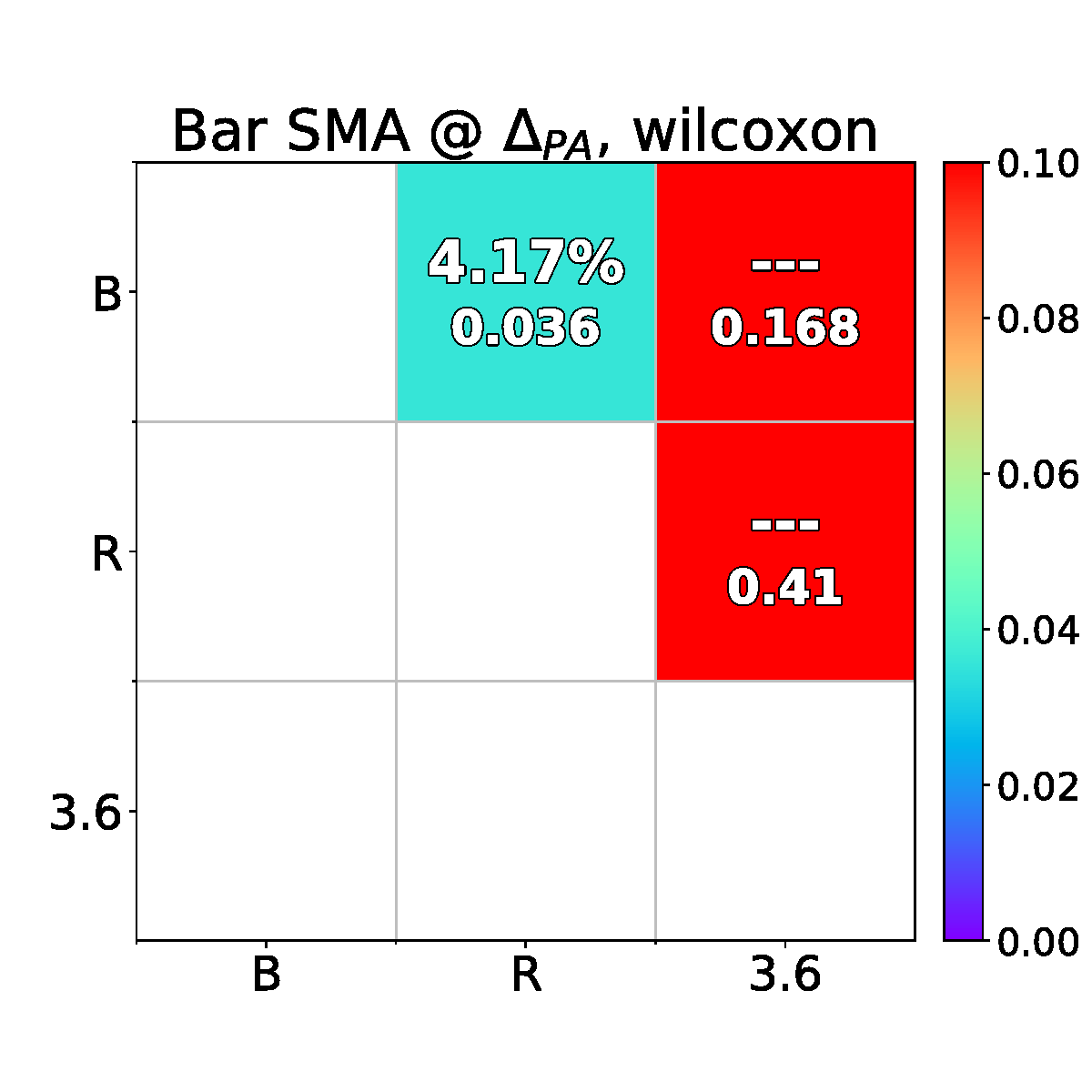}
\includegraphics[scale=0.24]{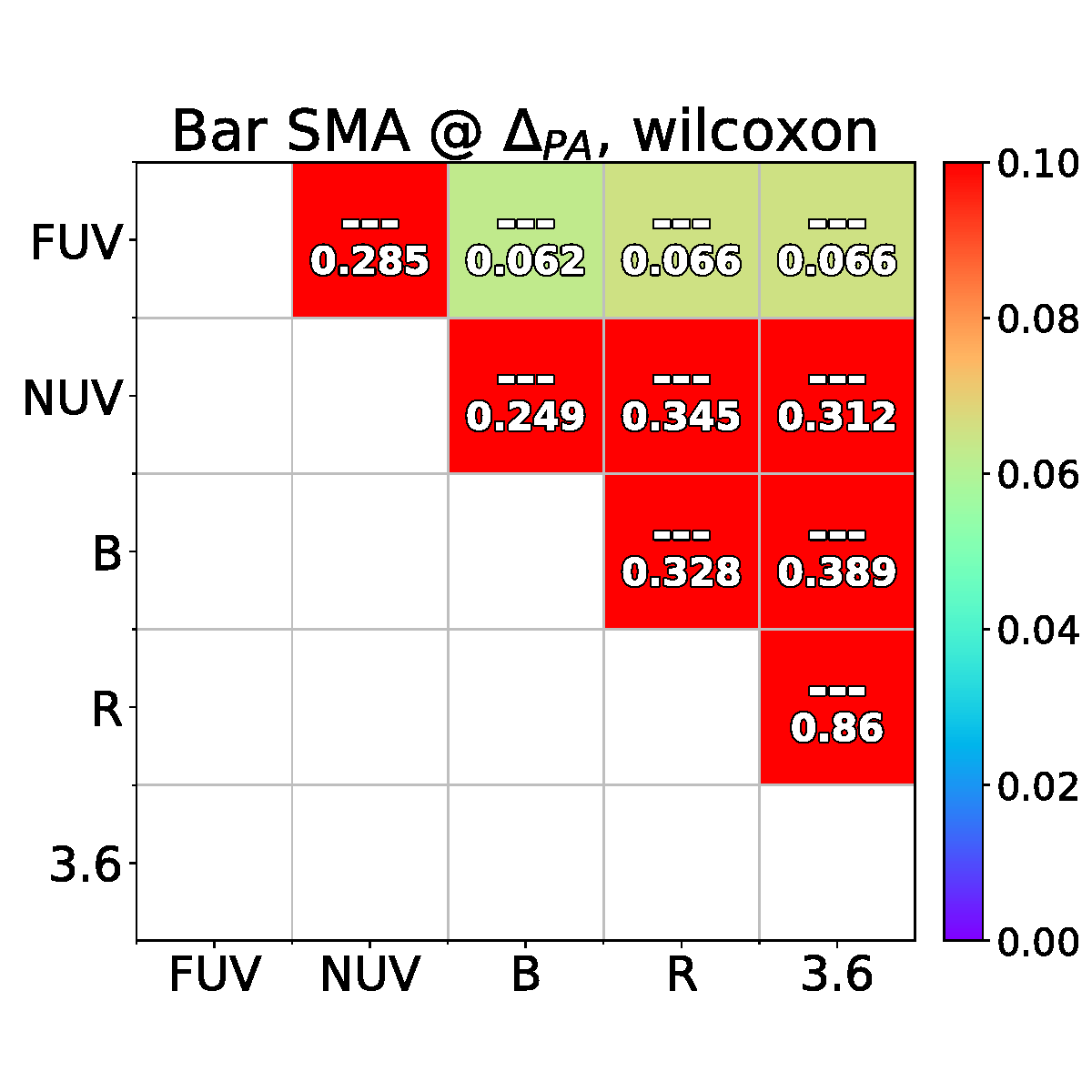}
\linebreak
\vskip -.3in
\includegraphics[scale=0.24]{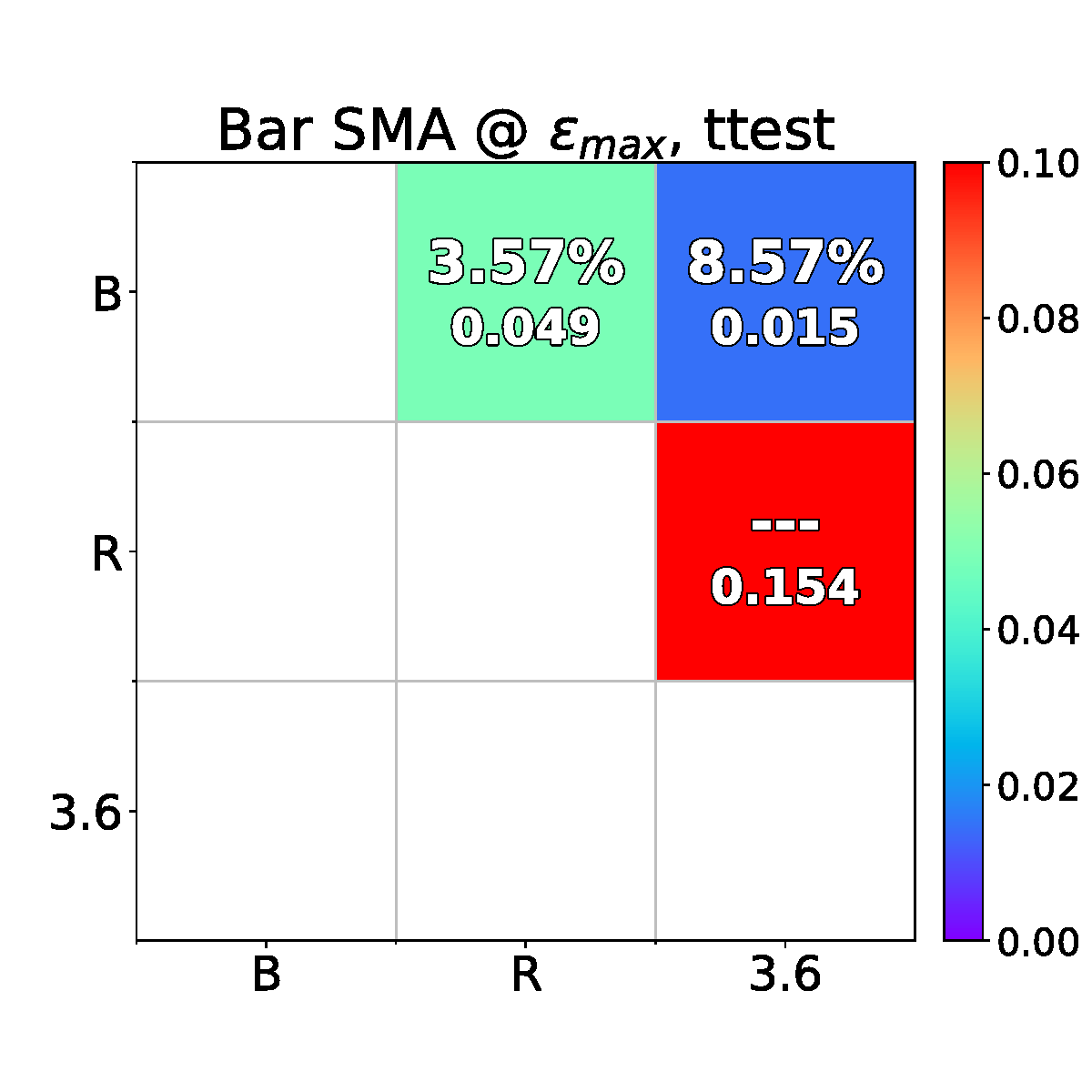}
\includegraphics[scale=0.24]{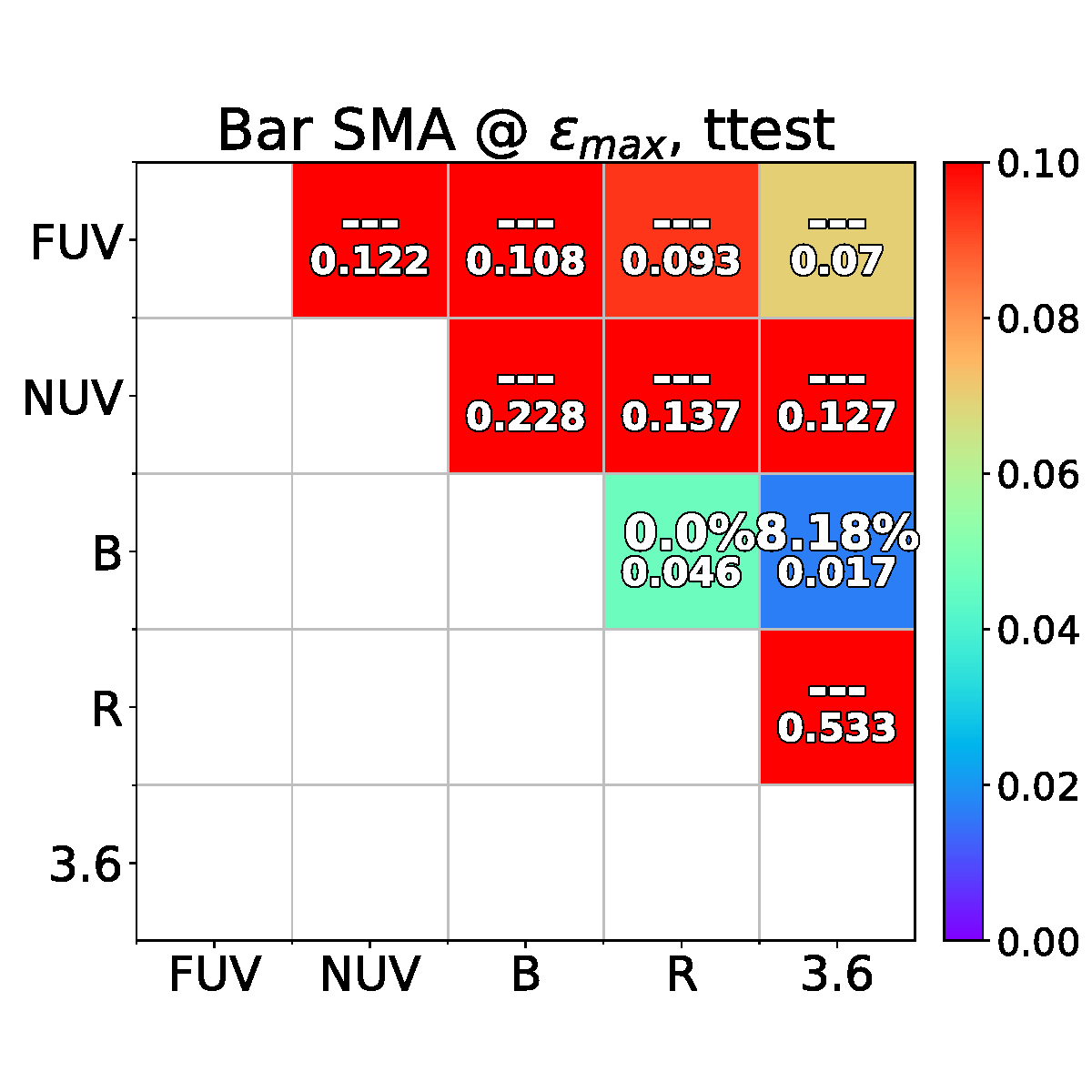}
\linebreak
\vskip -.3in
\includegraphics[scale=0.24]{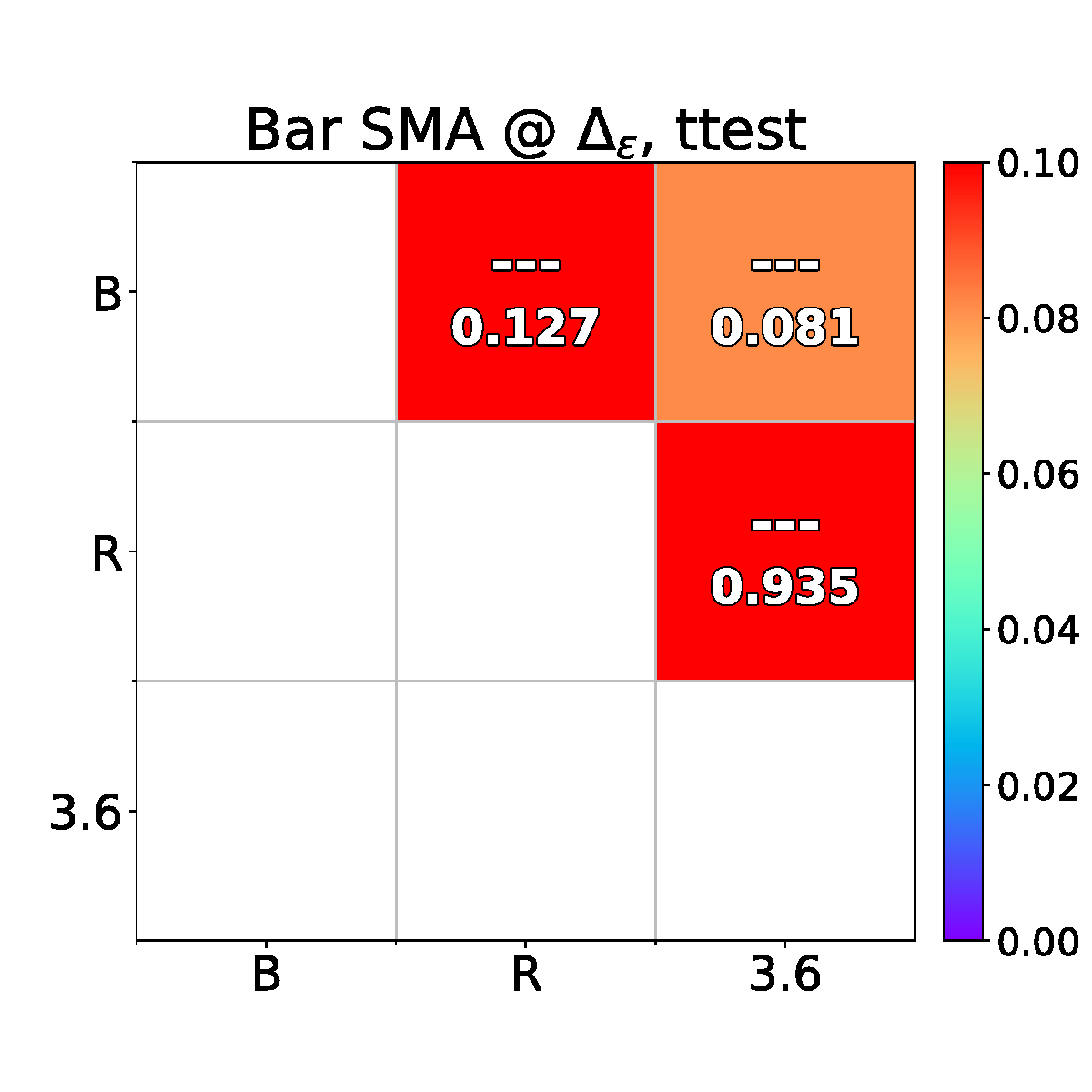}
\includegraphics[scale=0.24]{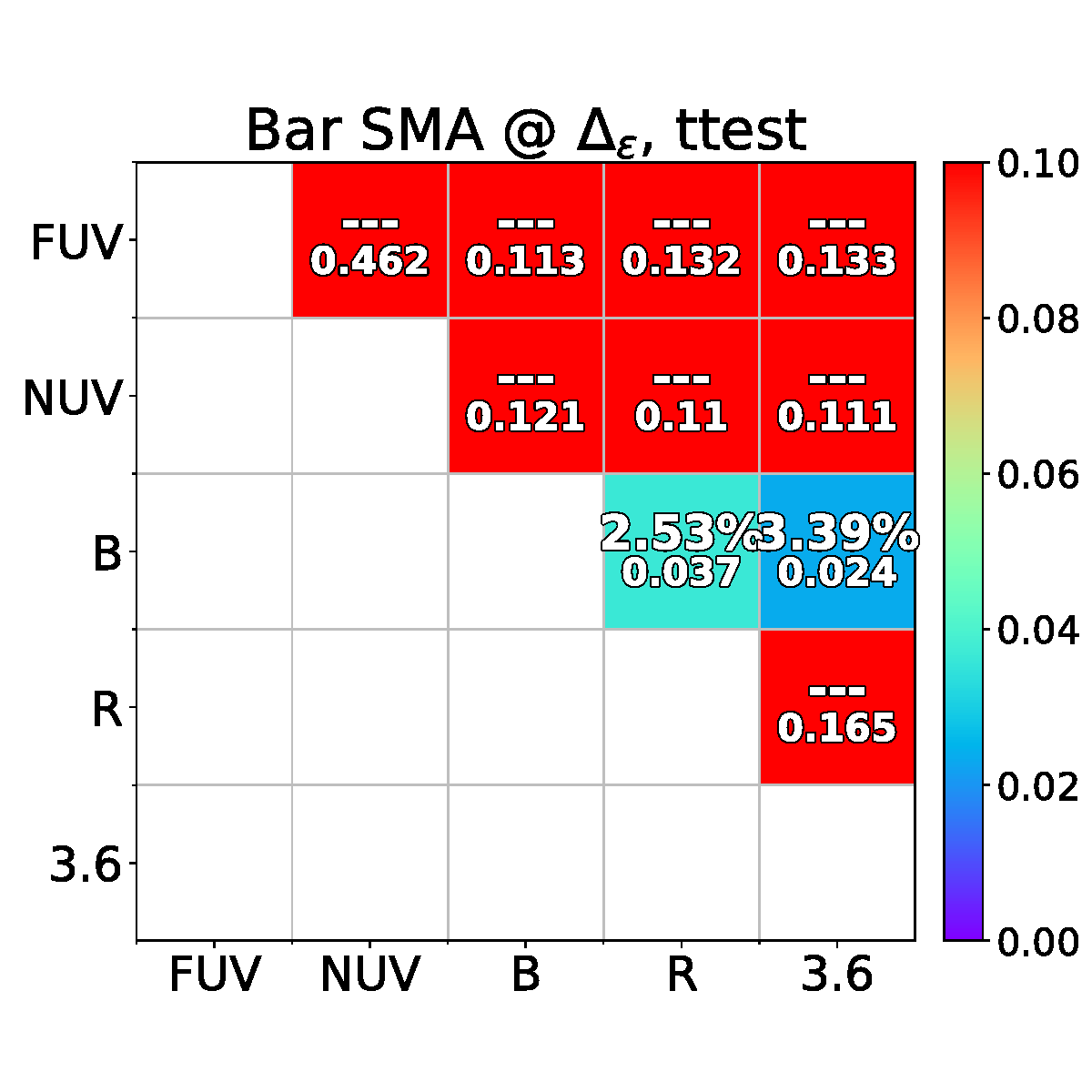}
\linebreak
\vskip -.3in
\includegraphics[scale=0.24]{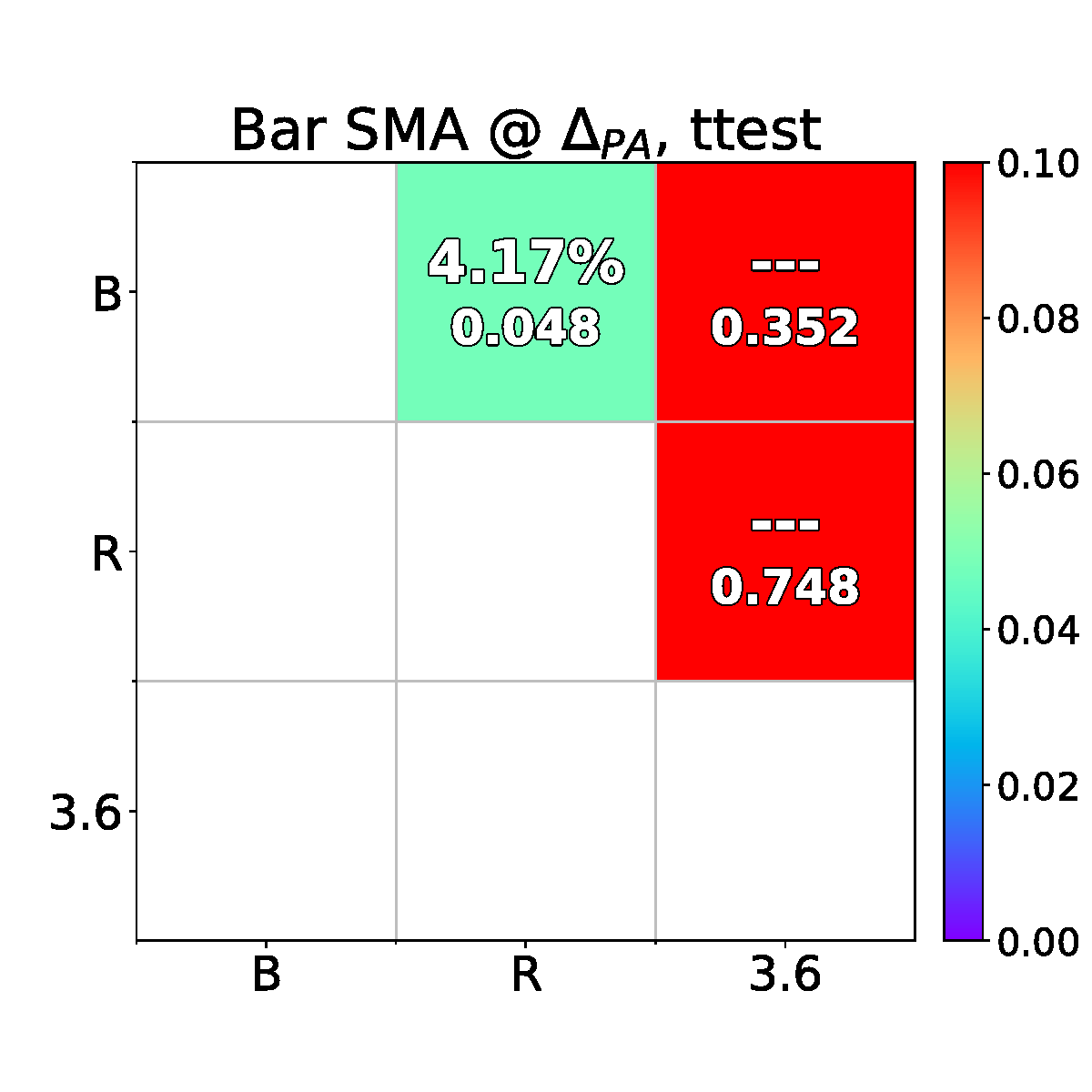}
\includegraphics[scale=0.24]{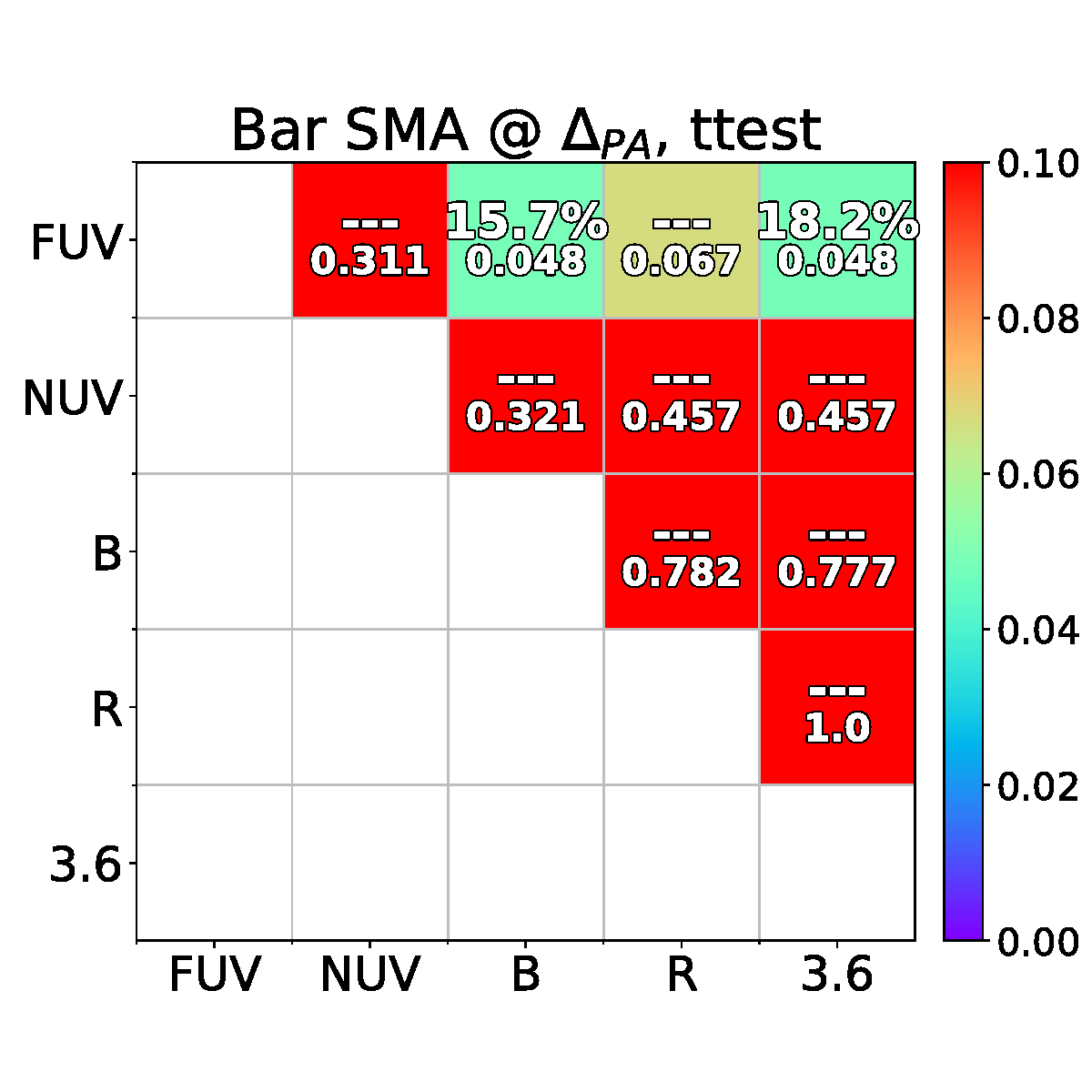}
\linebreak
\vskip -.3in
\includegraphics[scale=0.24]{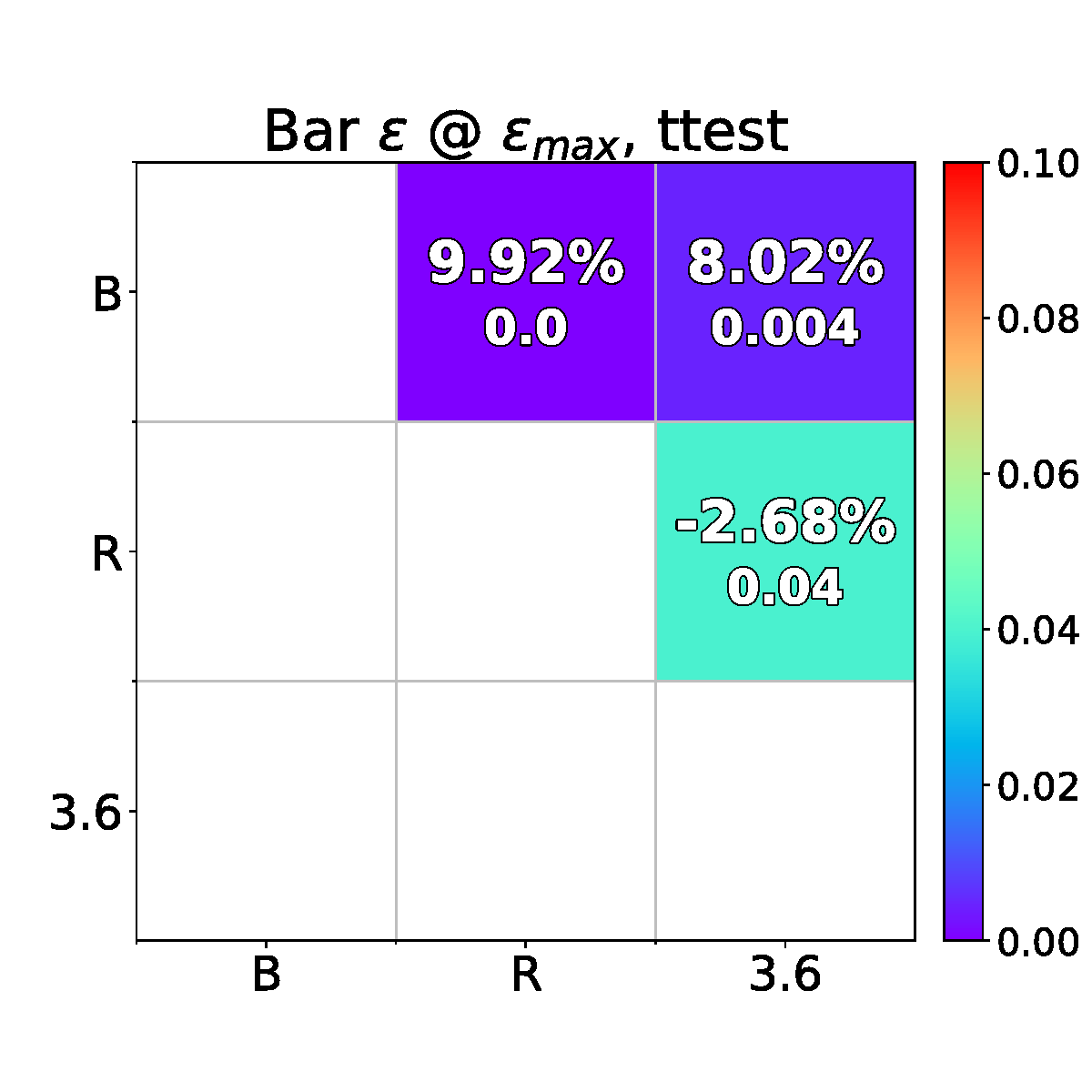}
\includegraphics[scale=0.24]{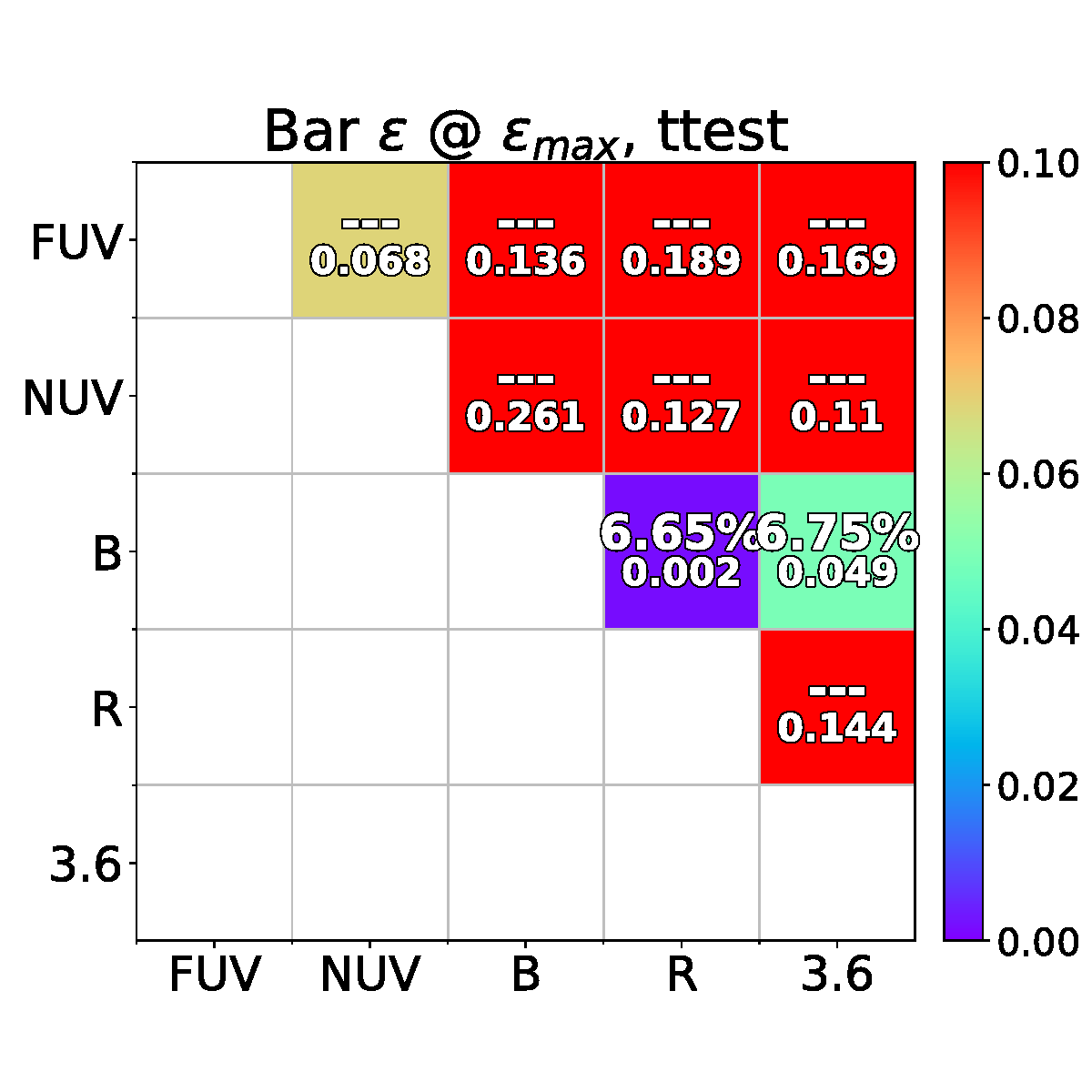}

\caption{Comparison of bar ellipticity and bar length measurements in different bands, with significance of each comparison tested with the paired t-test and the Wilcoxon test, for both the high-resolution and the low-resolution study.}
\label{appendixC_stats_fig}
\end{figure*}

We summarize in Fig.\,\ref{appendixC_stats_fig} our bar length and bar strength multi-band comparison, with the statistical significance of these results indicated by the paired t-test and Wilcoxon test for both the {\it higher-resolution optical-through-MIR} and the {\it low resolution UV-through-MIR} study. The results for the Wilcoxon test with bar lengths measured by the location of the bar ellipticity maximum ($a_{\epsilon \rm{max}}$) are presented on Tables \ref{wilcoxon_table_hires} and \ref{wilcoxon_table_lores}. We here show the two alternative methods for bar length measurement that we explored: $a_{\Delta\epsilon}$ and  $a_{\Delta \rm{PA}}$.




%
%

\newpage
\end{document}